\documentclass[printer, traditabstract]{aa}
\usepackage{setspace}

\usepackage{caption}

\usepackage[english]{babel}
\usepackage{graphicx}
\usepackage{subfigure}

\usepackage{txfonts} 
\usepackage{textcomp}
\usepackage{natbib}
\usepackage{amssymb}
\usepackage{graphicx}
\usepackage{tabularx}
\usepackage[english]{babel}
\usepackage[T1]{fontenc}
\usepackage{multirow}
\def\HI{H{\,\small I}}

\newcommand{\kms}{$\,$km$\,$s$^{-1}$}

\newcommand{\mJybeam}{mJy beam$^{-1}$}

\newcommand{\msunyr}{{${\rm M}_\odot$ yr$^{-1}$}}

\newcommand{\couno}{{CO(1-0)}}
\newcommand{\cotre}{{CO(3-2)}}
\def\HI{H{\,\small I}}
\def\HII{H{\,\small II}}

\def\emph#1{{\sl #1}}
\newcommand{\ltsima} {$\; \buildrel < \over \sim \;$}
\newcommand{\gtsima} {$\; \buildrel > \over \sim \;$}
\newcommand{\lta} {\lower.5ex\hbox{\ltsima}}
\newcommand{\gta} {\lower.5ex\hbox{\gtsima}}

\newcommand{\sauron}{{\texttt {SAURON}}}
\newcommand{\atlas}{{ATLAS$^{\rm 3D}$}}

\begin{document}
\title{The HI absorption "Zoo"}

\author{K. Ger\'{e}b $^{1,2}$, F.M. Maccagni$^{1,2}$, R. Morganti$^{2,1}$, T.A. Oosterloo$^{2,1}$}
\authorrunning{Ger\'{e}b et al.}

\institute{$^{1}$Kapteyn Astronomical Institute, University of Groningen, P.O. Box 800, 9700 AV Groningen, The Netherlands \\ $^{2}$Netherlands Institute for Radio Astronomy (ASTRON), P.O. Box 2, 7990 AA Dwingeloo, The Netherlands \\ }

%\keywords{}

\abstract{We present an analysis of the \HI\ 21 cm absorption in a sample of  101 flux-selected radio AGN (S$_{\rm{1.4 \ GHz}}$ $>$ 50 mJy) observed with the Westerbork Synthesis Radio Telescope (WSRT). 
We detect \HI\ absorption in 32 objects (30\% of the sample). In a previous paper, we performed a spectral stacking analysis on the radio sources, while here we characterize the absorption spectra of the individual detections using the recently presented busy function (Westmeier et al. 2014).

The \HI\ absorption spectra show a broad variety of widths, shapes, and kinematical properties. The full width half maximum (FWHM) of the busy function fits of the detected \HI\ lines lies in the range 32 \kms\ $<$ FWHM $<$ 570 \kms, whereas the full width at 20\% of the peak absorption (FW20) lies in the range 63 \kms\ $<$ FW20 $<$ 825 \kms. The width and asymmetry of the profiles allows us to identify three groups: narrow lines (FWHM $<$ 100 \kms), intermediate widths (100 \kms $<$ FWHM $<$ 200 \kms), and broad profiles (FWHM $>$ 200 \kms).   
We study the kinematical and radio source properties of each group, with the goal of identifying different morphological structures of \HI. Narrow lines mostly lie at the systemic velocity and are likely produced by regularly rotating \HI\ disks or gas clouds. More \HI\ disks can be present among galaxies with lines of intermediate widths; however, the \HI\ in these sources is more unsettled.

We study the asymmetry parameter and blueshift/redshift distribution of the lines as a function of their width. We find a trend for which narrow profiles are also symmetric, while broad lines are the most asymmetric. Among the broadest lines, more lines appear blueshifted than redshifted, similarly to what was found by previous studies. Interestingly, symmetric broad lines are absent from the sample. We argue that if a profile is broad, it is also asymmetric and shifted relative to the systemic velocity because it is tracing unsettled \HI\ gas. In particular, besides three of the broadest (up to FW20 = 825 \kms) detections, which are associated with gas-rich mergers, we find three new cases of profiles with blueshifted broad wings (with FW20 $\gtrsim$ 500 \kms) in high radio power AGN. 

These detections are good candidates for being HI outflows. Together with the known cases of outflows already included in the sample (3C 293 and 3C 305), the detection rate of \HI\ outflows is 5$\%$ in the total radio AGN sample. Because of the effects of spin temperature and covering factor of the outflowing gas, this fraction could represent a lower limit. However, if the relatively low detection rate is confirmed by more detailed observations, it would suggest that, if outflows are a characteristic phenomenon of all radio AGN, they would have a short depletion timescale compared to the lifetime of the radio source. This would be consistent with results found for some of the outflows traced by molecular gas.

Using stacking techniques, in our previous paper we showed that compact radio sources have higher $\tau$, FWHM, and column density than extended sources.
In addition, here we find that blueshifted and broad/asymmetric lines are more often present among compact sources. In good agreement with the results of stacking, this suggests that unsettled gas is responsible for the larger stacked FWHM detected in compact sources. Therefore in such sources the \HI\ is more likely to be unsettled. This may arise as a result of jet-cloud interactions, as young radio sources clear their way through the rich ambient gaseous medium.}

\maketitle

\section{Introduction}\label{Intro}

Nuclear activity in radio AGN is thought to be connected with the presence and kinematical properties of the gas in the circumnuclear regions. Observational evidence clearly shows that interactions between AGN and their ambient gaseous medium do occur. Thus, such interplay is thought to be responsible for the balance between the feeding of the black hole and feedback processes. Neutral hydrogen (\HI\ 21 cm, indicated as \HI\ in the rest of the paper), is one of the components that may play a role in these processes.

Radio AGN are typically hosted by early-type galaxies \citep{Bahcall, Best2005}. In the nearby Universe our knowledge of the cold gas properties of early-type galaxies has increased in recent years thanks to projects like WSRT-\sauron\ \citep{Morganti2006, Oosterloo2010} and \atlas\ \citep{Serra, Young2011, Davis2013}. In radio-loud AGN, \HI\ absorption studies can be used to explore the presence and the kinematics of the gas. A number of \HI\ absorption studies have provided a better understanding of the \HI\ properties of radio galaxies \citep{Gorkom, Morganti2001, Vermeulen2003, Gupta2006, Curran2010, Emonts}.

In these studies, the morphology and the kinematics of \HI\ gas are found to be very complex in radio galaxies. Neutral hydrogen can trace rotating disks, offset clouds, and complex morphological structures of unsettled gas, e.g., infall and outflows. Van Gorkom et al. (1989) reported a high fraction of redshifted \HI\ detections in compact radio sources, and they estimated that infalling \HI\ clouds can provide the necessary amount of gas to fuel the AGN activity. Later work revealed that not just infalling gas, but also blueshifted, outflowing \HI\ is present in some AGN, and in particular in compact gigahertz-peaked spectrum (GPS) and compact steep spectrum (CSS) sources \citep{Vermeulen2003, Gupta2006}. The structure of compact sources often appears asymmetric in brightness, location, and polarization. Such disturbed radio source properties indicate dynamical interactions between the radio jets and the circumnuclear medium, and this process is likely to be the driver of fast \HI\ outflows that has been detected in a number of radio galaxies. All these observations are consistent with a scenario in which interactions between the radio source and the surrounding gas have an effect both on the gas and on the radio source properties. It is clear that one needs to disentangle all these phenomena in order to understand the intricate interplay between AGN and the gas. 

Because AGN and their host galaxies are known to have a broad range of complex \HI\ morphologies, kinematics, gas masses, and column densities, future large datasets will require robust methods to extract and analyze meaningful information that can be relevant for our understanding of the amount and conditions of the gas. Recently, \cite{Westmeier2014} presented the busy function (BF) for parametrizing \HI\ emission spectra. The BF is efficient in fitting both Gaussian and asymmetric profiles, therefore it is also suitable for fitting the wide variety of {\sl absorption} lines in our sample. 

In this paper, we use for the first time the BF to parametrize and describe the complex \HI\ absorption properties of a relatively large sample of 32 radio sources with \HI\ detections. The total sample of 101 sources was recently presented in~\cite{katinka}, hereafter referred to as \mbox{Paper \small{I}}. The main goal of \mbox{Paper \small{I}} was to carry out a spectral stacking analysis of the \HI\ absorption lines and to measure the co-added \HI\ signal of the sample at low $\tau$ detection limit. Stacking is very efficient at reproducing the global spectral properties, but it does not provide information on the distribution of these properties among the sample. Here, we present the detailed discussion of the \HI\ absorption busy fit parameters in relation to the results of stacking.

One interesting finding of the \HI\ absorption studies presented above is that there appears to be a trend between the \HI\ properties and the evolutionary stage of the radio source. CSS and GPS sources have been proposed to represent young (\ltsima $10^{4}$ yr) radio AGN \citep{Fanti1995, Readhead1996, Owsianik1998}. The high \HI\ detection rate in compact CSS and GPS sources has been interpreted as evidence for a relation between the recent triggering of the AGN activity and the presence of \HI\ gas \citep{Pihlstrom, Gupta2006, Emonts, Chandola}.

In \mbox{Paper \small{I}} we looked at the \HI\ properties of compact and extended sources using stacking techniques. We found that compact sources have higher detection rate and optical depth, and also larger profile width than extended sources. We argue that such \HI\ properties reflect the presence of a rich gaseous medium in compact sources, and that the larger FWHM of compact sources is due to the presence of unsettled gas. In the present paper, we use the BF to measure the \HI\ parameters of individual detections in compact and extended sources. We discuss these measurements in relation to the results of stacking from \mbox{Paper \small{I}}. 

Several examples from the literature show that \HI\ mass outflow rates of a few $\times 10$ M$_\odot$ yr$^{-1}$ are associated with fast ($\sim 1000$ \kms) jet-driven outflows \citep{Morganti2005, Kanekar2008, Morganti2013a, Tadhunter2014}, therefore such feedback effects are considered to have a major impact both on the star formation processes in galaxies and the further growth of the black hole. However, at the moment little is known about the frequency and lifetime of such \HI\ outflows in radio galaxies, and larger samples are needed to constrain the role and significance of outflows in the evolution of galaxies.
We have not found signatures of broad, blueshifted wings in the stacked spectra presented in \mbox{Paper \small{I}}, although this is likely due to the small size of the sample. Here, we use the busy fit parameters to identify and characterize new cases of \HI\ outflows.

In this paper the standard cosmological model is used, with parameters $\Omega_{m}$ = 0.3, $\Omega_{\Lambda}$ = 0.7 and $H_0$ = 70 km s$^{-1}$ Mpc$^{-1}$. 

\section{Description of the sample and observations}

As described in \mbox{Paper \small{I}}, the sample was selected from the cross-correlation of the Sloan Digital Sky Survey (SDSS, York et al. 2000) and Faint Images of the Radio Sky at 20 cm (FIRST, Becker et al. 1995) catalogs. In the redshift range $0.02 < z < 0.23$, 101 sources were selected with peak flux $S_{\rm{1.4 \ GHz}}$ $>$ 50 mJy in the FIRST catalog. The corresponding radio power distribution of the AGN lies in the range 10$^{23}$ -- 10$^{26}$ W Hz$^{-1}$.

The observations were carried out with the Westerbork Synthesis Radio Telescope (WSRT). Each target was observed for 4 hours. In the case of 4C +52.37, we carried out 8 hour follow-up observations in order to increase the \HI\ sensitivity in the spectra. This will be  discussed in Sec. \ref{Outflows}. A more detailed description of the observational setup and the data reduction can be found in \mbox{Paper \small{I}}. 

Because our sample is solely flux-selected, we can expect to have a mix of radio sources with various types of host galaxies. In Table \ref{table:detections} we summarize the characteristics of the detected sources. Each source is given an identification number, which is used throughout the paper. The radio galaxy sample consists of compact (CSS, GPS, and unclassified) and extended sources. The size of the radio sources varies between 4 pc and 550 kpc. Besides radio galaxies, we also find optically blue objects with $g - r$ $<$ 0.7 colors. These blue objects are associated with different types of radio sources, for example gas-rich mergers (\mbox{UGC 05101}, \mbox{UGC 8387}, and \mbox{Mrk 273}, no. 7, no. 19, and no. 22 respectively), Seyfert galaxies (no. 21) and QSOs (Quasi Stellar Objects, no. 14). To make the AGN sample homogeneous in the selection of early-type radio galaxies, in \mbox{Paper \small{I}} we excluded these sources from the stacking analysis. In this paper, these objects are excluded from the overall analysis of the sample and are separately discussed in Sec. 4.3.

In \mbox{Paper \small{I}}, we have separated the sample in compact and extended radio sources based on the NRAO VLA Sky Survey (NVSS) major-to-minor axis ratio vs.\ the FIRST peak-to-integrated flux ratio. Compact sources are defined as having NVSS major-to-minor axis ratio $<$ 1.1 and FIRST peak-to-integrated flux ratio $>$ 0.9. Most of the extended sources have NVSS major-to-minor axis ratio $> 1.1$ and FIRST peak-to-integrated flux ratio $<$ 0.9. The same classification is used here.

\begin{table*} 
\caption{
 Characteristics of the \HI\ detections. For each detected source we list: 1) identifier that will be used in the text; 2) sky coordinates; 3) alternative name of the sources; 4) SDSS redshift; 5) 1.4 GHz flux density; 6) 1.4 GHz radio power; 7) radio morphology; 8) compact extended classification; 9) peak optical depth; 10) column density. The radio morphology abbreviations in column 7 are as follows:  CSO: Compact Symmetric Object,  CSS: compact steep spectrum source, CJ: Core-Jet, CX: complex morphology, U: unresolved, QSO: quasar, FSRQ: Flat Spectrum Radio Quasar. While in column 8, the radio sources are classified as follows: C: compact, E: extended, M: merger or blue galaxy}

  \begin{center}
  
\scalebox{0.75}{  
    \begin{tabular}{l c c c c c c c c c c c c}

         \hline

no. & RA, Dec &    Name     &  z          & S$_{\rm{1.4 \ GHz}} $ & P$_{\rm{1.4 \ GHz}}$  &  Radio  & Compact & $\tau_{peak}$ & N(\HI)  \\        
                          &   & &               & mJy              & W Hz$^{-1}$ &            Morphology    & Extended  &  &$ 10^{18}$ ($T_{\rm{spin}}/c_{\rm{f}}$) cm$^{-2}$ \\ \\
\hline	

1	&	 07h57m56.7s +39d59m36s	&	B3 0754+401	&	0.066	&	92	&	23.98	&	CSS	&	C	&	0.042	&	9.4	\\
2	&	 08h06m01.5s +19d06m15s	&	 2MASX J08060148+1906142	&	0.098	&	142	&	24.54	&	-	&	E	&	0.099	&	26.9	\\
3	&	08h09m38.9s +34d55m37s	&	B2 0806+35	&	0.082	&	142	&	24.38	&	CJ	&	E	&	0.009	&	1	\\
4	&	08h36m37.8s +44d01m10s	&	B3 0833+442	&	0.055	&	134	&	23.99	&	CSO?	&	C	&	0.016	&	1.9	\\
5	&	08h43m07.1s +45d37m43s	&	B3 0839+458	&	0.192	&	331	&	25.54	&	CSO	&	C	&	0.273	&	34.5	\\
6	&	09h09m37.4s +19d28m08s	&	Mrk 1226	&	0.028	&	63	&	23.05	&	FSRQ?	&	C	&	0.119	&	23.3	\\
7	&	09h35m51.6s +61d21m11s	&	UGC 05101	&	0.039	&	148	&	23.73	&	-	&	M	&	0.073	&	53.6	\\
8	&	 10h20m53.7s +48d31m24s	&	4C +48.29	&	0.053	&	82	&	23.74	&	X-shaped	&	E	&	0.05	&	9.3	\\
9	&	 10h53m27.2s +20d58m36s	&	J105327+205835	&	0.052	&	79	&	23.72	&	-	&	C	&	0.023	&	3.9	\\
10	&	11h20m30.0s +27d36m11s	&	2MASX J112030+273610	&	0.112	&	177	&	24.76	&	-	&	C	&	0.147	&	15.6	\\
11	&	12h02m31.1s +16d37m42s	&	2MASX J12023112+1637414	&	0.119	&	82	&	24.48	&	-	&	C	&	0.042	&	7.4	\\
12	&	12h05m51.4s +20d31m19s	&	NGC 4093 - MCG +04-29-02	&	0.024	&	80	&	23.01	&	-	&	C	&	0.034	&	5.2	\\
13	&	12h08m55.6s +46d41m14s	&	B3 1206+469	&	0.101	&	69	&	24.25	&	-	&	E	&	0.052	&	2.6	\\
14	&	12h32m00.5s +33d17m48s	&	 B2 1229+33	&	0.079	&	94	&	24.16	&	FR II	&	M	&	0.034	&	5.6	\\
15	&	 12h47m07.3s +49d00m18s	&	4C +49.25	&	0.207	&	1140	&	26.15	&	CSS	&	C	&	0.002	&	0.5	\\
16	&	12h54m33.3s +18d56m02s	&	2MASX J125433+185602	&	0.115	&	76	&	24.42	&	CSO	&	C	&	0.068	&	6.3	\\
17	&	13h01m32.6s +46d34m03s	&	2MASX J13013264+4634032	&	0.206	&	97	&	25.07	&	-	&	C	&	0.018	&	3.2	\\
18	&	13h17m39.2s +41d15m46s	&	B3 1315+415	&	0.066	&	246	&	24.42	&	CX	&	C	&	0.031	&	7	\\
19	&	13h20m35.3s +34d08m22s	&	IC 883, UGC 8387	&	0.023	&	97	&	23.07	&	-	&	M	&	0.162	&	78.8	\\
20	&	13h25m13.4s +39d55m53s	&	SDSS J132513.37+395553.2	&	0.076	&	37	&	23.71	&	-	&	C	&	0.053	&	9.3	\\
21	&	13h40m35.2s +44d48m17s	&	IRAS F13384+4503	&	0.065	&	36	&	23.57	&	CJ	&	M	&	0.26	&	20.3	\\
22	&	13h44m42.1s +55d53m13s	&	Mrk 273	&	0.037	&	132	&	23.63	&	-	&	M	&	0.091	&	86.2	\\
23	&	 13h52m17.8s +31d26m46s	&	3C 293	&	0.045	&	3530	&	25.23	&	FR I	&	E	&	0.057	&	14.5	\\
24	&	14h22m10.8s +21d05m54s	&	2MASX J142210+210554	&	0.191	&	84	&	24.94	&	-	&	C	&	0.048	&	10.1	\\
25	&	 14h35m21.7s +50d51m23s	&	 2MASX J14352162+5051233	&	0.099	&	141	&	24.55	&	U	&	C	&	0.013	&	4.7	\\
26	&	14h49m21.6s +63d16m14s	&	3C 305 - IC 1065	&	0.042	&	2500	&	25.01	&	FR I	&	E	&	0.005	&	1.5	\\
27	&	15h00m34.6s +36d48m45s	&	2MASX J150034+364845	&	0.066	&	61	&	23.81	&	QSO	&	C	&	0.19	&	35.3	\\
28	&	15h29m22.5s +36d21m42s	&	2MASX J15292250+3621423	&	0.099	&	38	&	23.97	&	-	&	C	&	0.075	&	25.1	\\
29	&	16h02m46.4s +52d43m58s	&	4C +52.37	&	0.106	&	577	&	25.22	&	CSO	&	C	&	0.015	&	6.7	\\
30	&	16h03m32.1s +17d11m55s	&	NGC 6034	&	0.034	&	278	&	23.87	&	-	&	E	&	0.066	&	5.6	\\
31	&	16h03m38.0s +15d54m02s	&	 Abell 2147	&	0.109	&	100	&	24.49	&	FSRQ	&	C	&	0.125	&	50.7	\\
32	&	16h12m17.6s +28d25m47s	&	2MASX J161217+282546	&	0.053	&	78	&	23.72	&	-	&	C	&	0.061	&	7.8	\\

\end{tabular} }
\label{table:detections}	
\end{center}	 
\end{table*} 		

\begin{figure}
\begin{center}
\includegraphics[trim = 20 175 20 200, clip,width=.5\textwidth]{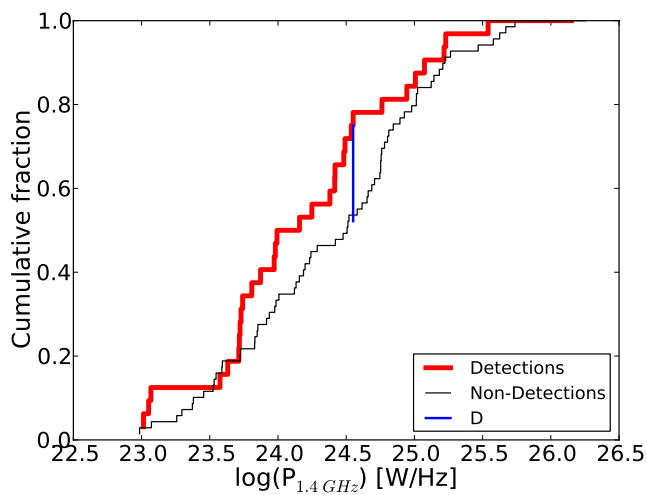}
\caption{The 1.4 GHz radio power cumulative fraction of detections and non-detections in the sample of 101 AGN. We measure D = 0.228 for 32 detections and 69 non-detections. The source parameters for detection and non-detections are listed in Table \ref{table:detections} and Table \ref{table:non-detections}.}\label{fig:KS_RadioPower}
\end{center}
\end{figure}
%\noindent

\section{Results}\label{results}

\begin{figure*}
\begin{center}
		\includegraphics[trim = 0 0 30 30, clip,width=.33\textwidth]{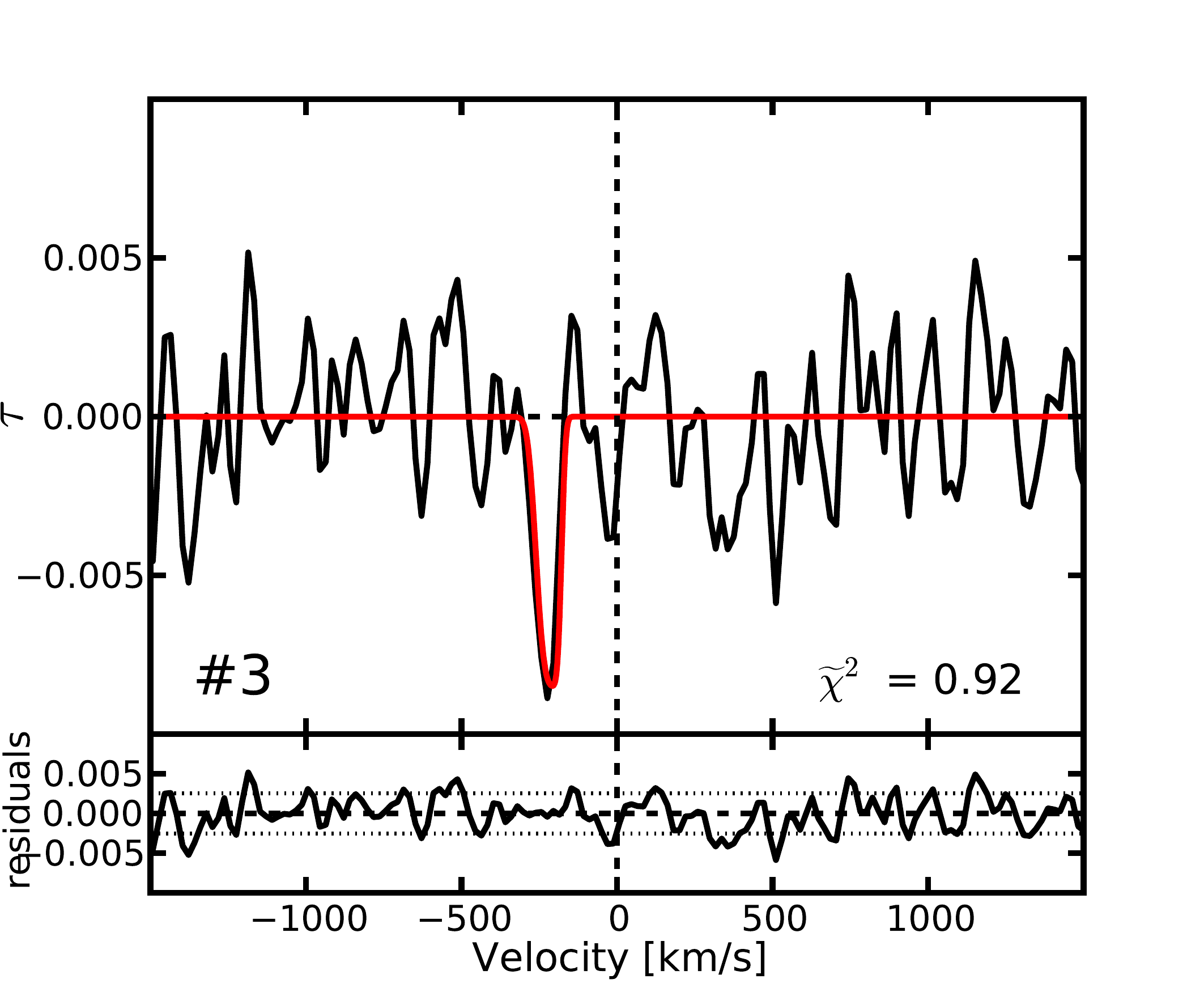}
		\includegraphics[trim = 0 0 30 30, clip,width=.33\textwidth]{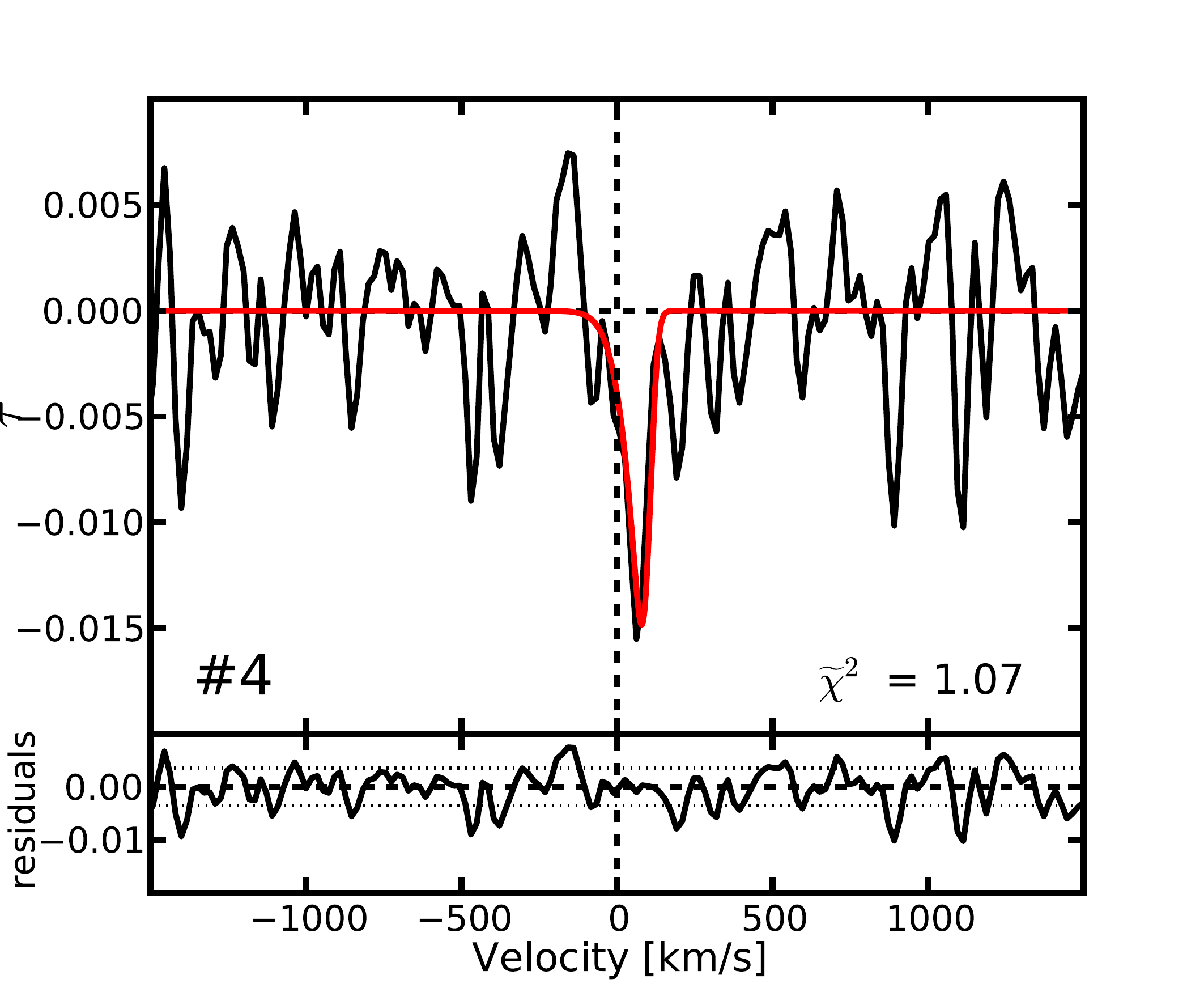}
		\includegraphics[trim = 0 0 30 30, clip,width=.33\textwidth]{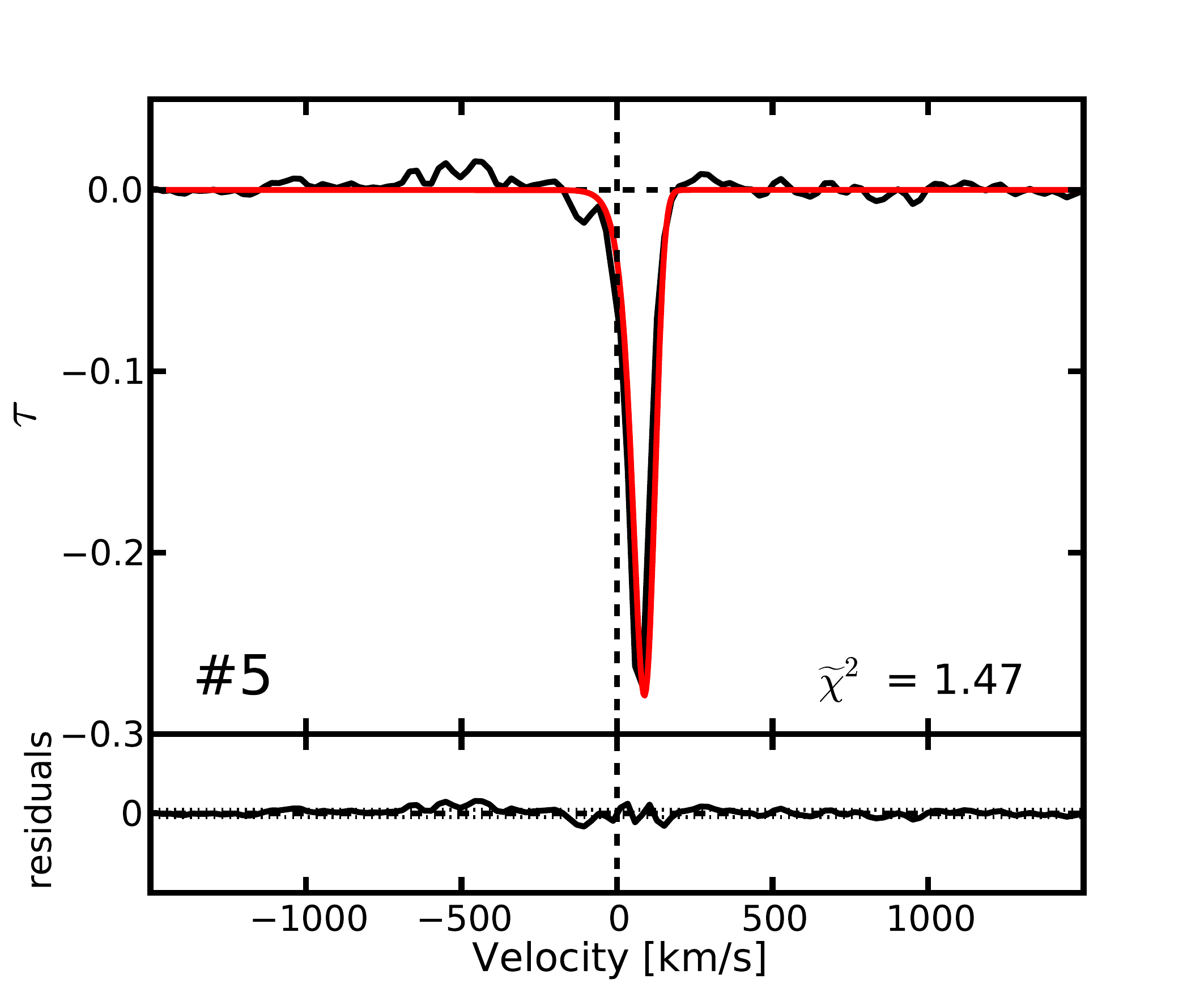}
		\includegraphics[trim = 0 0 30 30, clip,width=.33\textwidth]{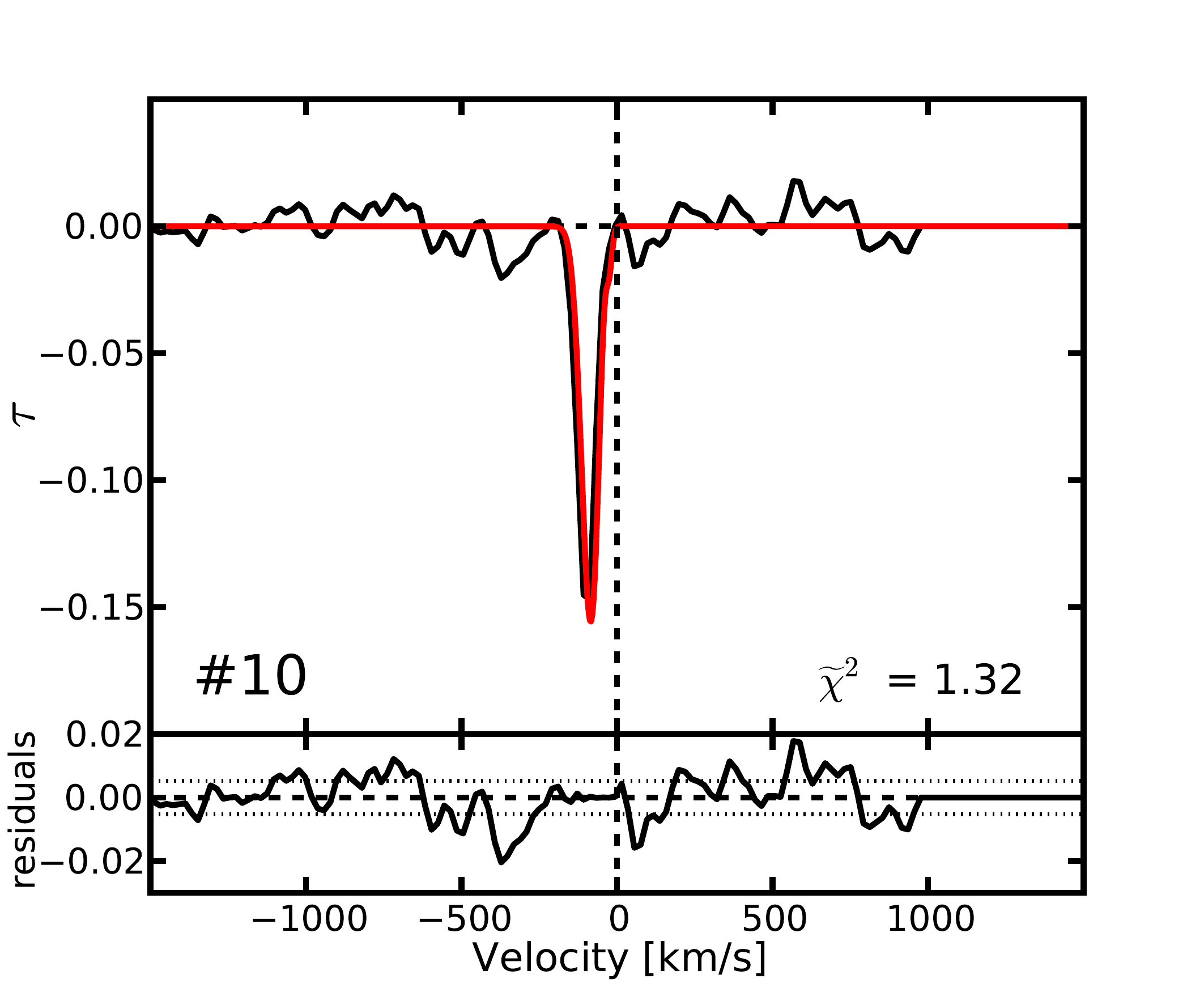}
		\includegraphics[trim = 0 0 30 30, clip,width=.33\textwidth]{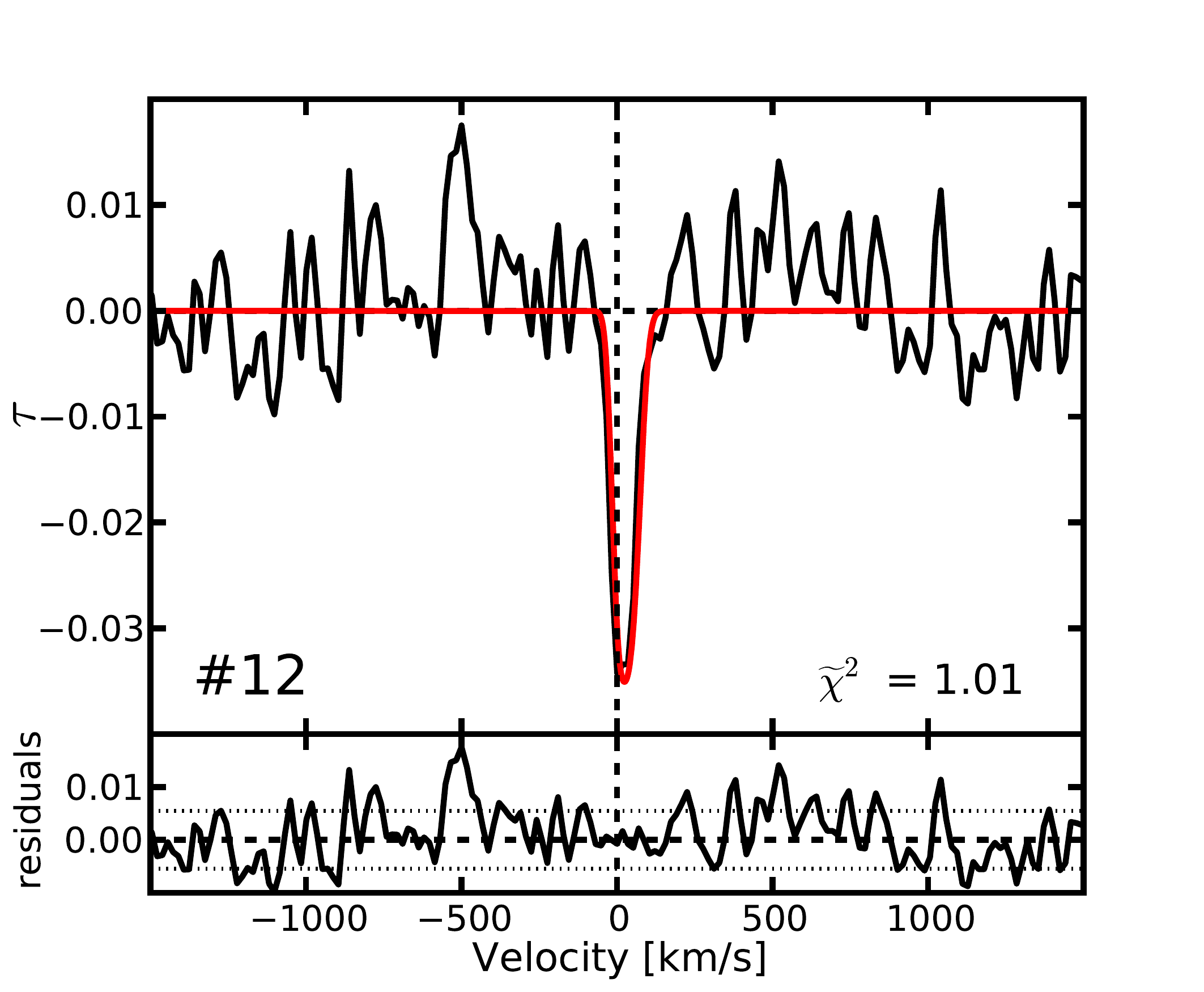}		
		\includegraphics[trim = 0 0 30 30, clip,width=.33\textwidth]{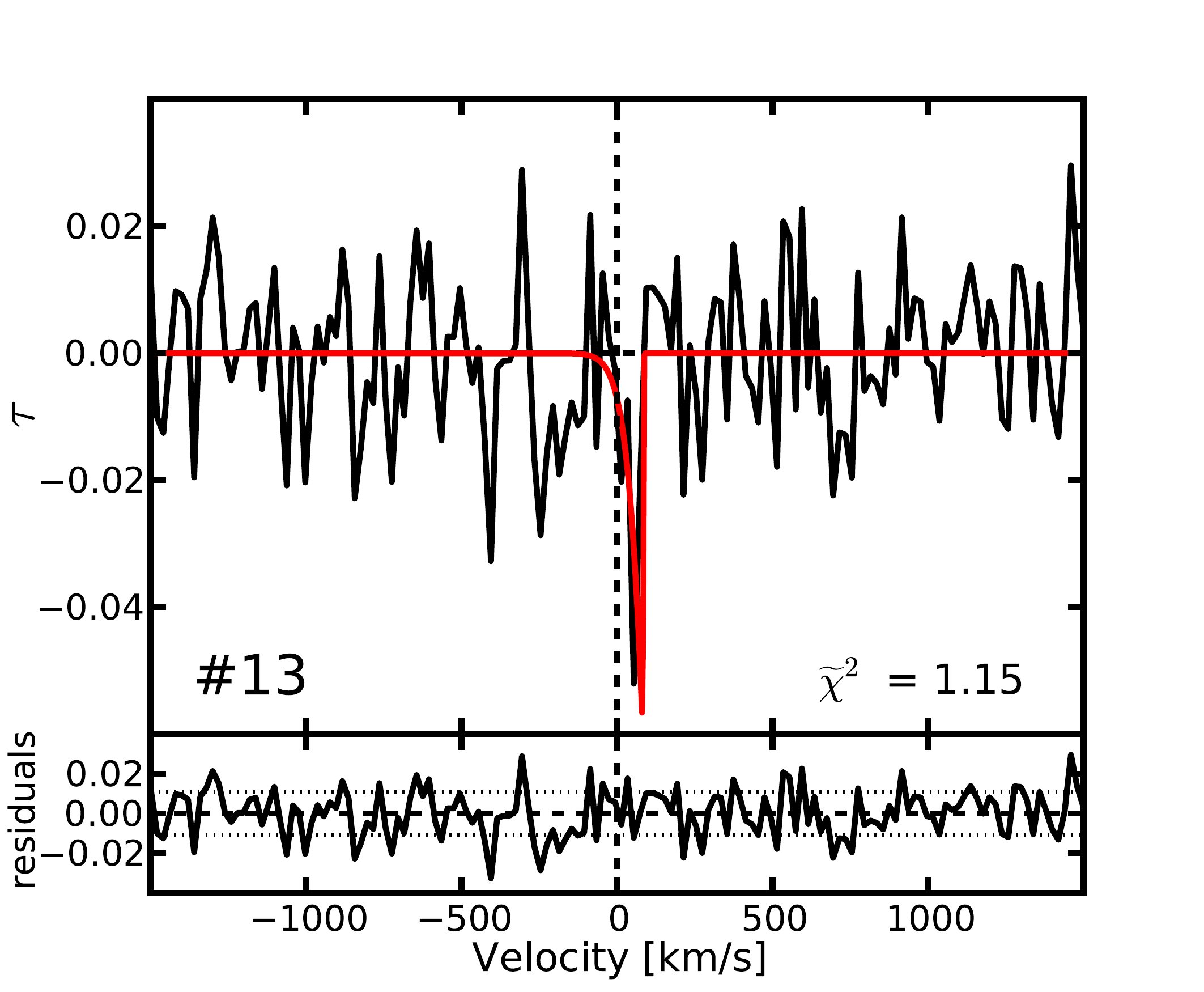}
		\includegraphics[trim = 0 0 30 30, clip,width=.33\textwidth]{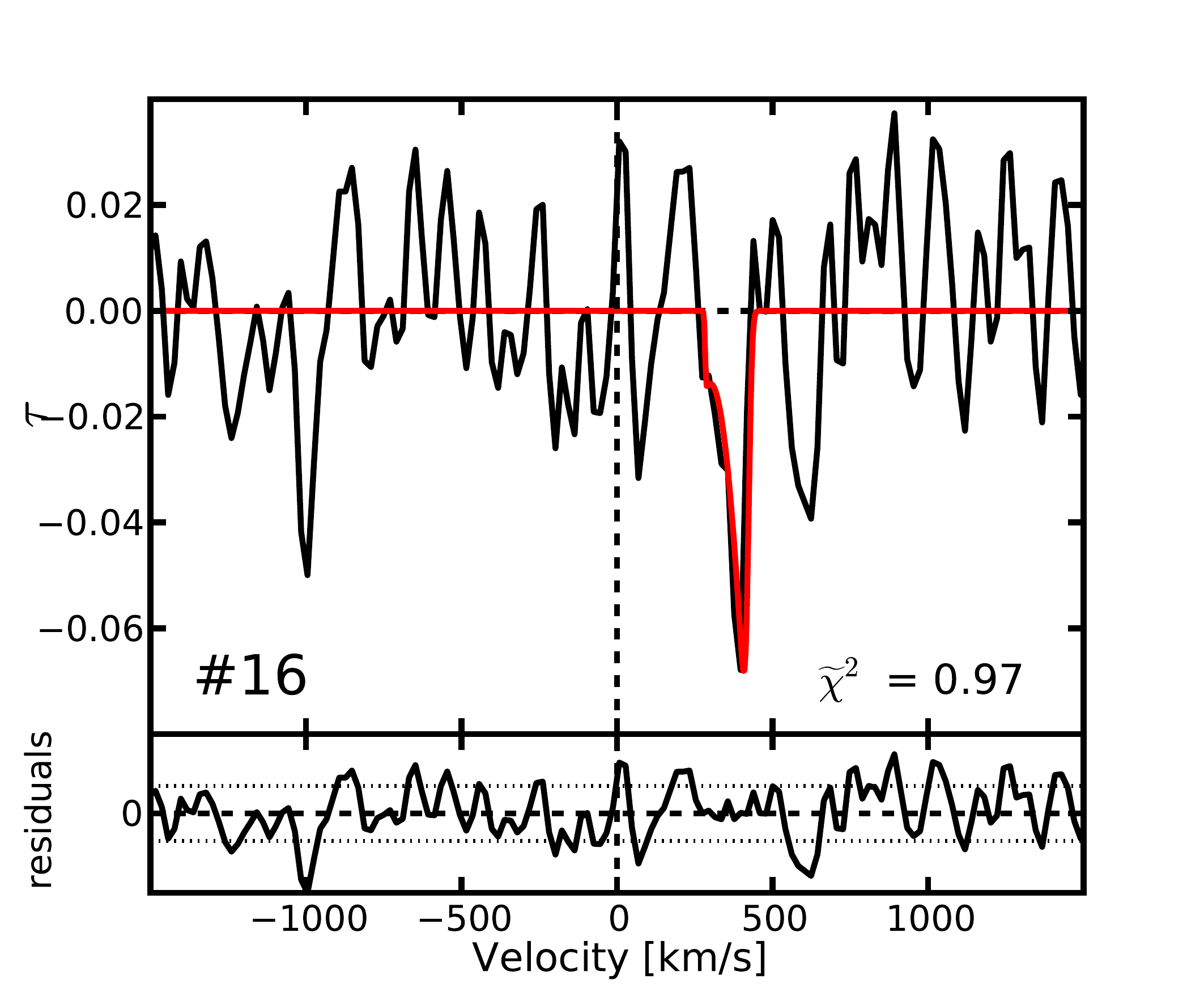}
		\includegraphics[trim = 0 0 30 30, clip,width=.33\textwidth]{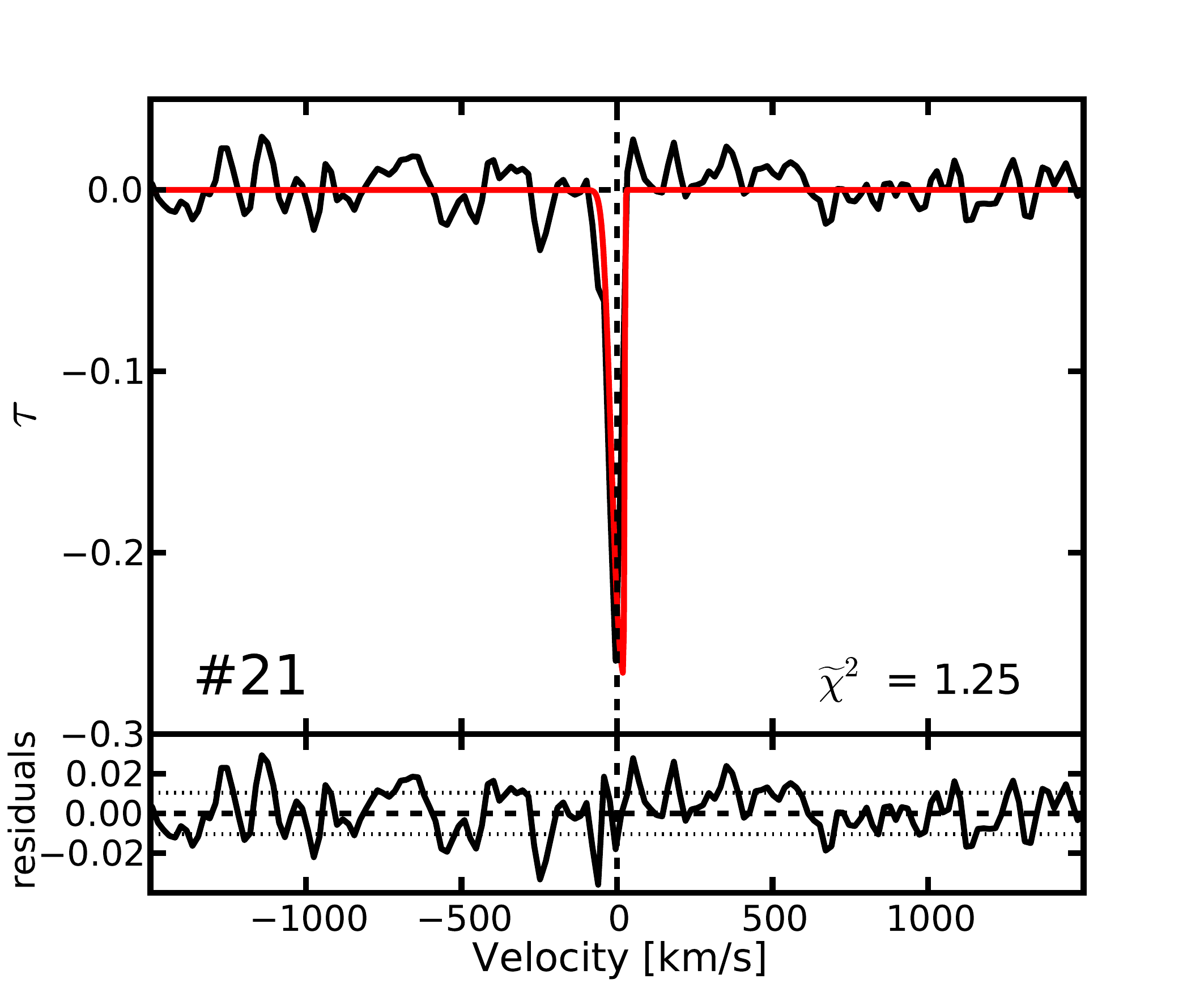}
		\includegraphics[trim = 0 0 30 30, clip,width=.33\textwidth]{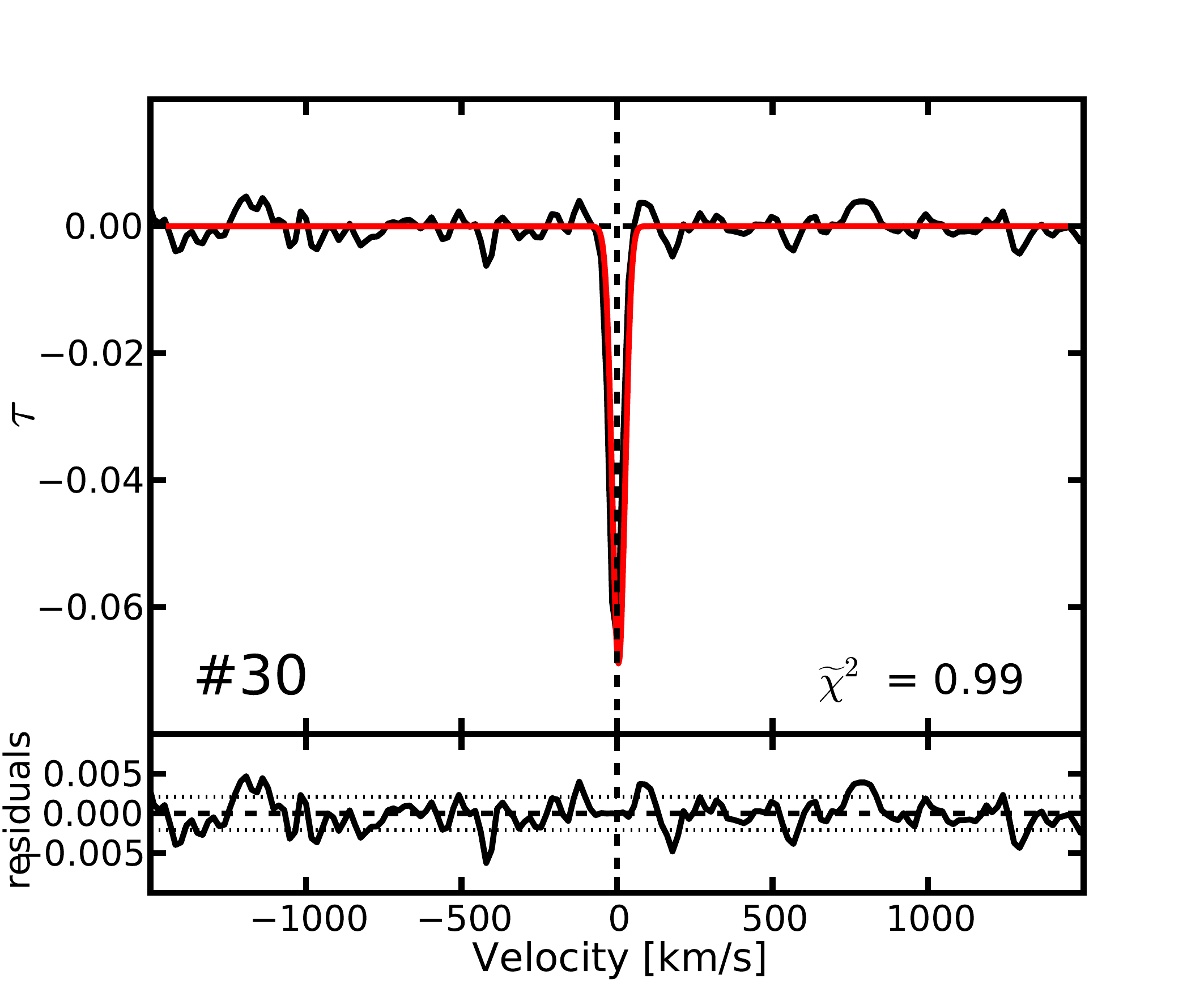}
		\includegraphics[trim = 0 0 30 30, clip,width=.33\textwidth]{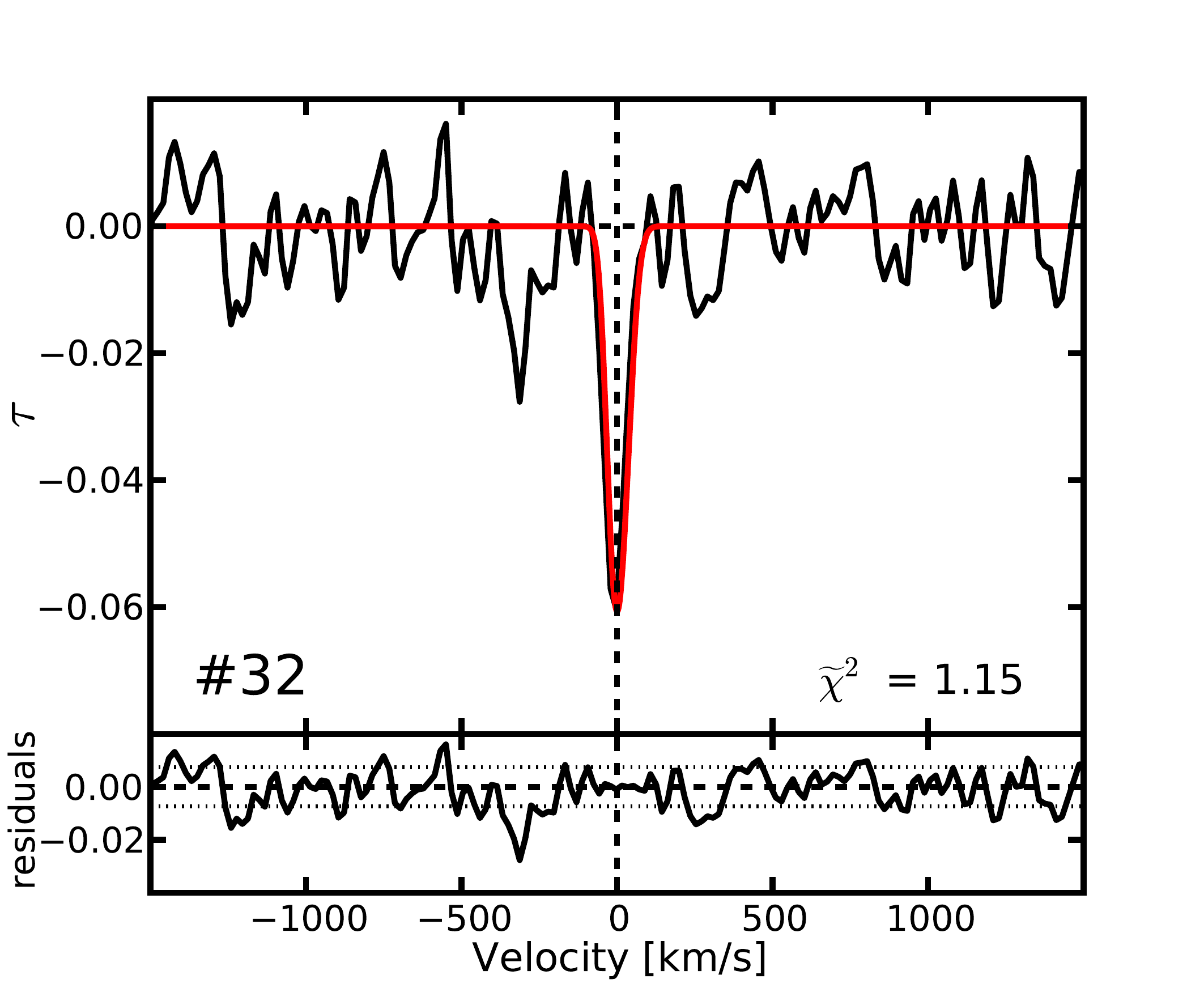}	
		\caption{(a) - \HI\ profiles (10 detections) in the narrow region with FWHM $<$ 100 \kms. The data are shown in \emph{black}, the BF fit is shown in \emph{dashed red}. The residuals of the fit are plotted in \emph{black} in the bottom boxes, along with the $\pm1$-$\sigma$ noise level (horizontal dotted lines).}\label{Profiles1}
\end{center}
\end{figure*}

We detect \HI\ absorption at the $\geq3-\sigma$ level in 32 of the observed galaxies, and 24 of these are new detections (see Table~\ref{table:detections} and notes on individual sources in Appendix \ref{notes}). All the spectra have been extracted against the radio core of the sources. The \HI\ profiles in Fig. \ref{Profiles1} show a variety of complex shapes and kinematics, really an \HI\ "Zoo".
The \HI\ lines are separated in three groups based on the profile analysis that will be discussed in Sec. \ref{Sec:busy} and Sec. \ref{sec:BFcharcteristics}. As we mention in \mbox{Paper \small{I}}, the $\tau$ and N(\HI) range of detections is quite broad, with the inferred \HI\ column densities spanning two orders of magnitude, 10$^{17}$ -- 10$^{19}$ ($T_{\rm{spin}}/c_{\rm{f}}$) cm$^{-2}$, where $T_{\rm{spin}}$ is the spin temperature and $c_{\rm f}$ is the covering factor of the gas. Where the degenerate $T_{\rm{spin}}/c_{\rm{f}}$ can be measured, i.e., in damped Lyman-$\alpha$ absorbers, the observable quantities of the $T_{\rm{spin}}/c_{\rm{f}}$ vary from 60 K to $\sim$10$^4$ K \citep{Kanekar2003, Curran2007}. As a result, the inferred column density may be in error by $\sim$3 orders of magnitude depending on $T_{\rm{spin}}/c_{\rm{f}}$. Because of these uncertainties, here, we provide the column densities in units of $T_{\rm{spin}}/c_{\rm{f}}$.

Non-detections and the corresponding 3-$\sigma$ upper limits are presented in Table \ref{table:non-detections} of Appendix \ref{Table:NonDet}. The N(\HI) upper limits for non-detections were calculated by assuming FWHM = 100 \kms, following the analysis of the profile widths in Sec. \ref{Sec:busy}. Below, we study in more detail the width, asymmetry parameters, and the blueshift/redshift distribution of the detections using the BF. 

In \mbox{Paper \small{I}} we show that statistically, detections and non-detections have a similar distribution of continuum flux, implying that detections in our sample are not biased toward brighter sources. 
Here, in Fig. \ref{fig:KS_RadioPower} we present the radio power distribution of detections and non-detections. According to the Kolmogorov-Smirnov test, the significance level that the two distributions are different is only 10$\%$, implying that statistically detections and non-detections have a similar radio power distribution. The largest difference between the two distributions (D) is measured at $\sim10^{24.6}$ W Hz$^{-1}$.     

Additional notes on the individual detections are presented in Appendix \ref{notes}.

\addtocounter{figure}{-1}   
\begin{figure*}
\begin{center}
		\includegraphics[trim = 0 0 30 30, clip,width=.33\textwidth]{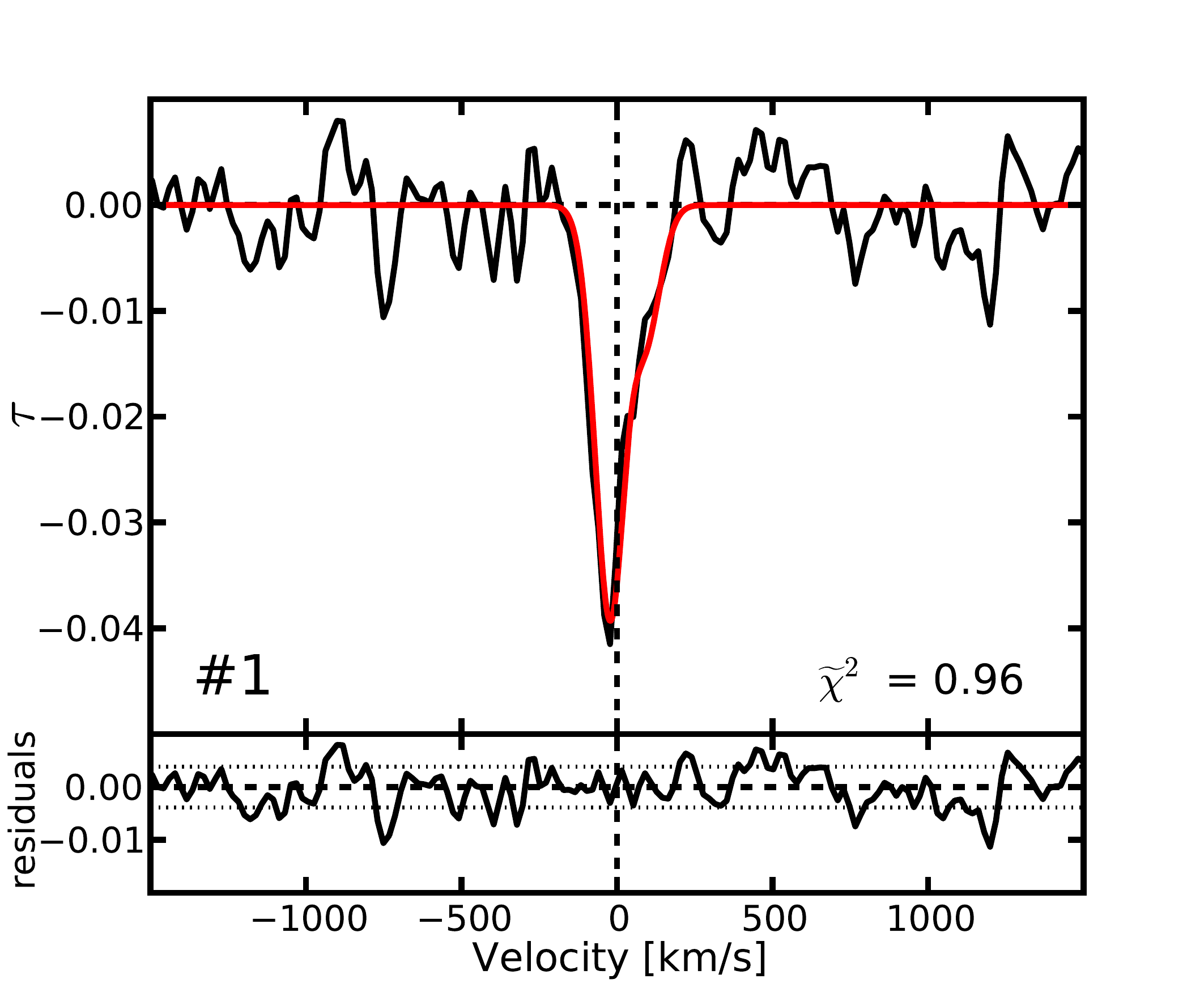}
		\includegraphics[trim = 0 0 30 30, clip,width=.33\textwidth]{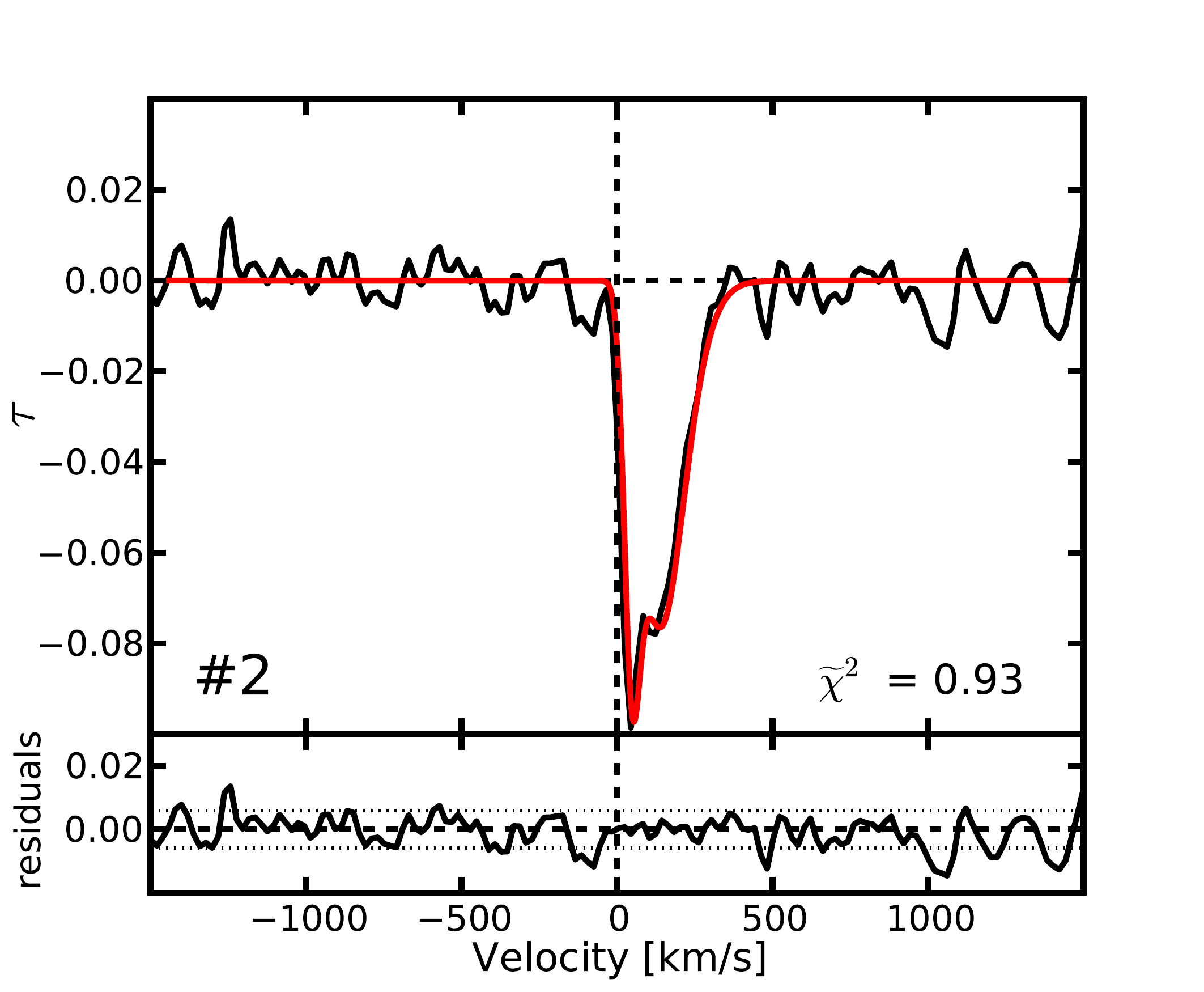}
	    	\includegraphics[trim = 0 0 30 30, clip,width=.33\textwidth]{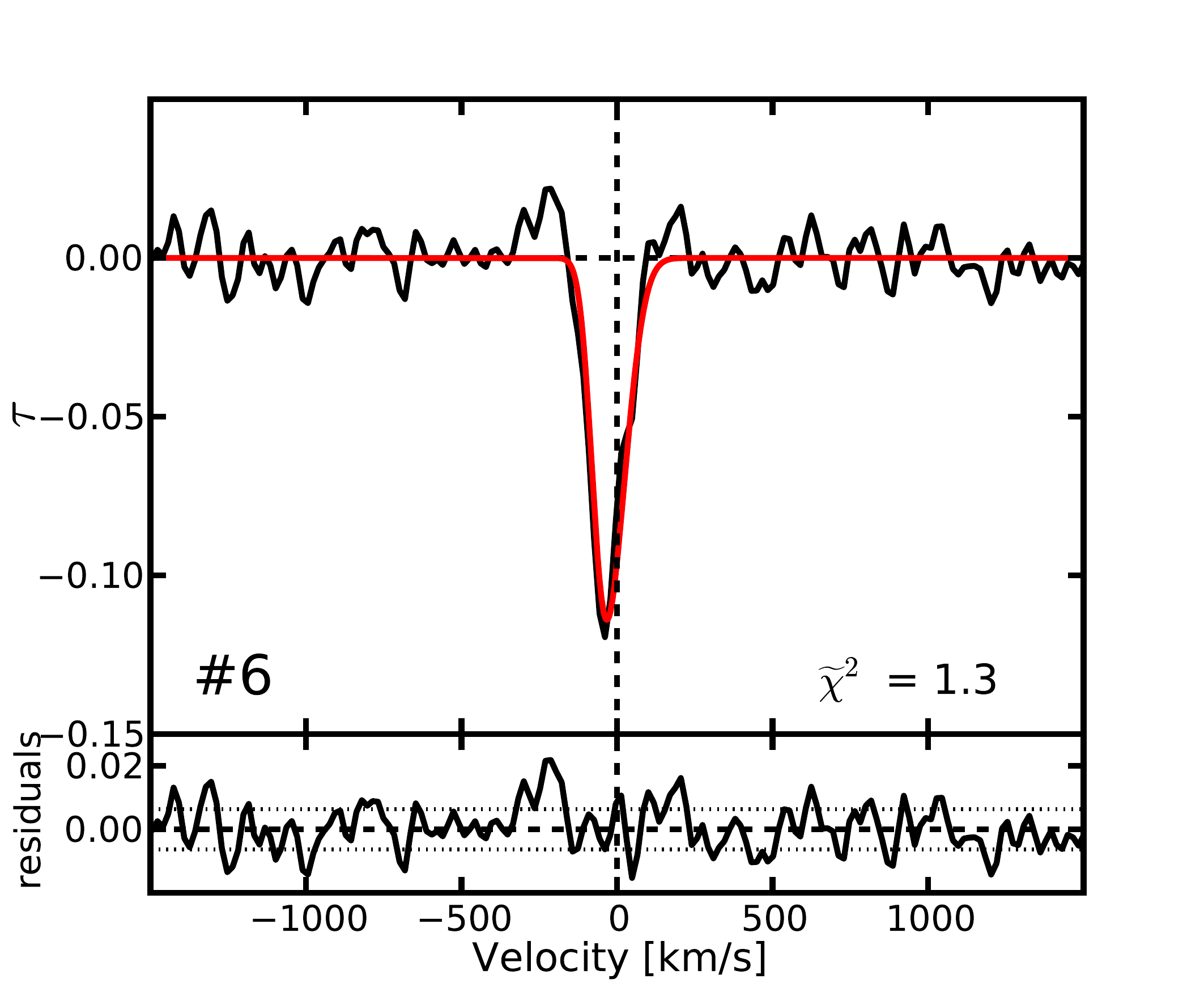}
		\includegraphics[trim = 0 0 30 30, clip,width=.33\textwidth]{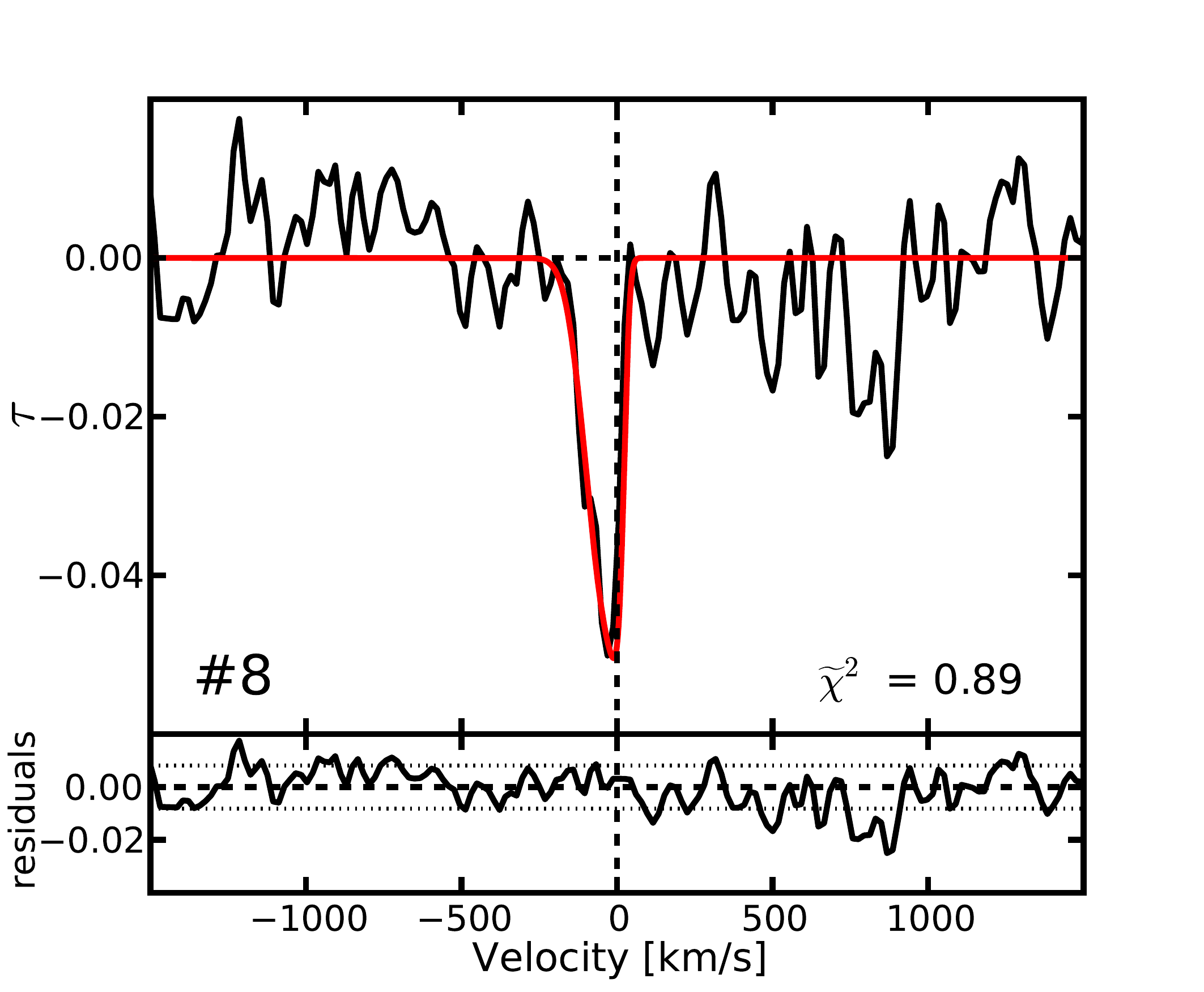}
		\includegraphics[trim = 0 0 30 30, clip,width=.33\textwidth]{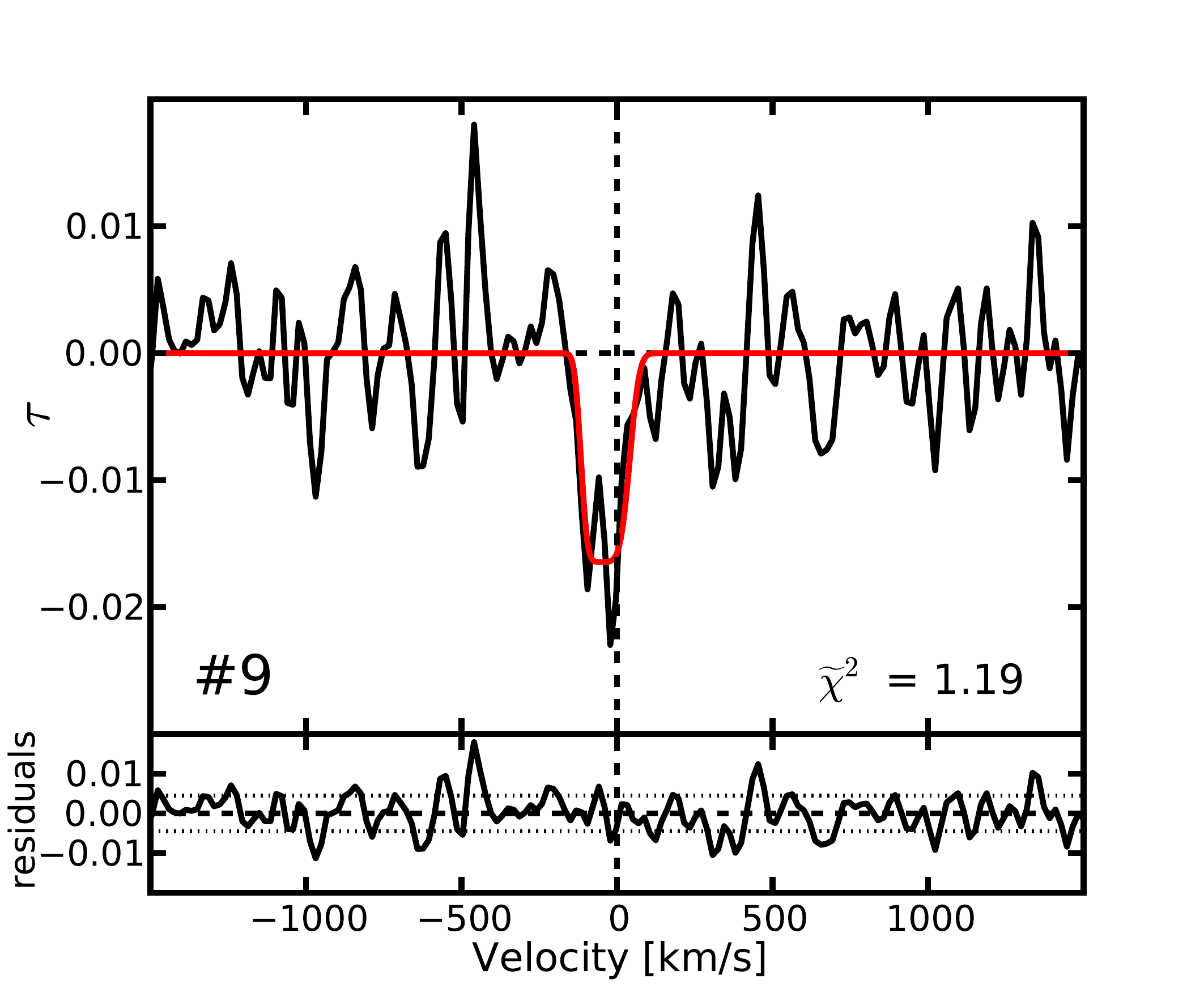}
		\includegraphics[trim = 0 0 30 30, clip,width=.33\textwidth]{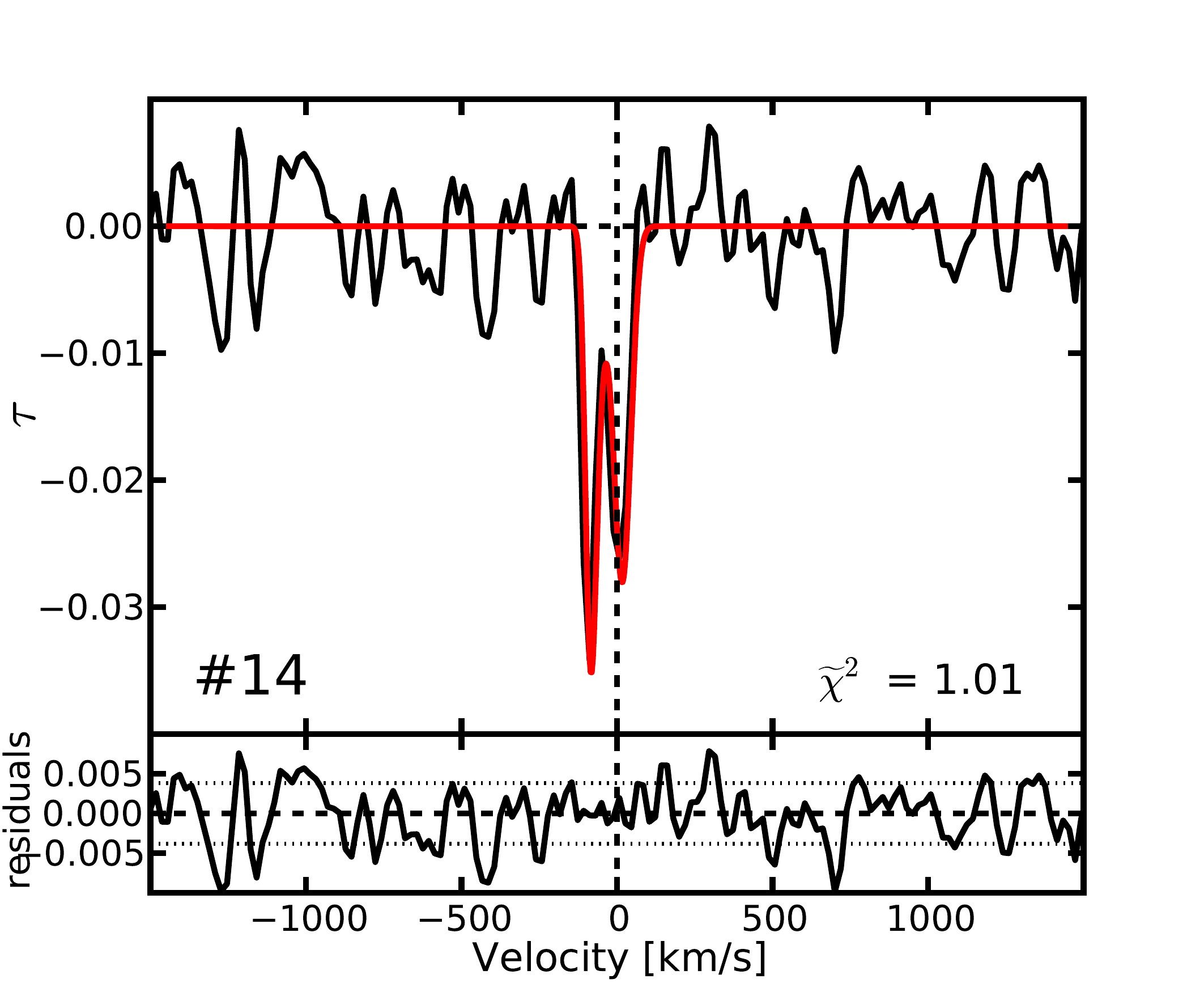}
		\includegraphics[trim = 0 0 30 30, clip,width=.33\textwidth]{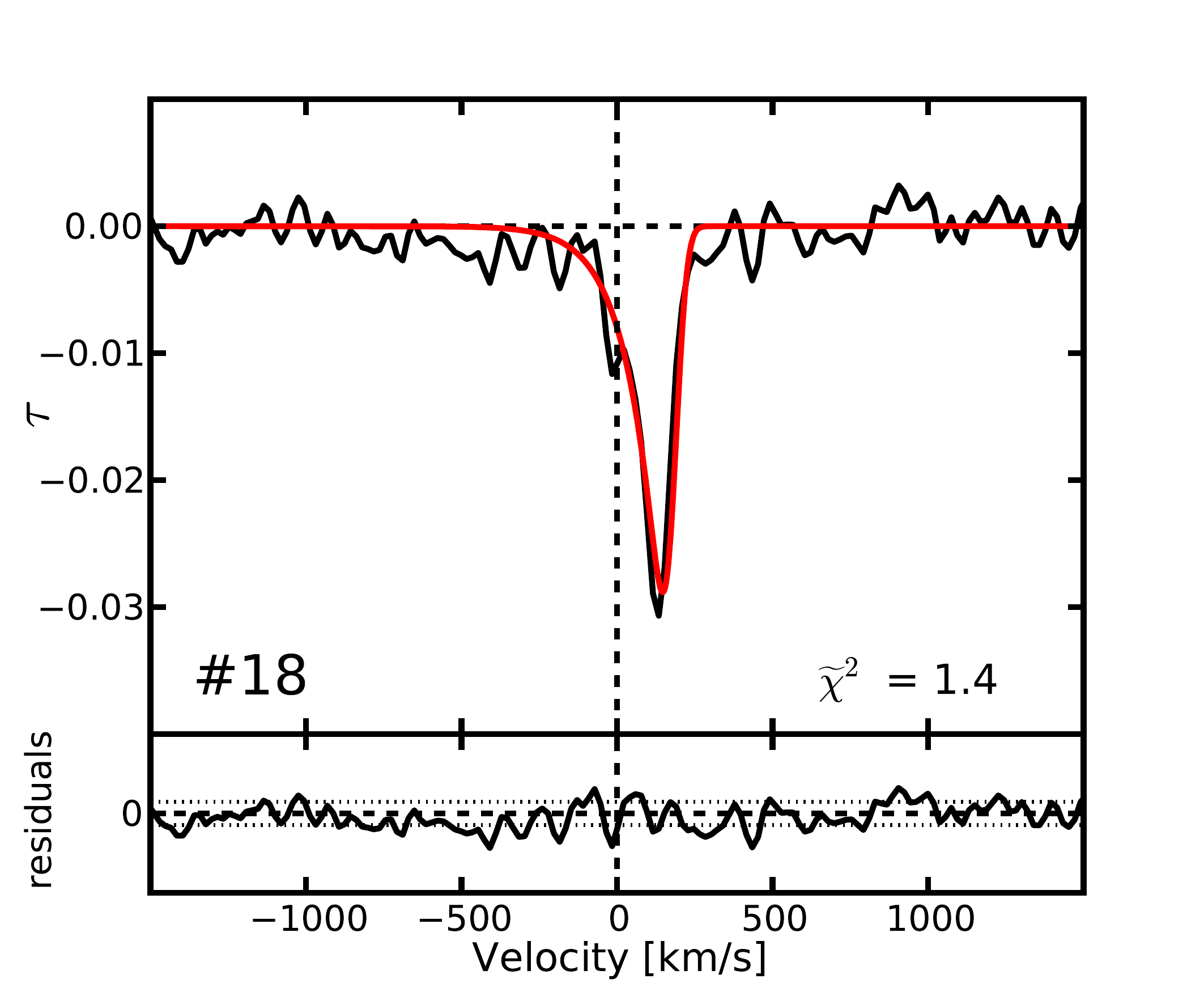}
		\includegraphics[trim = 0 0 30 30, clip,width=.33\textwidth]{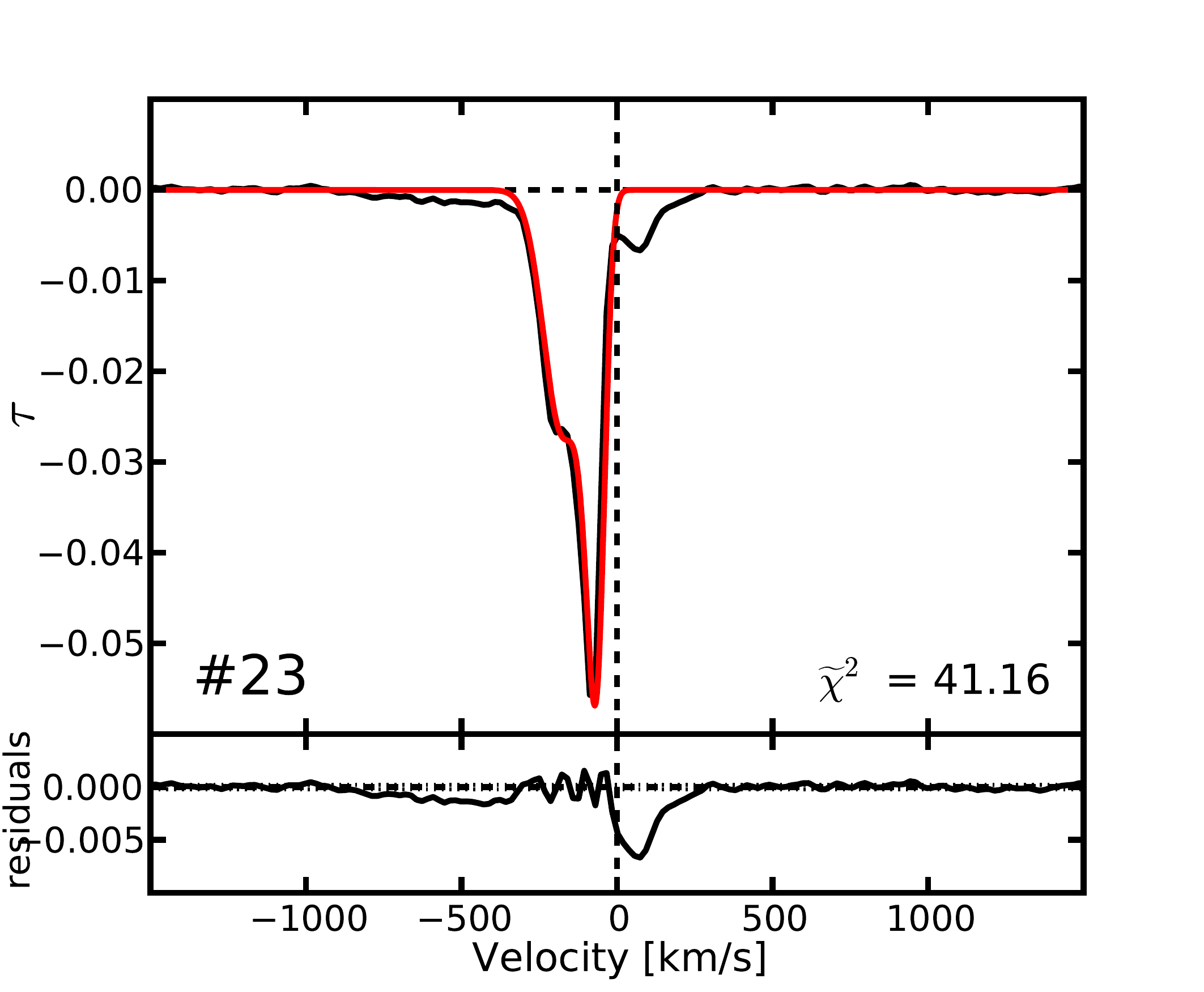}		
		\includegraphics[trim = 0 0 30 30, clip,width=.33\textwidth]{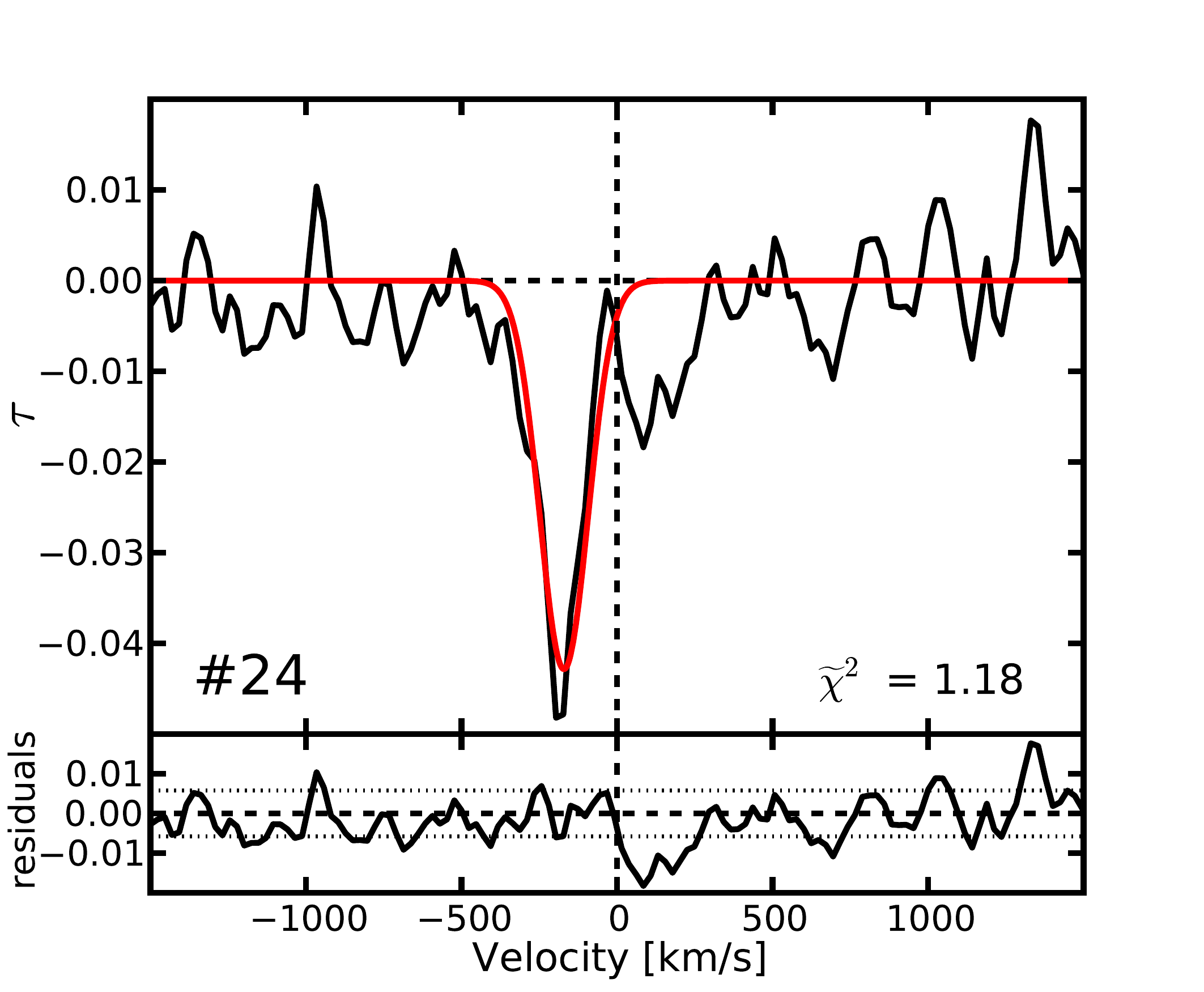}
		\includegraphics[trim = 0 0 30 30, clip,width=.33\textwidth]{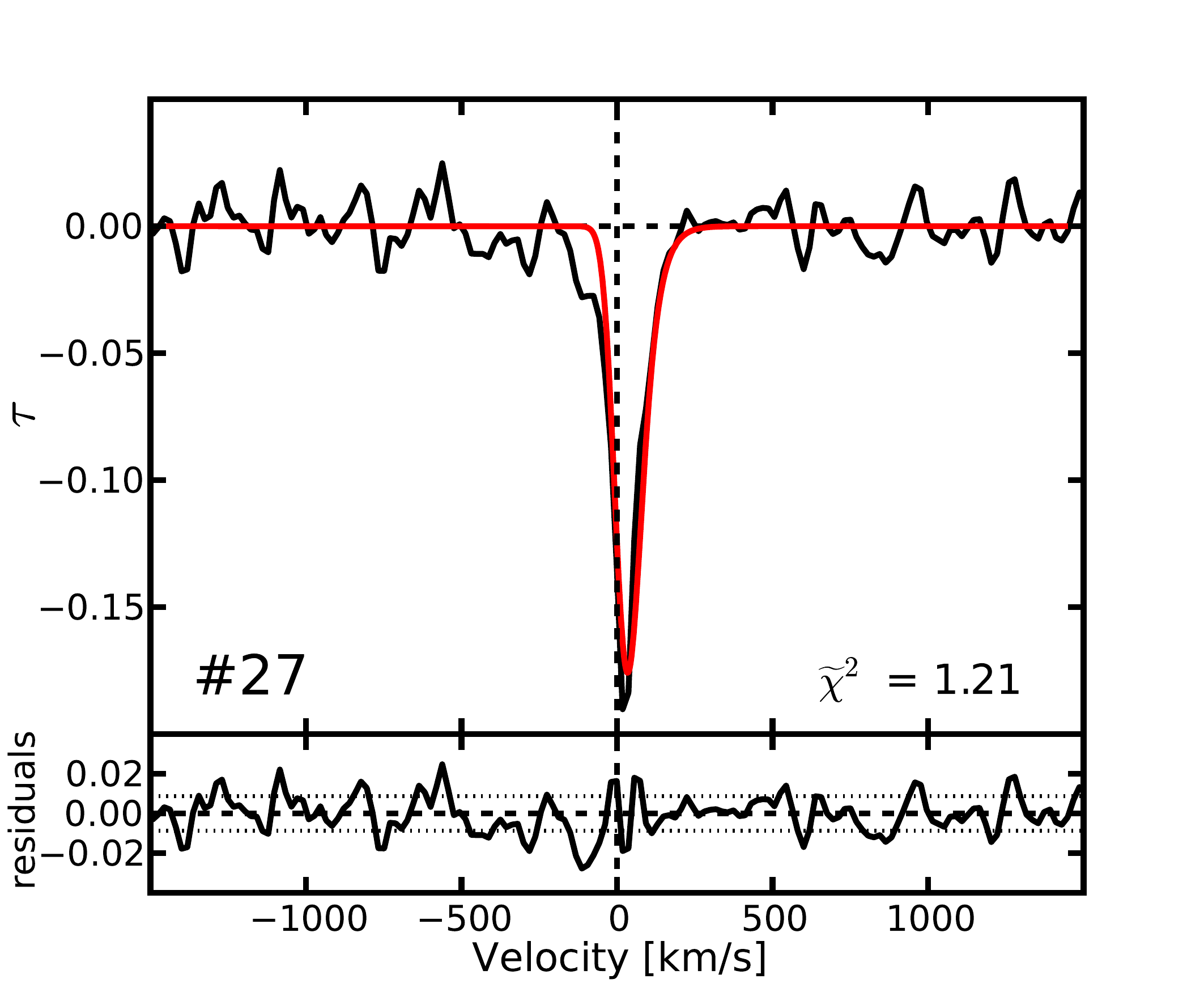}
		\caption{(b) - \HI\ profiles (11 detections) in the intermediate width region at 100 \kms $<$ FWHM $<$ 200 \kms.}\label{Profiles2}
\end{center}
\end{figure*}

\addtocounter{figure}{-1}    
\begin{figure*}
\begin{center}
		\includegraphics[trim = 0 0 30 30, clip,width=.33\textwidth]{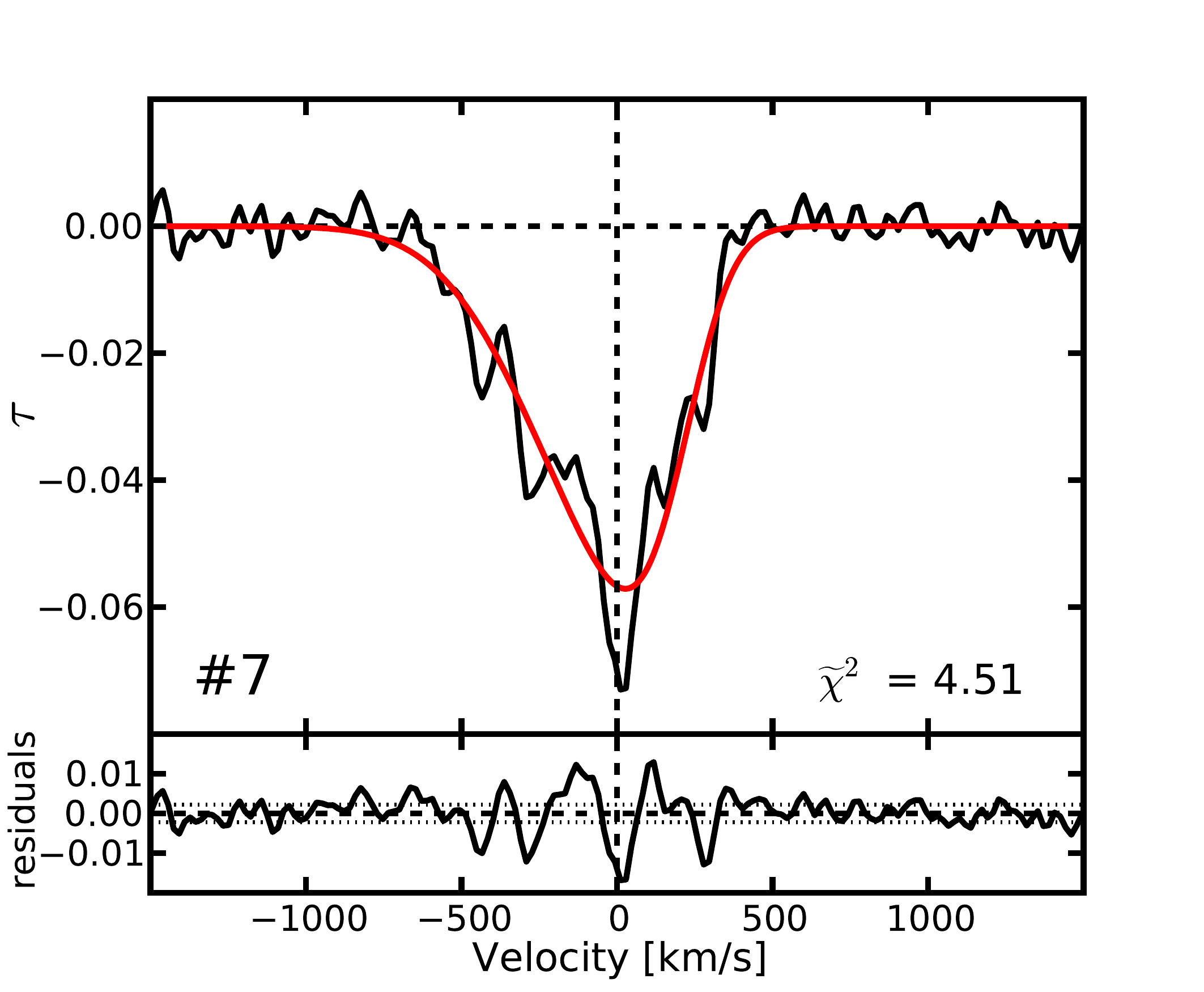}
		\includegraphics[trim = 0 0 30 30, clip,width=.33\textwidth]{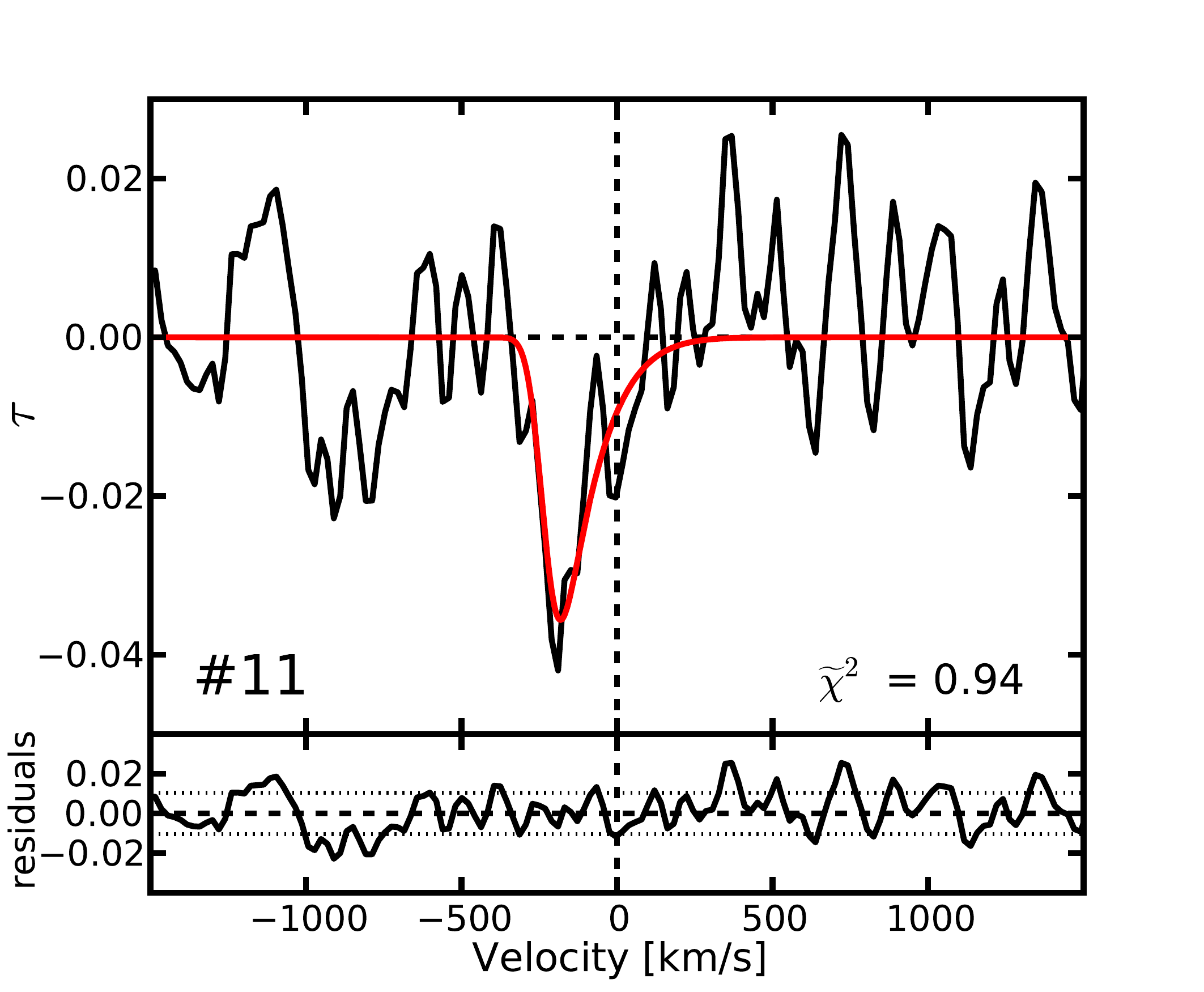}
		\includegraphics[trim = 0 0 30 30, clip,width=.33\textwidth]{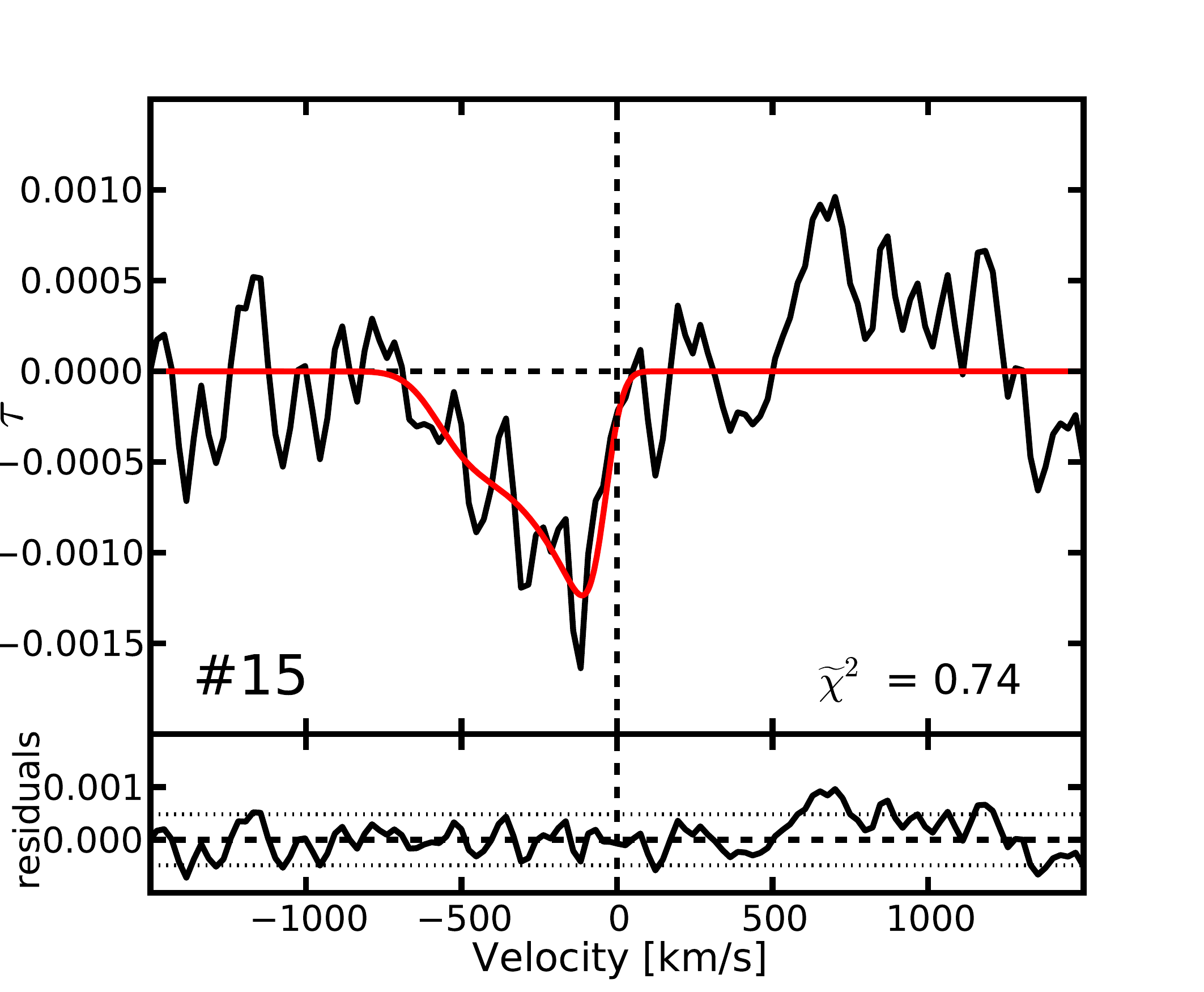}
		\includegraphics[trim = 0 0 30 30, clip,width=.33\textwidth]{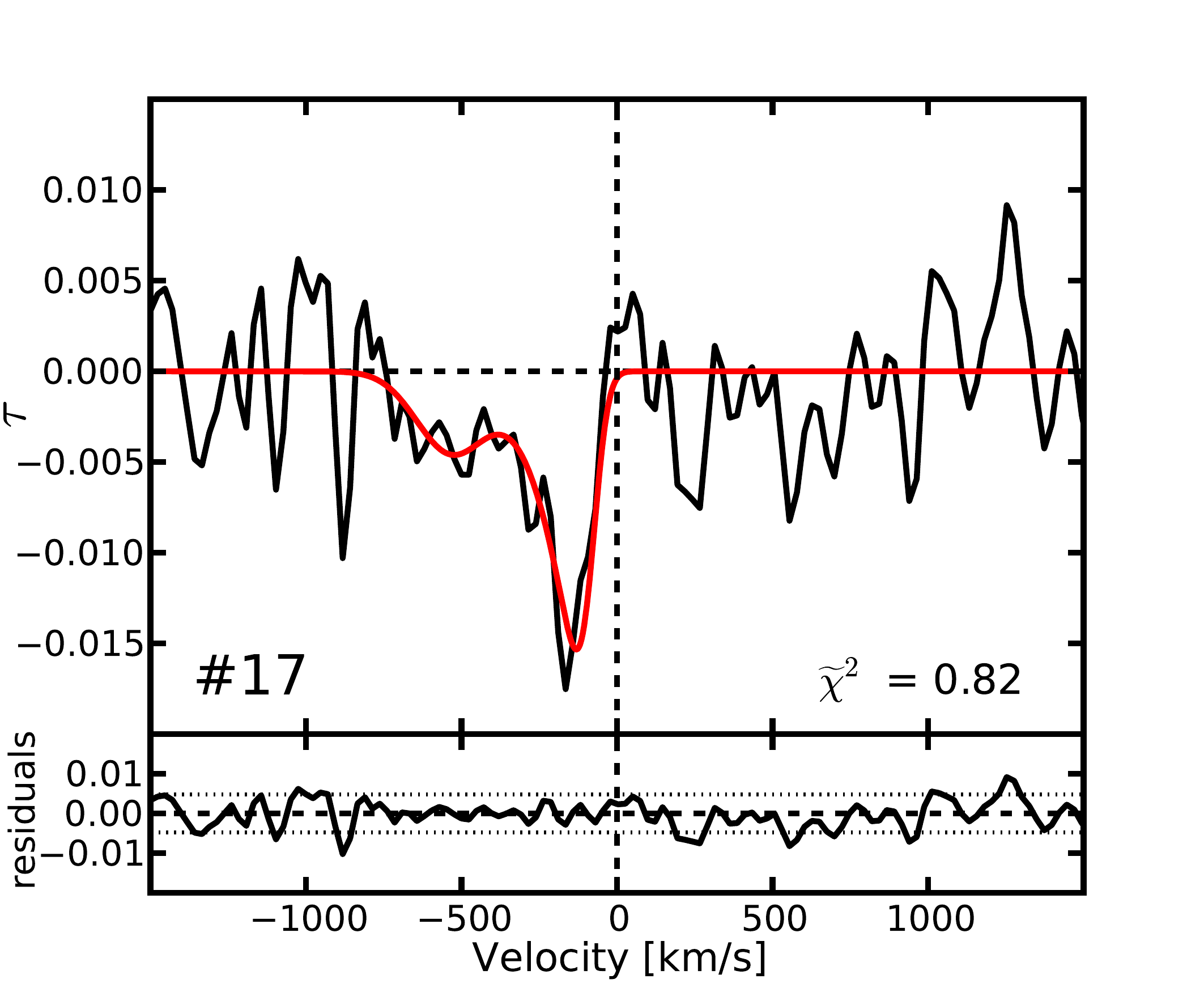}		
		\includegraphics[trim = 0 0 30 30, clip,width=.33\textwidth]{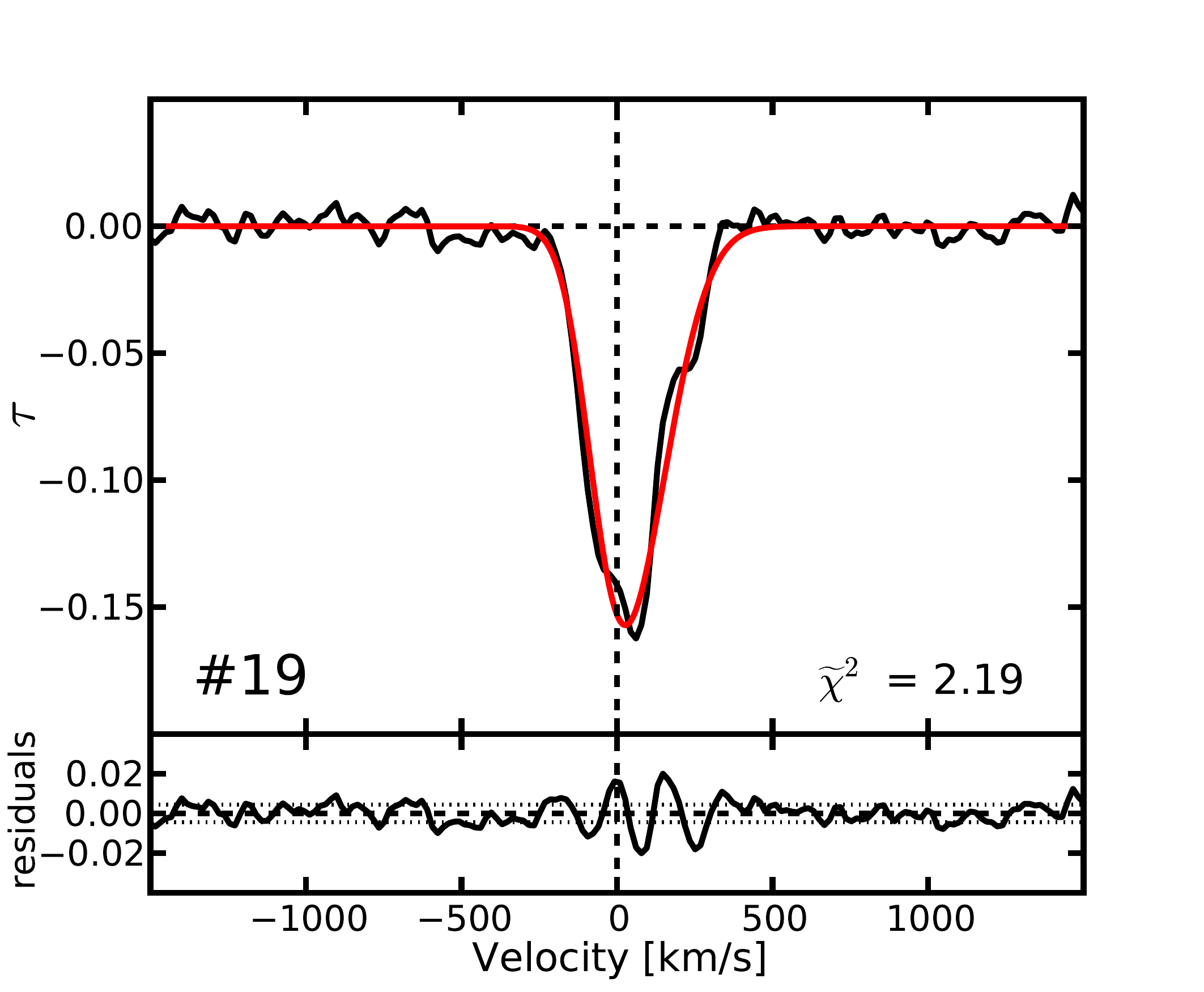}
		\includegraphics[trim = 0 0 30 30, clip,width=.33\textwidth]{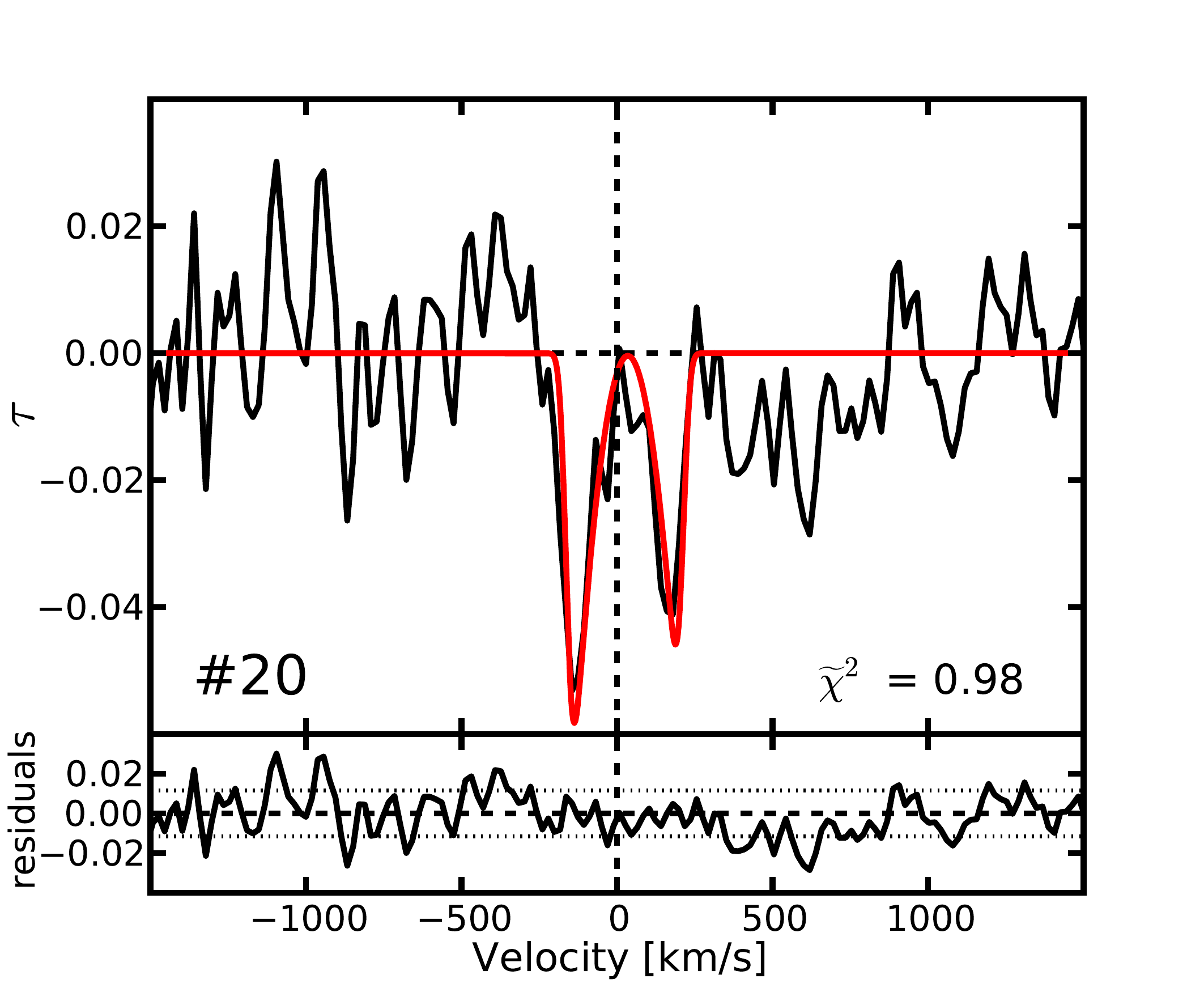}			
		\includegraphics[trim = 0 0 30 30, clip,width=.33\textwidth]{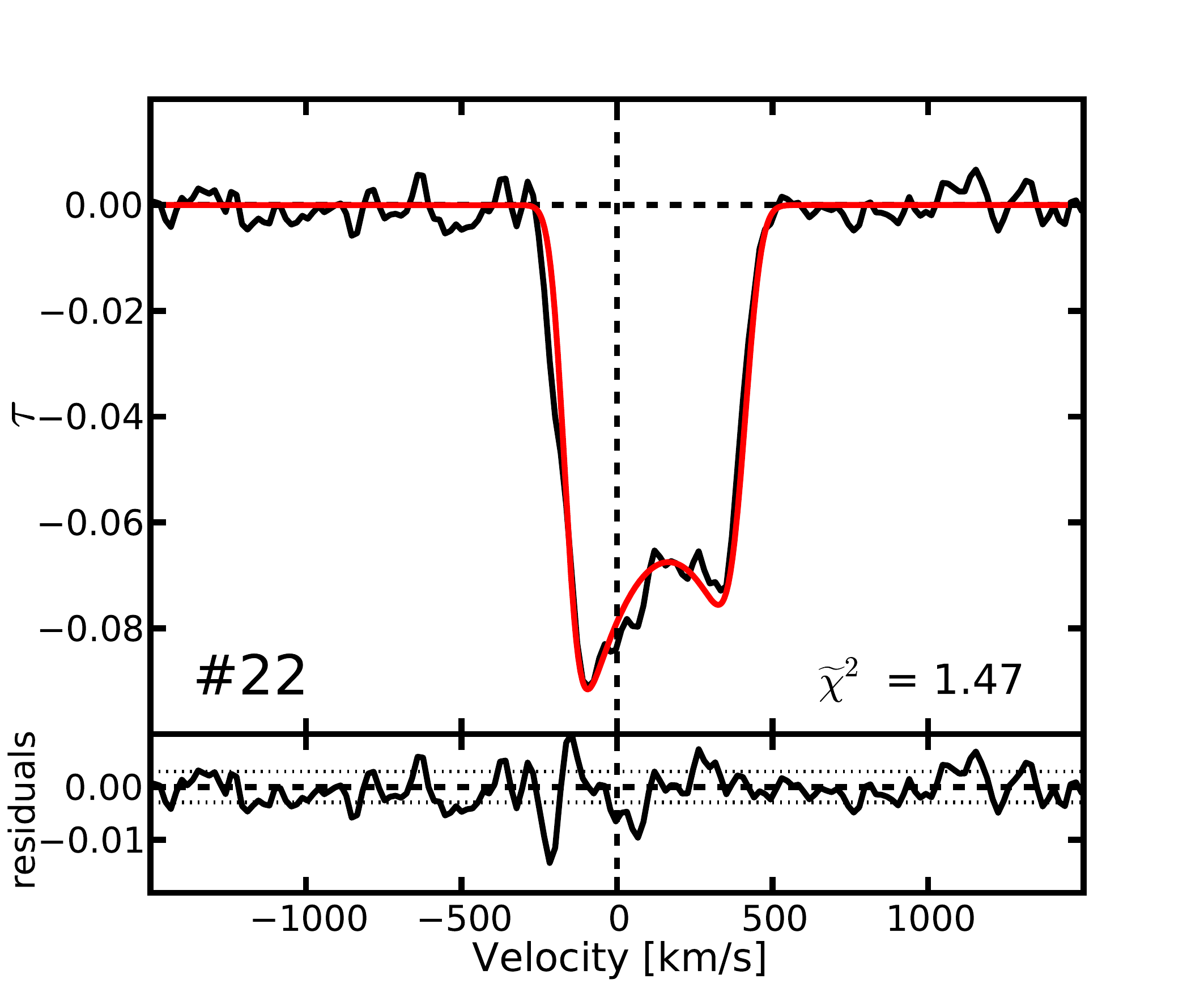}			
		\includegraphics[trim = 0 0 30 30, clip,width=.33\textwidth]{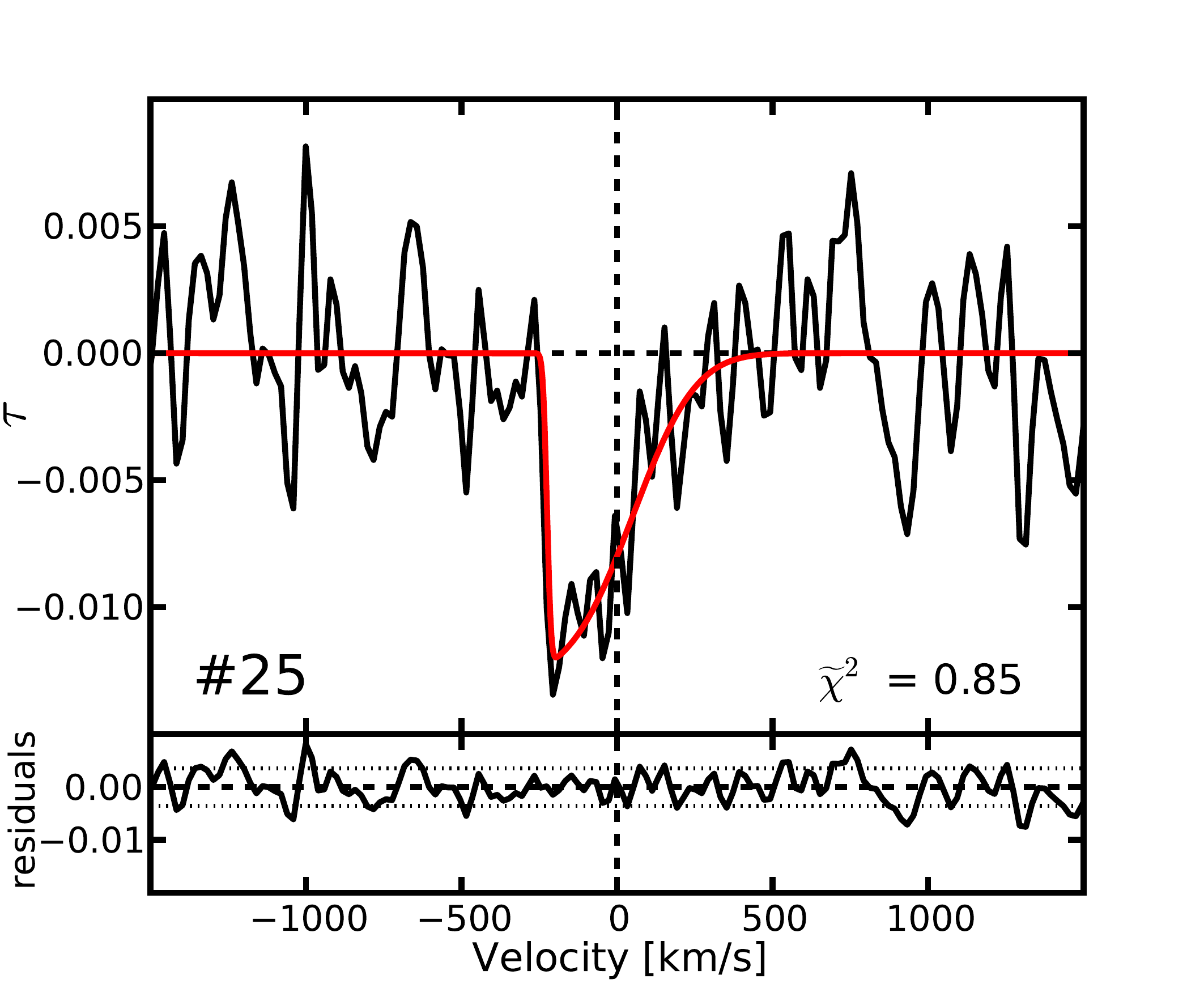}
       		 \includegraphics[trim = 0 0 30 30, clip,width=.33\textwidth]{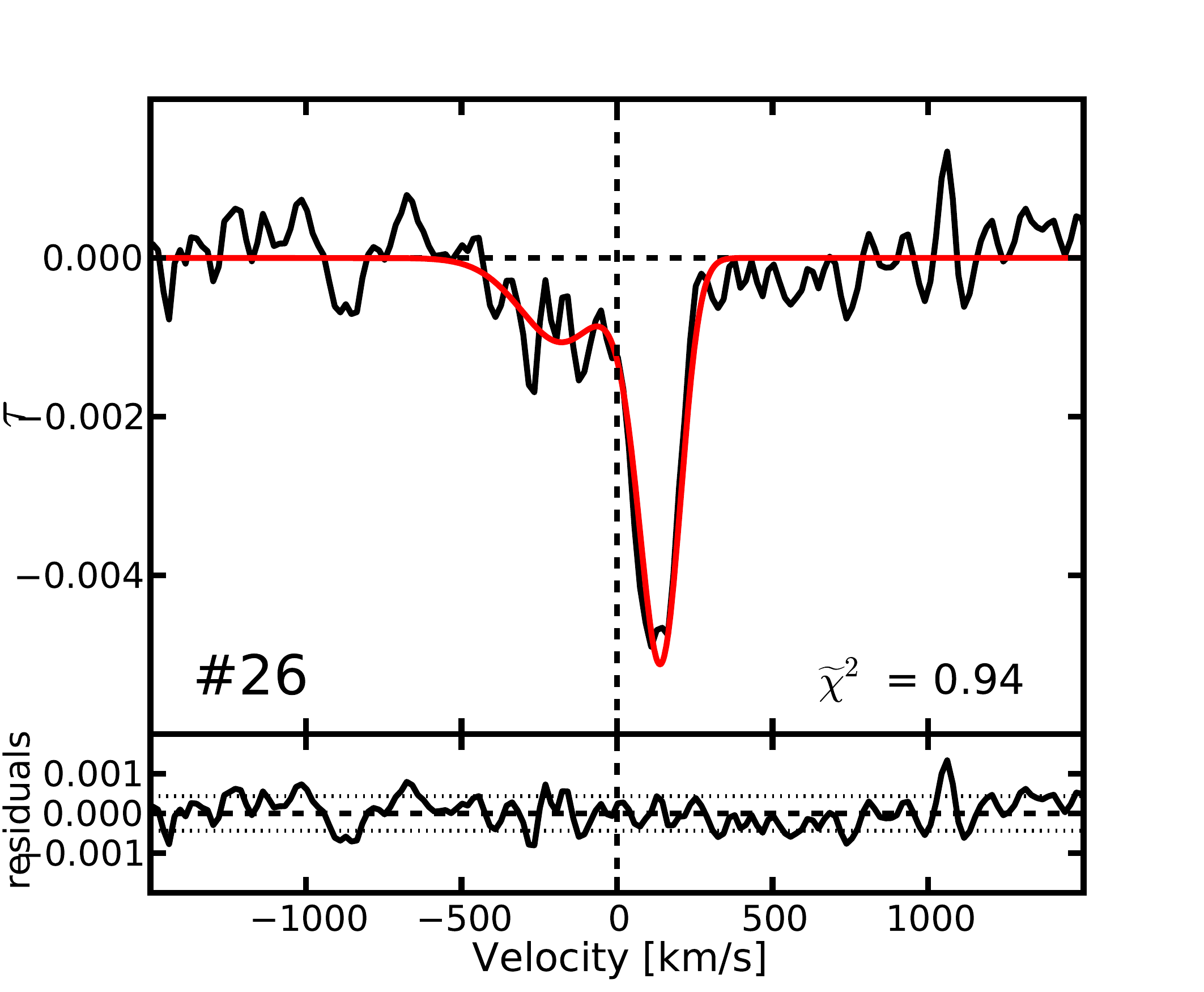}
		\includegraphics[trim = 0 0 30 30, clip,width=.33\textwidth]{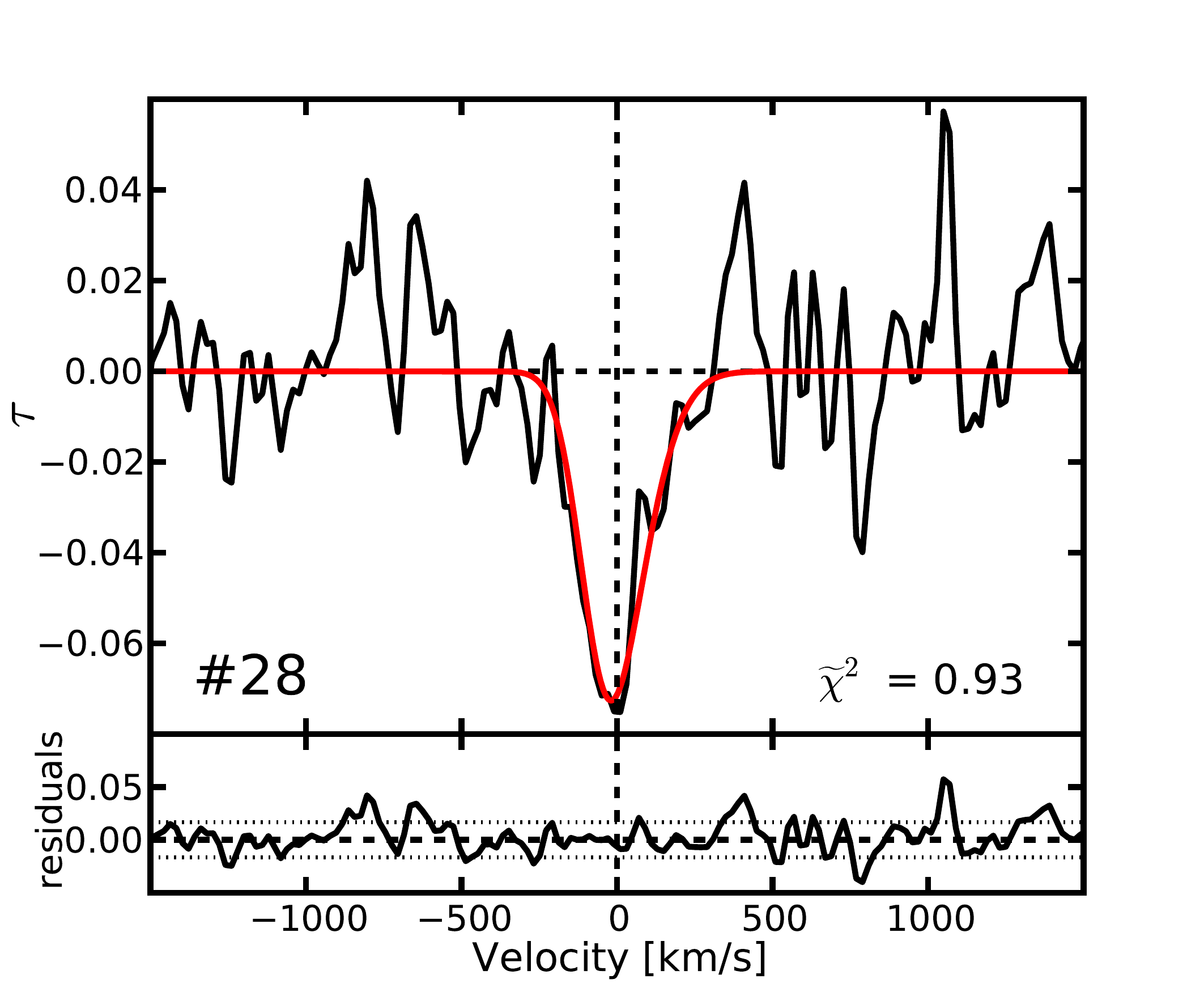}
		\includegraphics[trim = 0 0 30 30, clip,width=.33\textwidth]{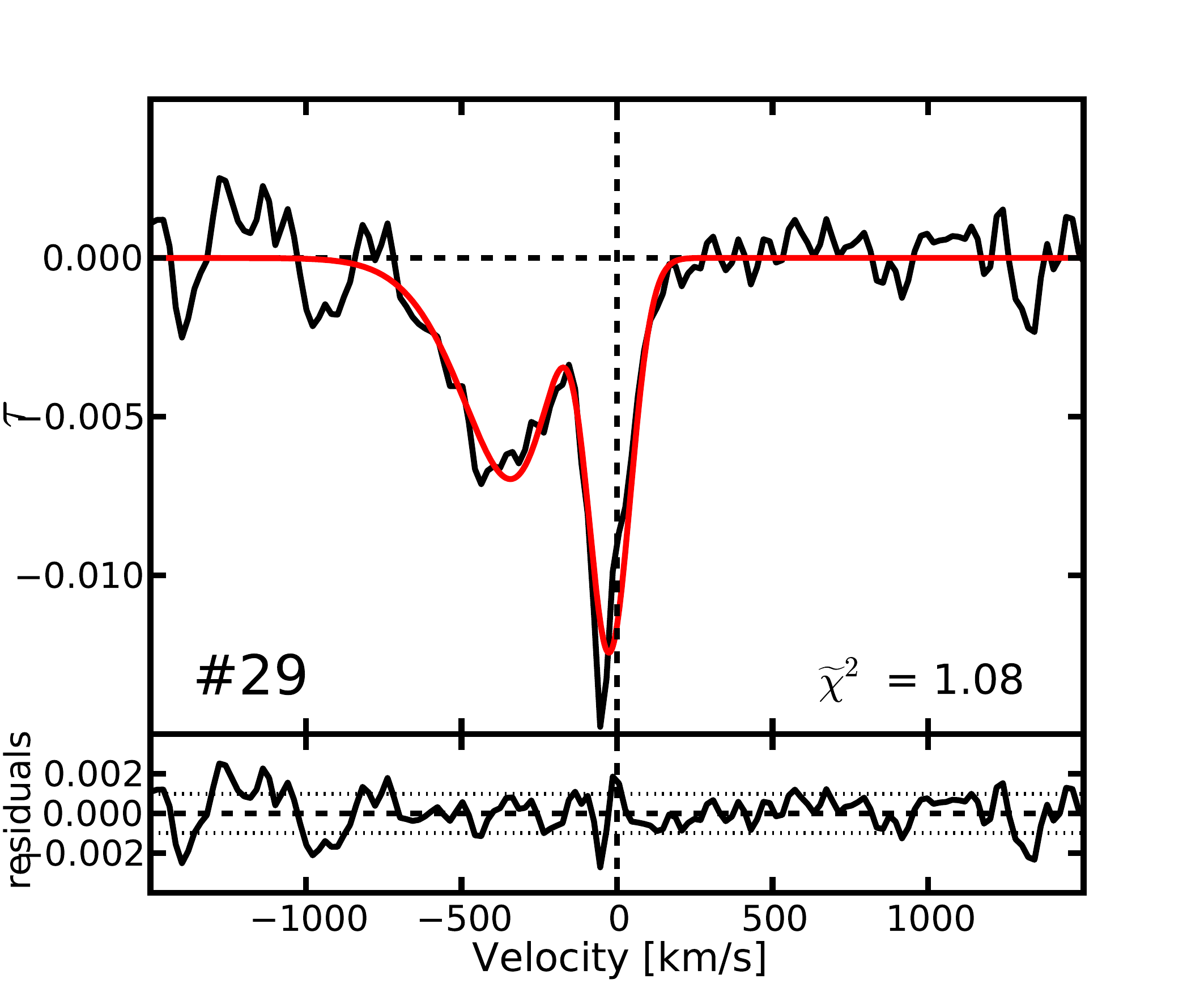}
		\includegraphics[trim = 0 0 30 30, clip,width=.33\textwidth]{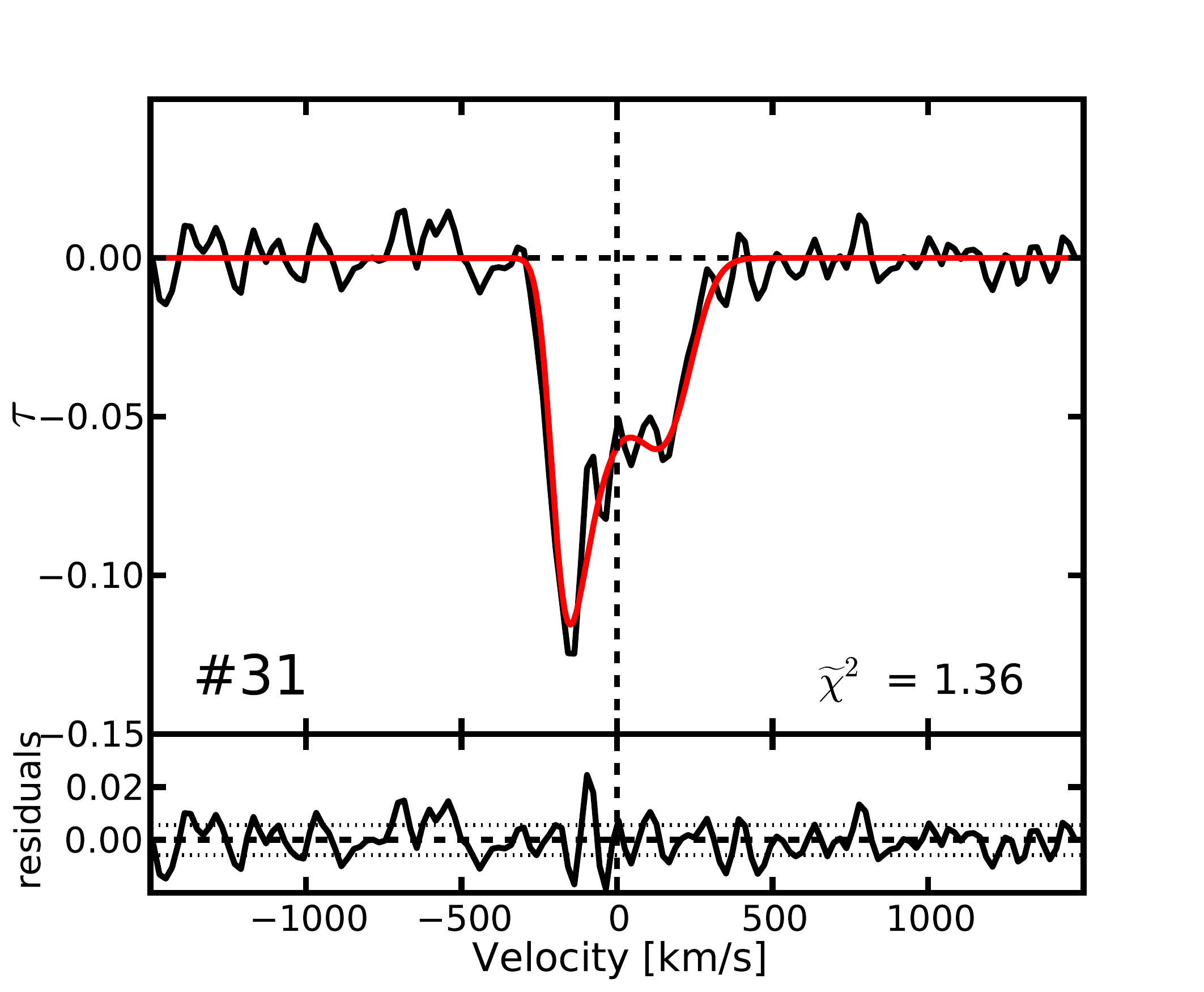}
		\caption{(c) - \HI\ profiles (11 detections) in the broad width region at FWHM $>$ 200 \kms.}\label{Profiles3}
\end{center}
\end{figure*}

\subsection{Fitting complex \HI\ absorption profiles with the BF}\label{Sec:busy}

The sample of absorption lines in Fig. \ref{Profiles1} is very heterogeneous in terms of line shapes and widths. Thus, it is crucial to develop a uniform method to characterize the properties of this variety of \HI\ profiles. So far, Gaussian fitting has been widely used to derive absorption line properties, e.g., the width of the profile, and to determine the presence of multiple components \citep{Vermeulen2003, Gupta2006, Curran2011}. When multiple peaked profiles occur, like in our absorption sample, Gaussian fitting methods have the disadvantage of having to make an \emph{a priori} assumption on the physical conditions of the gas, by choosing the number of components to be fitted. In the case of \HI\ integrated emission profiles, an alternative solution has been proposed by \cite{Westmeier2014}: the BF fitting method. The BF is a heuristic, analytic function, given by the product of two error functions and a polynomial factor:

\begin{equation}
\begin{array}{lp{0.8\linewidth}}
$B(x)=$&$\frac{a}{4}\times({\rm erf}[b_1\{w+x-x_e\}]+1)$\\
&$\times({\rm erf}[b_2\{w-x+x_e\}]+1)\times(c|x-x_p|^n+1)$.
\end{array}
\end{equation}

The main advantage of this function is that a proper combination of the parameters can fit a wide variety of line profiles. When $c=0$, the BF may well approximate a Gaussian profile, while if $c\neq0$, for different values of $b_1$ and $b_2$, an asymmetric double horn profile can be reproduced. With the same function it is then possible to fit single and double-peaked profiles, while with Gaussian fitting more functions are needed for the fitting of multiple lines.
\cite{Westmeier2014} applied the BF fit to integrated emission lines of the HI Parkes All Sky Survey (HIPASS) sample, but such a method has never been applied to absorption lines before. Here, we used the \emph{C++} code provided by \cite{Westmeier2014} to fit our absorption profiles and estimate the width, asymmetry, and blue/red-shift of the profiles. The FWHM and the full width at 20\% maximum (FW20) of the lines are determined measuring the width of the fitted BF profiles at $50\%$ and at $20\%$ of the peak intensity. The line centroid is measured at the middle point of the width at $20\%$.  These parameters are summarized in Table \ref{table:BF} for all the absorption lines of our sample. The BF fit allows us to measure these general characteristics of the lines in a uniform way through the entire sample, regardless of the different line shapes and signal-to-noise ratios of the spectra. On average, the chi-square fit values are close to unity, and the residuals of the fit are consistent with the noise. However, the properties of the lines that we can uniformly measure are limited by the different S/N values among the spectra and by the quality of the fits. For example, in a few cases the residuals of the fit are higher than the noise, thus our choice of measuring the width at 20\% peak intensity (instead of lower, e.g., 10\% level) depends on the sensitivity of the spectra. Nevertheless, with respect to previous surveys on \HI\ absorption lines (e.g., \cite{Gorkom,Gallimore,Vermeulen2003,Gupta2006,Curran2011}), this is the first time that the FWHM is used in combination with the FW20 to describe the properties of the lines.

The BF fails to fit the profiles in source no. 23 and in source no. 4. In the latter, both emission and absorption are present in the spectra. Hence, we extracted a new spectrum from the cube at a location still close to the nuclear region, where the emission is not very strong. This spectrum can be successfully fitted by the BF. In source no. 23, instead, the fit has a high $\chi^2$ value because of the high S/N ratio of the spectrum. This allows us to detect two faint and complex features (see Appendix \ref{notes} for further details), which cannot be fitted by the BF. Nevertheless the BF fit allows us to measure the main parameters of the line (FWHM, FW20, asymmetry, and blue/red-shift of the profile), hence we include this source in our analysis.

\subsection{Characterization of the profiles with BF parameters}\label{sec:BFcharcteristics}

\begin{figure}
\begin{center}
\includegraphics[trim = 20 0 0 20, clip,width=.52\textwidth]{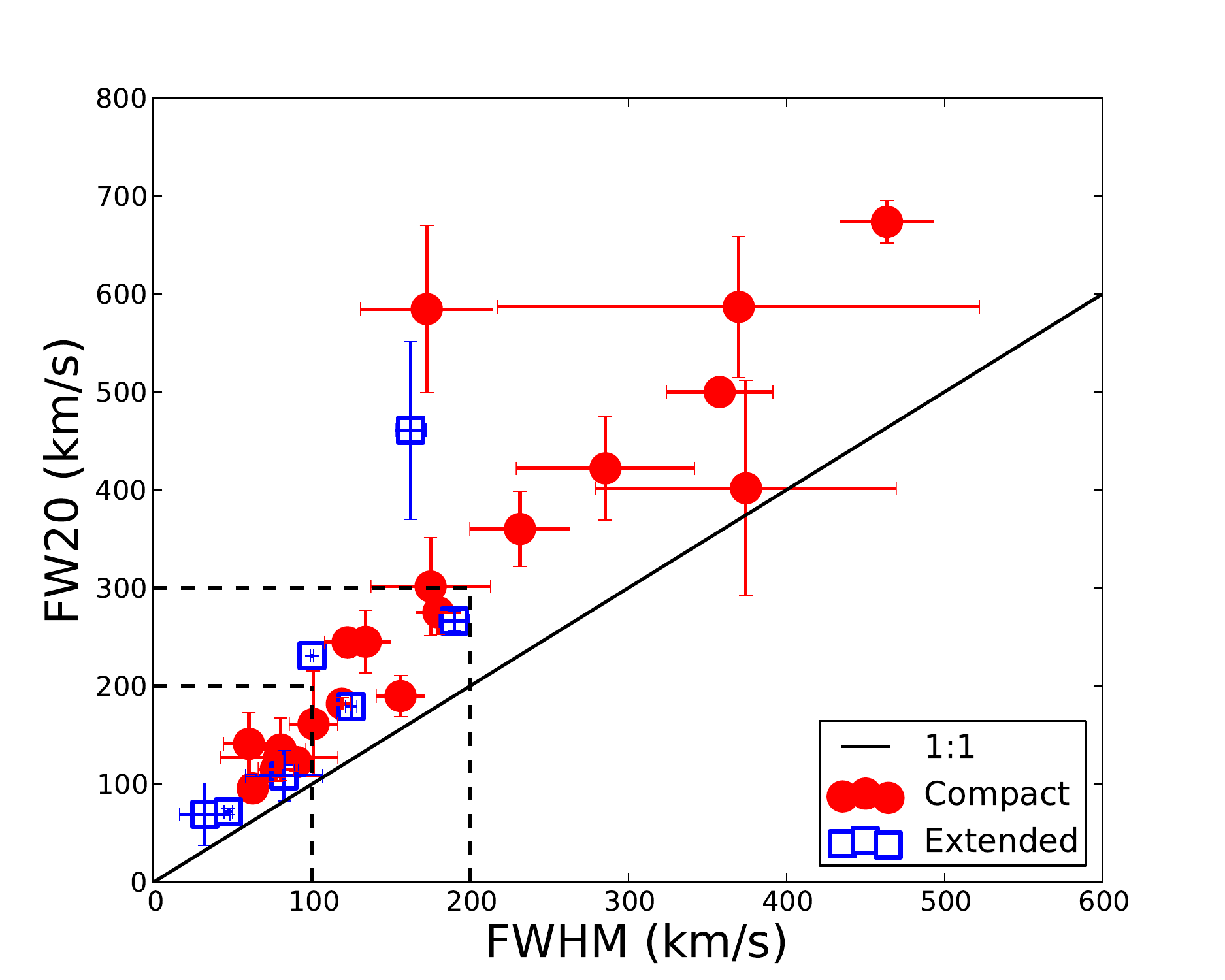}
\caption{The FWHM and FW20 width distribution of the lines we measured using the BF. We use red circles for lines originated by compact radio sources, and blue empty squares in the case of extended sources.}\label{width}
\end{center} 
\end{figure}

The profile width distribution of our sample is presented in Fig. \ref{width}. We detect a broad range of widths between 32 \kms\ $<$ FWHM $<$ 570 \kms\ and 63 \kms\ $<$ FW20 $<$ 825 \kms. Following a visual inspection, we find that broader profiles are more complex than narrow lines. We clearly see a separation in shape with increasing width, it appears that we can separate three groups, representing physically different \HI\ structures. 
The first group consists of narrow single components, the second group of broader profiles with two (or more) blended components. Whereas, among the third group of the broadest profiles, the blended lines hint that the \HI\ has multiple kinematical components. These groups are presented and separated at the dashed lines in Fig. \ref{width}, where we show FW20 vs. FWHM of the lines for the compact and extended sources. The grouping of the \HI\ profiles in Fig. \ref{Profiles1} is also based on this selection. This separation is further supported by the asymmetry analysis below.

\begin{figure*}
\begin{center}
\includegraphics[trim = 10 0 0 0, clip,width=.45\textwidth]{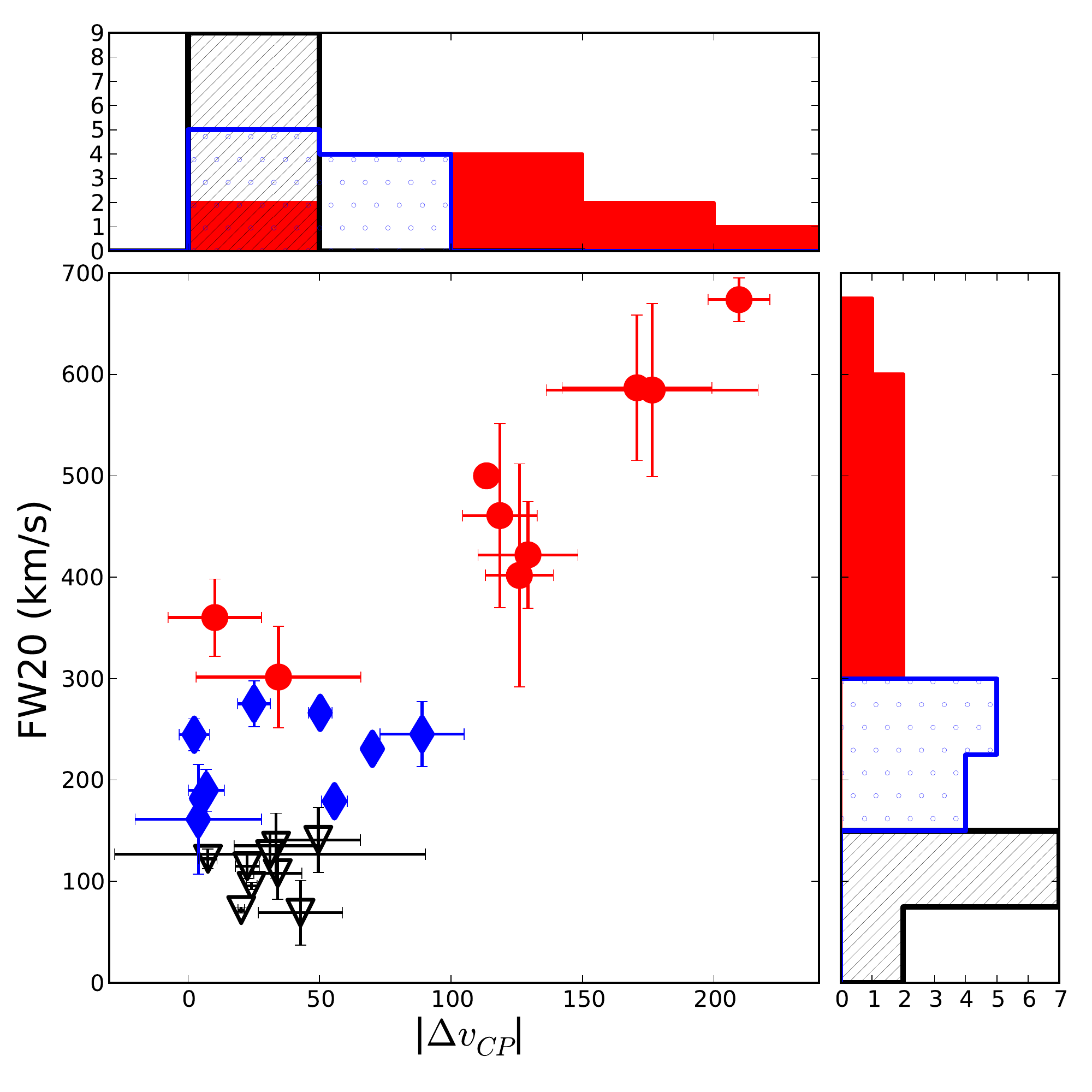}
\includegraphics[trim = 10 0 0 0, clip,width=.45\textwidth]{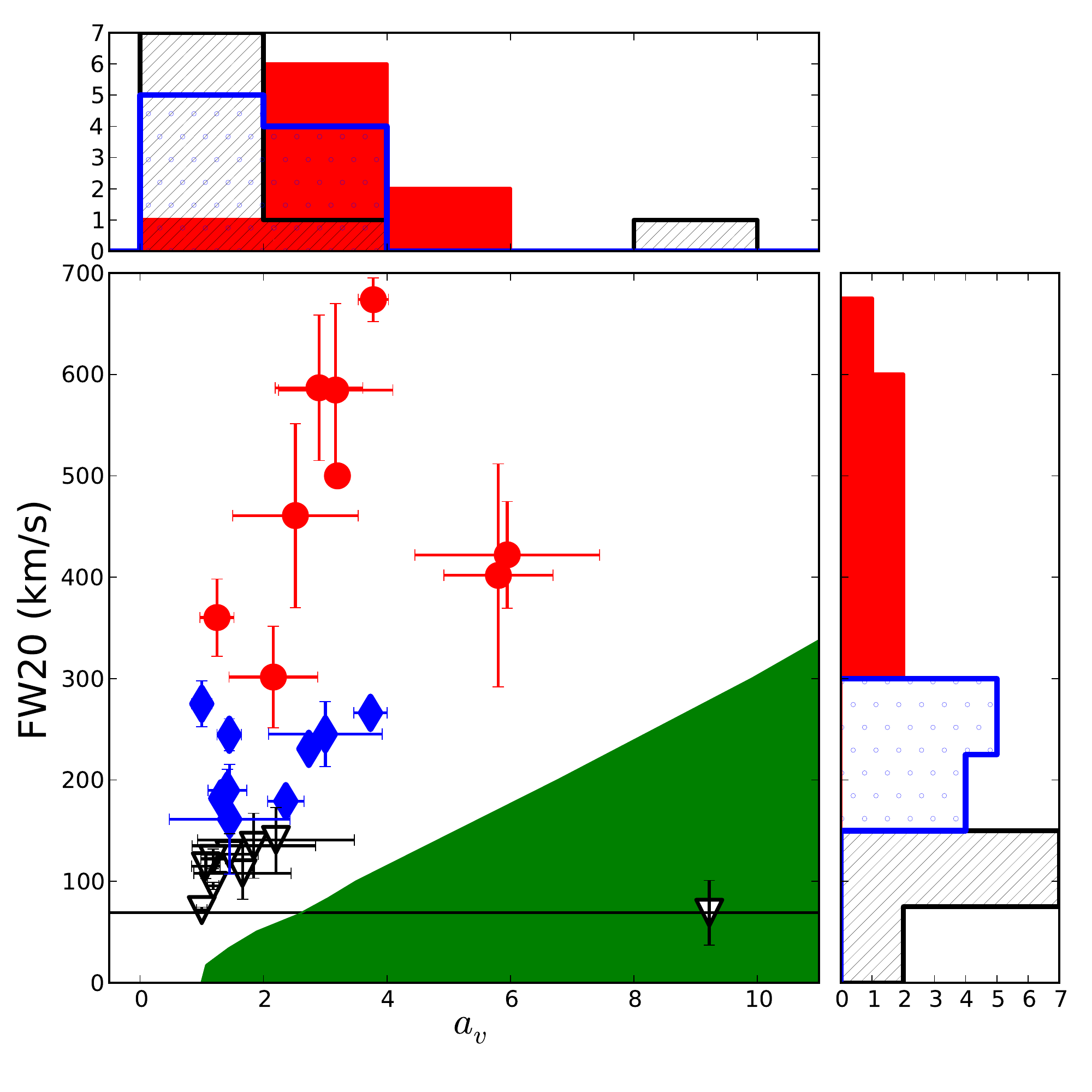}
\caption{1. (left): asymmetry ($|\Delta v_{\rm{CP}}|$)  vs. FW20 distribution of the line profiles. 2. (right): normalized asymmetry  ($\rm a_{\rm v}$) vs. FW20 width distribution. The classification of the groups is based on the width regions from Fig. \ref{width}. Black triangles mark narrow lines with FW20 $<$ 200 \kms, blue diamonds mark the middle width region with 200 \kms $<$ FW20 < 300 \kms and red circles indicate broad line detections with FW20 $>$ 300 \kms. The green region shows the asymmetry values which cannot be measured because of the resolution of our spectra. {In the histograms, the hatched region indicates narrow width profiles, the dotted hatched region marks intermediate lines while red bars mark broad lines. }
 }\label{asymmetry}
\end{center} 
\end{figure*}

In order to quantify the asymmetry of the detected lines, we derive the asymmetry parameter as the $\Delta v_{\rm{CP}} = v_{\rm{Centroid}} - v_{\rm{HI \ Peak}}$ velocity offset between the centroid and the peak intensity of the \HI\ line. In Fig. \ref{asymmetry} (left) we show the absolute value of the asymmetry distribution as function of the FW20 profile width. We find that in narrow profiles at FW20 $< 200$ \kms, the offset between the centroid and the \HI\ peak is $<$ 50 \kms. In the group with 200 \kms\ $<$ FW20 $<$ 300 \kms, the asymmetry parameters are larger, with up to 100 \kms\ difference between the line centroid and the \HI\ peak. Broad detections at FW20 $>$ 300 \kms\ have the most asymmetric profiles, with $|\Delta v_{\rm{CP}}|$ parameters larger than a few $\times$ 100 \kms. Among broad lines with FW20 $>$ 300 \kms\ there are almost no symmetric profiles detected in Fig. \ref{asymmetry} (left). Narrow lines cannot yield large values of $|\Delta v_{\rm{CP}}|$ by construction. Hence we consider a measurement of the asymmetry which is naturally normalized by the width of the line

\begin{equation}
 a_{v}=\max\bigg(\frac{v_{{\rm\small FW20}\ R}- v_{\rm{HI} \ Peak}}{ v_{\rm{HI} \ Peak}- v_{{\rm\small FW20}\ B}},\bigg(\frac{v_{{\rm\small FW20}\ R}- v_{\rm{HI} \ Peak}}{v_{\rm{HI} \ Peak}- v_{{\rm\small FW20}\ B}}\bigg)^{-1}\bigg),
\end{equation}

\noindent where $v_{\rm{HI} \ Peak}$ is the position of the peak, while $v_{{\rm\small FW20}\ R}$ and $v_{{\rm\small FW20}\ B}$ are the velocities at $20\%$ of the peak flux, on the red and blue side of the line with respect to the position of the peak. This asymmetry parameter has been widely used in characterizing $\rm{Ly\alpha}$ emission lines \citep{Rhoads2003,Kurk2004}. Here, we consider the maximum value between the velocity ratio and its reciprocal to obtain an asymmetry value always $>$ 1, independently if one line is more asymmetric towards its red or its blue side. The result is shown in Fig. \ref{asymmetry} (right). The three different groups of profiles maintain different distributions in the asymmetry value: narrow lines are mostly symmetric and the asymmetry increases as the lines become broader. Broad lines have a very different distribution compared to the other two groups. Symmetric broad lines (with FW20 $>$ 300\kms) are absent. The green shaded region shows how this asymmetry measurement is limited by the velocity resolution of our spectra. For a narrow asymmetric line, we cannot measure high asymmetries, since the convolution with the resolution element turns asymmetric narrow profiles into symmetric ones. A Kendall-tau test can give a statistical measure of the correlation between the asymmetry and the FW20 of the line. For the overall sample (excluding source $\#13$ which falls in the green region), we measure $\tau=0.51$ and a $p=0.0003$,  thus indicating the presence of a correlation at the $>3-\sigma$ significance level.

To expand on the analysis of the line asymmetries, we derive the velocity offset of the \HI\ peak (with respect to the systemic optical velocity) to quantify the blueshift/redshift distribution of the main, deepest \HI\ component.  A detection is classified as blueshifted/redshifted if the velocity offset of the line is larger than $\pm$ 100 \kms. In Fig. \ref{Offset_HIPeak}, when considering the entire sample no clear correlation is seen between the velocity offset and the width distribution of the lines, while selecting only the broad lines (FWHM>$300$ \kms) they all seem to be blueshifted . \\

Using the line parameters of the objects in the three ranges of widths (see Fig. \ref{width}), we aim to identify \HI\ structures belonging to different morphological groups: disks, clouds in radial motion (outflows, in fall), and other unsettled gas structures, for example gas-rich mergers.\footnote{Following what has been found by the detailed (and spatially resolved) studies of single objects}
In general, lines of a few $\times$ 100 \kms\ at the systemic velocity can be due to gas regularly rotating in a disk-like structure, e.g., high resolution observations show that the main (deep) absorption component in 3C 293 is associated with an \HI\ disk \citep{Beswick2004}. 
However, the origin of \HI\ profiles with broader width $\gtrsim500$ \kms\ has to involve other physical processes, e.g., disturbed kinematics due to mergers, outflows, in order to accelerate the gas to such high velocities. 
We expect to find \HI\ disks in the narrow region, while broader, asymmetric profiles, for example outflows must belong to the broadest group with large FW20 in our sample. The nature of the gas structures in the three groups is discussed in Sec. \ref{Discussion}.

\begin{table}
  \caption{Busy fit parameters of the \HI\ absorption lines. The horizontal lines separate the three groups as in Fig. \ref{Profiles3} (a), (b), (c). Sources classified as `mergers' in Table~\ref{table:detections} are marked by \small(M).}\label{table:BF}
  
   \begin{center}
 \scalebox{0.9}{  
     \begin{tabular}{l c c c c c}
\hline															     

 no. & FWHM           & FW20           &    Centroid     &    $v_{\rm{HI} \ Peak}$   \\ 
			     & (km s$^{-1}$) & (km s$^{-1}$) & (km s$^{-1}$) & (km s$^{-1}$) \\
\hline
3	&	82	$\pm$	24	&	108	$\pm$	26	&	-243	&	-209	\\
4	&	80	$\pm$	16    &	135	$\pm$ 13     &	44	&	78	\\
5	&	79	$\pm$	37	&	127	$\pm$	20	&	58	&	89	\\
10	&	62	$\pm$	1	&	96	$\pm$	4	&	-108	&	-83	\\
12	&	90	$\pm$	9	&	122	$\pm$	10	&	18	&	25	\\
13	&	32	$\pm$	16	&	69	$\pm$	32	&	34	&	80	\\
16	&	60 $\pm$ 16	&	141	$\pm$ 32 	&	356	&	406	\\
21$^{\rm\small{(M)}}$	&	43 $\pm$ 19		&   63    $\pm$ 23		&	-23	&	19	\\
30	&	47	$\pm$	3	&	72	$\pm$	3	&	-16	&	4	\\
32	&	77	$\pm$	11	&	115	$\pm$	12	&	-21	&	2	\\

\hline

1	&	122	$\pm$	15	&	245	$\pm$	16	&	-22	&	-20	\\
2	&	190	$\pm$	9	&	266	$\pm$	10	&	104	&	54	\\
6	&	119	$\pm$	6	&	182	$\pm$	6	&	-39	&	-34	\\
8	&	125	$\pm$	4	&	179	$\pm$	11	&	-67	&	-12	\\
9	&	156	$\pm$	15	&	190	$\pm$	21	&	-58	&	-51	\\
14$^{\rm\small{(M)}}$	&	146	$\pm$	7	&	175	$\pm$	9	&	-50	&	-81	\\
18	&	134	$\pm$	32	&	245	$\pm$	75	&	59	&	148	\\
23	&	100	$\pm$	1	&	231	$\pm$	1	&	-142	&	-71	\\
24	&	180	$\pm$	14	&	275	$\pm$	23	&	-196	&	-171	\\
27	&	101	$\pm$	15	&	161	$\pm$	54	&	18	&	33	\\

\hline

7$^{\rm\small{(M)}}$	&	536	$\pm$	10	&	825	$\pm$	11	&	-78	&	26	\\
11	&	175	$\pm$	38	&	301	$\pm$	50	&	-148	&	-183	\\
15	&	370	$\pm$	152	&	586	$\pm$	72	&	-285	&	-114	\\
17	&	172	$\pm$	42	&	584	$\pm$	85	&	-309	&	-132	\\
19$^{\rm\small{(M)}}$	&	272	$\pm$	4	&	416	$\pm$	5	&	27	&	28	\\
20	&	376	$\pm$	95	&	401	$\pm$	110	&	-11	&	-137	\\
22$^{\rm\small{(M)}}$	&	570	$\pm$	2	&	638	$\pm$	2	&	85	&	-95	\\
25	&	286	$\pm$	56	&	422	$\pm$	53	&	-67	&	-196	\\
26	&	162	$\pm$	10	&	461	$\pm$	91	&	20	&	139	\\
28	&	232	$\pm$	32	&	360	$\pm$	38	&	-29	&	-19	\\
29	&	464	$\pm$	30	&	674	$\pm$	22	&	-256	&	-26	\\
31	&	358	$\pm$	34	&	500	$\pm$	7	&	-36	&	-149	\\

\end{tabular}}
\end{center}
\end{table}

\section{The nature of \HI\ absorption in flux-selected radio galaxies}\label{Discussion}

Using stacking techniques, in \mbox{Paper \small{I}} we have shown that in some of our galaxies \HI\ must be distributed in a flattened (disk) morphology, whereas \HI\ has a more unsettled distribution in other galaxies of our sample. This is in good agreement with the \atlas\ study \citep{Serra} of field early-type galaxies (ETGs). \atlas\ has shown that roughly half of the \HI\ detections in ETGs are distributed in a disk/ring morphology, and \HI\ has an unsettled morphology in the other half of the detected cases.

Here, our main goal is to use the BF parameters (see Table \ref{table:BF}) to identify such disks and unsettled structures. As mentioned in Sec. \ref{sec:BFcharcteristics}, depending on the different shapes of the profiles we expect to find different morphological structures in the three groups separated by the dashed lines in Fig. \ref{width}.

\begin{figure}[!b]
\begin{center}
\includegraphics[trim = 0 10 10 0, clip,width=.45\textwidth]{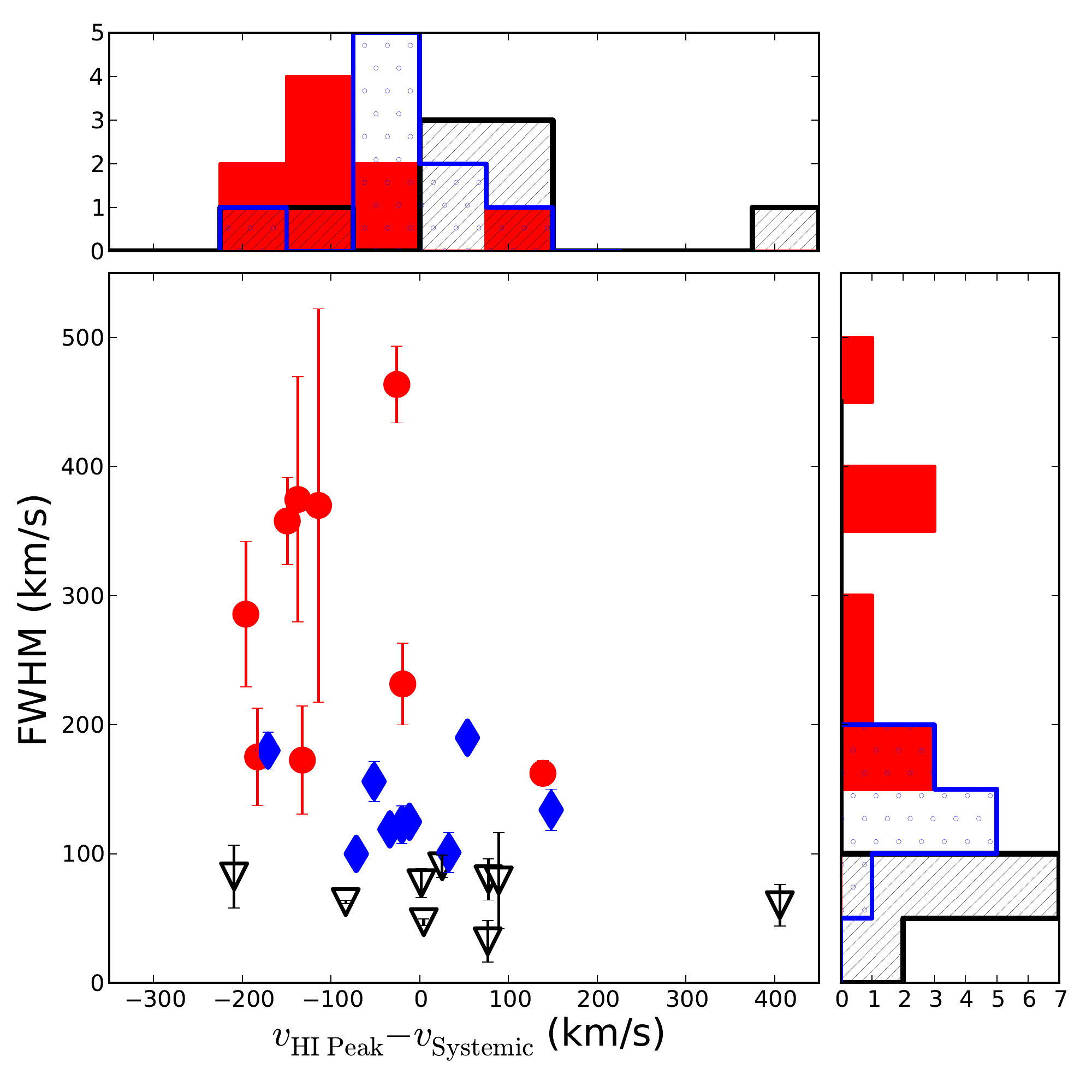}
\caption{Blueshift/redshift distribution of the \HI\ peak with respect to the systemic velocity vs. the FWHM of the lines of the sample. The symbols are the same as in Fig. \ref{asymmetry}}\label{Offset_HIPeak}
\end{center} 
\end{figure}

In Fig. \ref{Offset_HIPeak}, most of the narrow lines with FWHM $<$ 100 \kms\ (8 out of 9 sources; $8^{+3.95}_{-2.76}$ following Poisson statistics, see~\cite{gehrels1986}) are detected close to the systemic velocity with $v_{\rm{HI \ Peak}}$ $<$ $\pm$ 100 \kms. Narrow profiles at the systemic velocity are most likely produced by large scale disks, as seen in the case of the \atlas\ sample of early-type galaxies, where typical FWHM $<$ 80 \kms\ have been found for the \HI\ absorption lines.
Previously, \HI\ disks with similar profile characteristics have been observed in radio galaxies, e.g., in Cygnus A \citep{Conway1999, Struve2010}, Hydra A \citep{Dwara1995}.
Besides disks at the systemic velocity, for narrow lines we also see one case where the \HI\ peak is redshifted by $+$406 \kms\ (in source no. 16). Such narrow lines can be produced by infalling gas clouds with low velocity dispersion.
Similar cases of highly redshifted narrow lines have been detected before, e.g., in NGC 315 the narrow absorption is redshifted by $+$500 \kms. \cite{Morganti2009} found that the redshifted \HI\ line in NGC 315 is produced by a gas cloud at a few kpc distance from the nucleus.
In 4C 31.04, a neutral hydrogen cloud is detected with 400 \kms\ projected velocity towards the host galaxy \citep{Mirabel1990}, whereas in Perseus A the \HI\ absorption is redshifted by $\sim$3000 \kms\ \citep{Gorkom1983, Young1973}, and its nature is still unclear.

In Fig.\ref{Offset_HIPeak}, for the objects with intermediate widths of Fig. \ref{width}, we see that the \HI\ is still detected close to the systemic velocity in most of the cases (7 out of 9 sources, $7^{+3.77}_{-2.58}$ in Poisson statistics), while only two lines are blueshifted/redshifted more than $100$\kms. In Fig. \ref{Profiles2} (b) we see that multiple \HI\ components of unsettled gas make the \HI\ kinematics more complex in this group. These are indications that relatively large widths of 100 \kms\ $<$ FWHM $<$ 200 \kms\ (200 \kms\ $<$ FW20 $<$ 300 \kms\ ) can be produced by similar gas structures as narrow detections (disks, clouds), but with more complex kinematics.

Among the broadest lines with FWHM $>$ 200 \kms, instead, the main \HI\ component is blueshifted/redshifted in 7 out of 9 cases ($7^{+3.77}_{-2.58}$ in Poisson statistics). Hence, the blueshift/redshift distribution of the broadest lines is different from the other two groups at the $2-\sigma$ confidence level. As mentioned in Sec. \ref{sec:BFcharcteristics}, in Fig. \ref{asymmetry} there are no symmetric lines with width FWHM $>$ 200 \kms. Therefore, it is suggested that when a profile is broad, it is also asymmetric and mostly shifted with respect to the systemic velocity. The combination of these features may indicate that these profiles are tracing gas which is not simply rotating in a disk, but it is unsettled and may be interacting with the radio source.

\begin{figure}
\begin{center}
\includegraphics[trim = 0 10 10 0, clip,width=.45\textwidth]{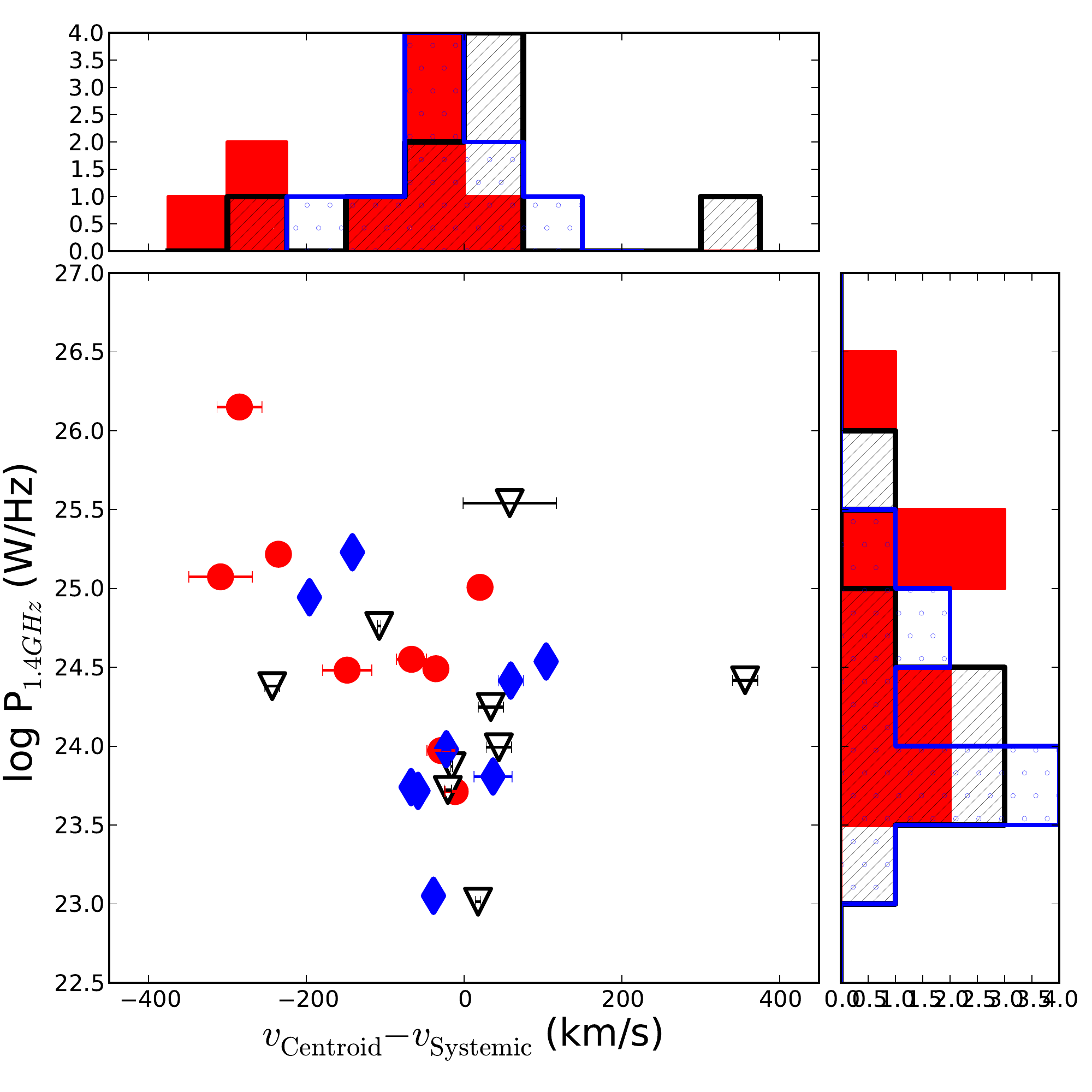}
\caption{Blueshift/redshift distribution of the \HI\ line centroid with respect to the systemic velocity vs. the radio power in the sample. 
The symbols are the same as in Fig. \ref{asymmetry} }\label{Regions}
\end{center} 
\end{figure}

Indeed, for these widths we find blueshifted, broad wings, e.g., in 3C 305, where both the kinematical and spatial properties of the \HI\ may indicate the presence of fast, jet-driven outflows \citep{Morganti2005}. In fact, when broad and blueshifted wings occur, the centroid velocity is a better measure of the line offset with respect to the systemic velocity, than the \HI\ peak. To test any connection of the radio power with the \HI\ gas motions, in Fig. \ref{Regions} we plot the velocity offset of the \HI\ centroid against the radio power of the AGN. Below, in Sec. \ref{Sec:interactingAGN} we discuss the gas and radio source properties of such blueshifted, broad lines.

\subsection{Are powerful AGN interacting with their ambient gaseous medium?}\label{Sec:interactingAGN}

In Fig. \ref{Regions} we consider the shift of the centroid of the line with respect to the systemic velocity: $v_{\rm{Centroid}} - v_{\rm{Systemic}}$. Overall, following Poisson statistics, $8^{+3.95}_{-2.76}$ profiles are blueshifted ($v_{\rm{Centroid}} - v_{\rm{Systemic}}$ < -100 \kms), fewer ($2^{+2.6}_{-1.3}$ sources) detections are redshifted ($v_{\rm{Centroid}} - v_{\rm{Systemic}}$ > +100 \kms), and $17^{+5.2}_{-4.1}$ are detected close to the systemic velocity. Thus, even if most of the \HI\ absorption lines are found at the systemic velocity (between $63\%^{+19\%}_{-15\%}$), many more lines appear to be blueshifted ($30\%^{+15\%}_{-10\%}$), then redshifted ($7.5\%^{+9.6\%}_{-4.8\%}$). The blueshift/redshift distribution of \HI\ absorption lines was previously studied by \cite{Vermeulen2003}, who found a similar trend in a sample of GPS and CSS sources. In the sample of \cite{Vermeulen2003}, 37\% of the profiles are blueshifted, and 16\% are redshifted with respect to the systemic velocity. A later study confirmed this trend, \cite{Gupta2006} reported a high, 65\% detection rate of blueshifted \HI\ profiles in GPS sources. These studies speculate that interactions between the radio source and the surrounding gaseous medium is the cause of the outflowing gas motions in higher luminosity sources. To further investigate if there is a significant anti-correlation between the blueshift/redshift distribution of the lines and the radio power of the sources, we carried out a Kendall-tau test over the entire sample. We measure $\tau=0.44$ and $p=0.0012$, thus indicating the presence of a correlation at the $\sim3-\sigma$ significance level.

\begin{figure}
\begin{center}
\includegraphics[trim = 0 10 10 0, clip,width=.45\textwidth]{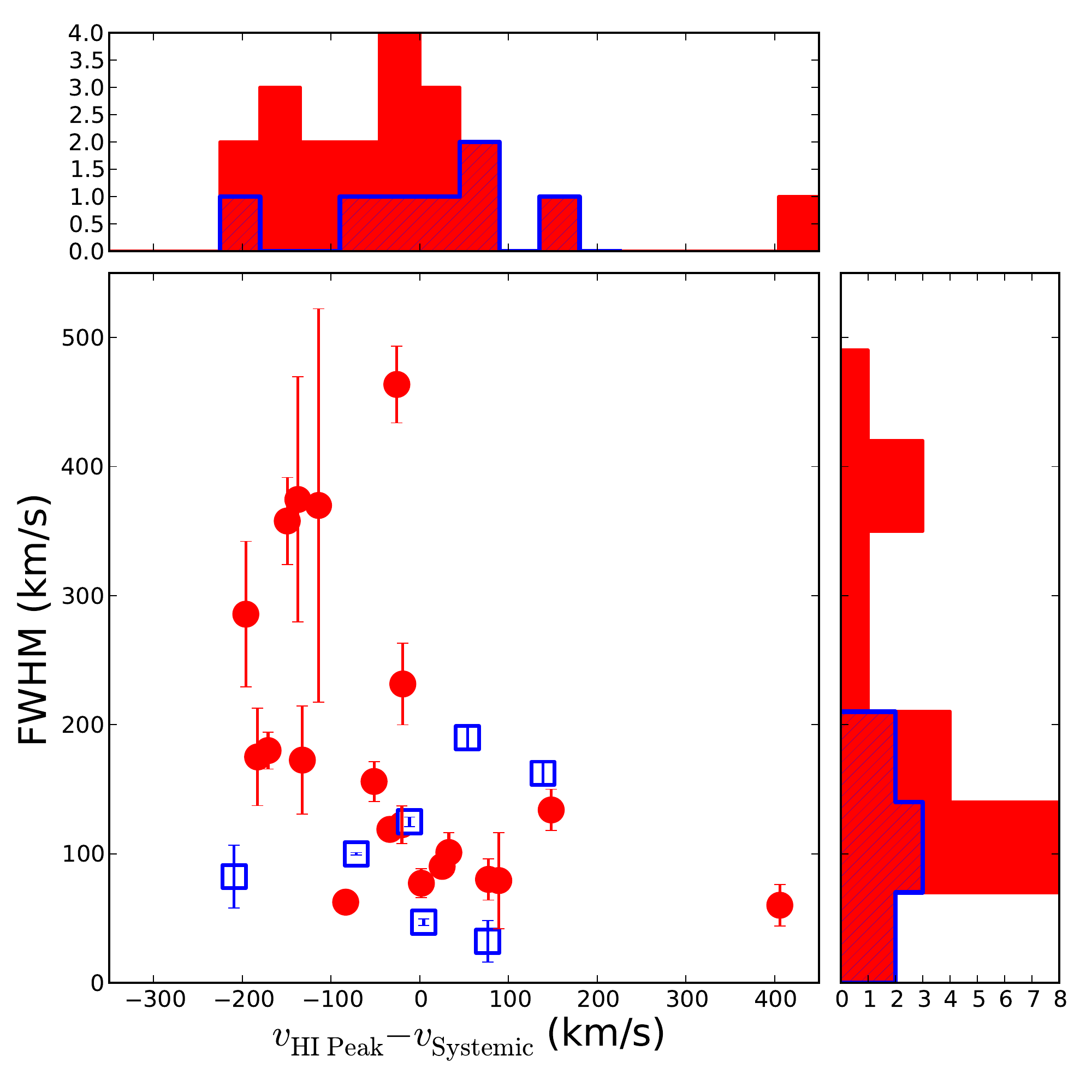}
\caption{Blueshift/redshift of the \HI\ peak with respect to the systemic velocity vs. FWHM distribution in compact (red circles) and extended (blue empty squares) sources.}\label{CompExt_BlueRedshift}
\end{center}
\end{figure}

The three most blueshifted ($v_{\rm{Centroid}} - v_{\rm{Systemic}}$ $<$ -250 \kms) profiles in Fig. \ref{Regions} are broad lines with FW20 $> 500$ \kms.
By identifier number these are no. 15, no. 17 and no. 29. These detections show similar kinematical properties as the outflows in 3C 293 and 3C 305 by displaying broad wings of blueshifted absorption. 

The \HI\ outflows in 3C 305 and 3C 293 are driven by powerful radio sources with log(P$_{1.4 \ \rm GHz}$) $>$ 25 W Hz$^{-1}$ (see Table \ref{table:detections}). It was estimated that in 3C 305 and 3C 293, the kinetic energy output of the jets is high enough to accelerate the gas to high velocities of about 1000 \kms\ \citep{Morganti2003, Morganti2005, Mahony2013}. 
In Fig. \ref{Regions} it appears that blueshifted, broad detections in our sample (in no. 15, no. 17 and no. 29) are likely to occur in high power radio galaxies with log(P$_{\rm{1.4 \ GHz}}$) $>$ 25 W Hz$^{-1}$, suggesting that their energy output is similar to that of 3C 305 and 3C 293. 
These are indications that interactions with the powerful radio source may be driving \HI\ outflows in these sources, and we discuss this possibility in more detail in Sec. \ref{Outflows}.

A broad blueshifted line can also trace phenomena other than outflows. For example, no. 31 has a very broad asymmetric \HI\ profile with multiple peaks, and it is an early-type galaxy in the Abell cluster. The shape of the profile may be indicative of complex gas motions within the galaxy cluster. \HI\ in absorption in clusters has been detected before in Abell 2597 \citep{ODea1994, Taylor1999} and in Abell 1795 \citep{Bemmel2012}.

\subsection{Fraction and timescale of candidate \HI\ outflows}\label{Outflows}

Because \HI\ outflows are very faint, with typical optical depth of $\tau$ = 0.01, until now only a handful of confirmed \HI\ outflows are known \citep{Morganti1998, Morganti2003, Morganti2013a, Kanekar2008, Tadhunter2014}. In our sample, besides the already known \HI\ outflows of 3C 305 \citep{Morganti2005} and 3C 293 \citep{Mahony2013}, we also have three new absorption lines, where in addition to the main \HI\ component (deep \HI\ detection close to the systemic velocity) a blueshifted shallow wing is seen: no. 15, no. 17, and no. 29.

Source 4C +52.37 (or source no. 29) is a CSS source from the CORALZ sample (COmpact Radio sources at Low redshift, see more in Sec. \ref{sec:CompExt}), and we find a broad blueshifted wing in this galaxy with FW20 = 674 \kms\ (see Table \ref{table:BF}). The dataset of the first set of observations of this source is highly affected by RFI. Thus, in order to verify our detection, we carried out follow-up observations of 4C +52.37, and confirm the presence of the wing by the second set of observations. Source no. 15 and no. 17 share similar kinematical properties, showing broad lines of almost $\sim$590 \kms\ FWHM. Even though in no. 15 the main, deeper component is not as prominent as in other two cases, \cite{Saikia2003} detected higher degree of polarization asymmetry in this object (4C +49.25). \cite{Saikia2003} argue that such polarization properties can indicate interactions of the radio source with clouds of gas which possibly fuel the AGN.

Based on their radio continuum properties and \HI\ kinematics (see Sec. \ref{Sec:interactingAGN}), we consider sources no. 15, no. 17, and no. 29 the best candidates for hosting jet-driven \HI\ outflows. However, more sensitive and higher resolution observations are needed to verify that these detections are indeed jet-driven, and to estimate how much of the energy output is concentrated in the jets.

\begin{figure*}
\begin{center}
\includegraphics[trim = 10 0 0 0,clip,width=.45\textwidth]{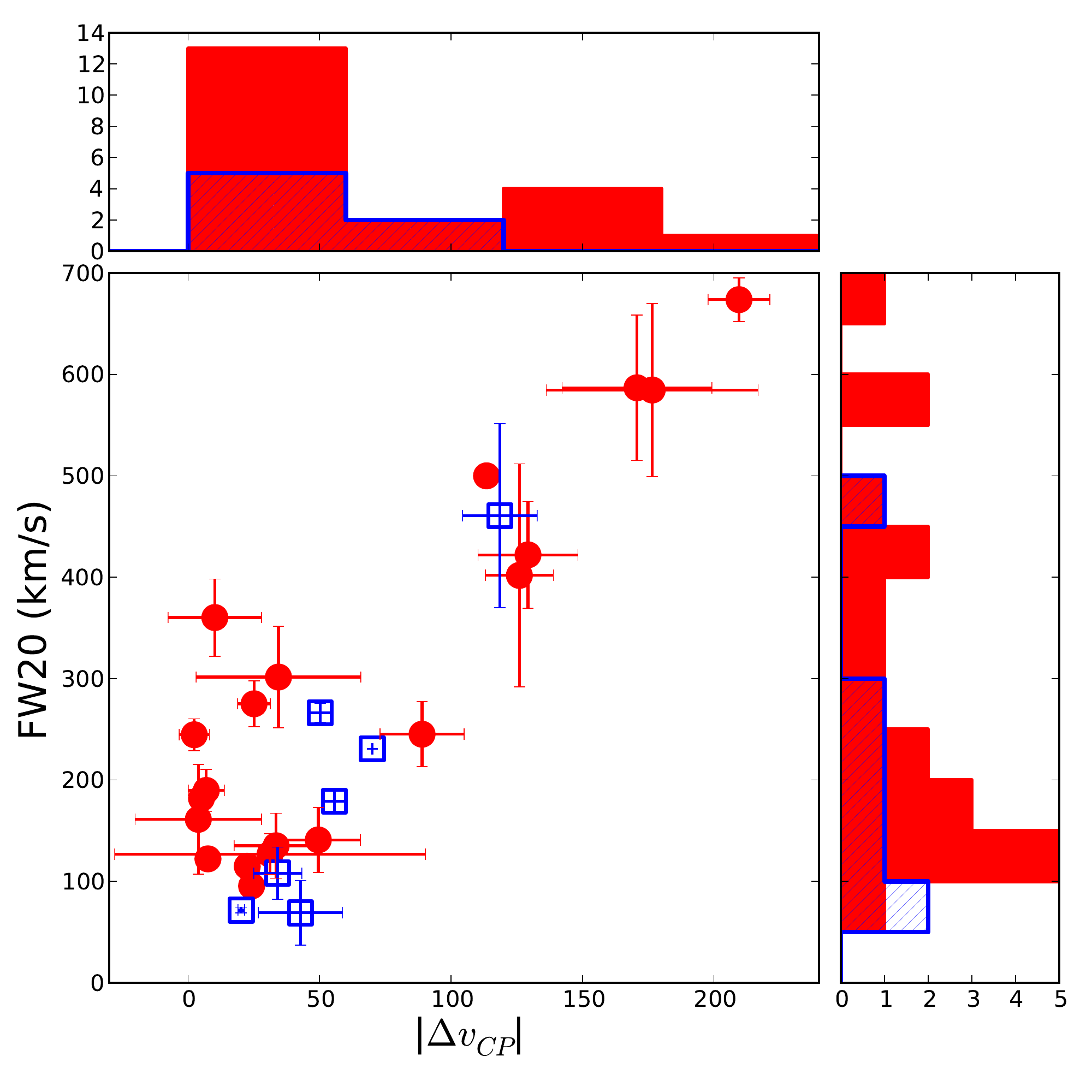}
\includegraphics[trim = 10 0 0 0,clip,width=.45\textwidth]{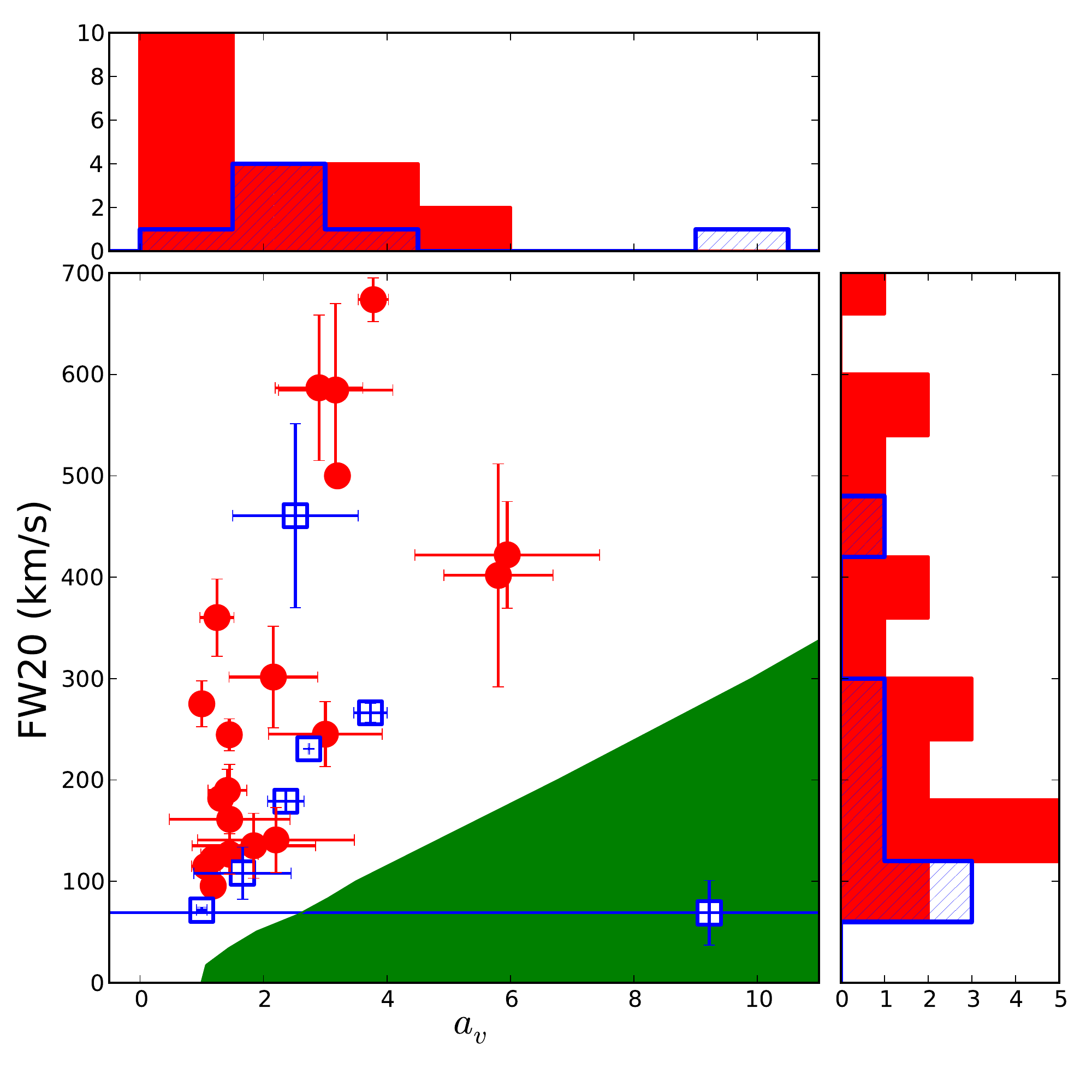}
\caption{1. (left): Asymmetry parameter vs. FW20 distribution of the lines of compact (red circles) and extended (blue empty squares) sources. 2. (right) Normalized asymmetry parameter vs. FW20 width distribution of the same groups. The green region shows the asymmetry values which cannot be measured because of the resolution of our spectra.
In the histograms compact sources are marked by red bars, while the hatched blue region indicates extended sources.
}\label{CompExt}
\end{center}
\end{figure*}

Including the two already known outflows in 3C~305, 3C~293, the detection rate of outflows, at the sensitivity of our observations, is  $15\%$ among all sources where absorption is detected, or $5\%$ in the total AGN sample.

Two caveats should be considered when discussing the detection rate.
If the spin temperature of the gas associated with the outflow is higher than for the quiescent \HI\ (as indeed found in some cases, e.g., PKS~1549-79, \citealt{Holt2006}), the outflows would be harder to detect. It is, however, worth remembering that gas with high column density can be present in the outflows and, in fact, the most massive component is found to be cold molecular gas (i.e., high density gas). Thus, the column density of the \HI\ associated with the outflow may be above the detection limit, even for higher spin temperatures.

The second caveat concerns the covering factor of the outflowing \HI\ which may, under unfavorable circumstances, reduce the strength and/or the amplitude of the absorption. Indeed, \cite{Holt2008} identified, for a sample of young radio sources, a dependence between the observed velocities of the outflows (derived from the optical emission lines) and the source orientation with respect to the observer's line of sight.  

Following all this, the detected 5\% may represent only a lower limit of the occurrence of \HI\ outflows.
 
It is interesting to note that, if we assume that every radio source goes through a phase of outflow during their typical lifetime (between a few $\times 10^7$ and $10^8$ yr, \citealt{Parma1999, Parma2007}), the derived detection rate may suggest that \HI\ outflows represent a relatively short phase (on average a few Myr) in the life of the galaxy (for each cycle of radio-loud activity). Although this result is affected by the caveats listed above, it is worth noting that similar conclusions have been drawn from observations of the molecular gas \citep{Alatalo2011, Morganti2013b, Cicone2014, Guillard2014}.

\noindent
\subsection{Gas rich mergers}\label{mergers}
 Among our detections we find three sources (UGC 05101, UGC 8387, Mrk 273) which are embedded in gas-rich merging systems. In these cases, both the AGN and the enhanced star-forming regions are likely to be the origin of the radio emission, hence we excluded these objects from the previous analysis. Their \HI\ line would belong to the group of broad profiles we presented above. FW20 is equal to 825 \kms, 416 \kms, and 638 \kms\ respectively. The lines are multi-peaked. The profiles are shown in Fig. \ref{Profiles3} (c) under the identifier numbers no. 7, no. 19, and no. 22. Their BF parameters are summarized in Table~\ref{table:BF}.   
 
In Table \ref{table:detections}, gas-rich mergers have the highest integrated optical depth, with corresponding column densities (5 - 8) $\times$ $ 10^{19}$ ($T_{\rm{spin}}/c_{\rm{f}}$) cm$^{-2}$, reflecting extreme physical conditions of the gas in merging galaxies. Even though the presence of AGN is not always clear in merging systems, there exist tentative signs that the presence of gas has an effect on the growth of merging BHs. Very Long Baseline Array (VLBA) observations of the \HI\ absorption in Mrk 273 by \cite{Carilli2000} show that the broad \HI\ profile is the result of several co-added components in this source (see notes on individual sources in Appendix \ref{notes}). In particular, \cite{Carilli2000} detected an infalling gas cloud towards the south-eastern component (SE) of Mrk 273, indicative of BH feeding processes. Considering the broad and multi-peaked nature of the \HI\ in UGC 8387 and UGC 05101, these sources likely have similar gas properties to Mrk 273, e.g., \HI\ absorption originating from several unsettled components. With our low-resolution observations we cannot distinguish between the different absorbing regions in merging systems, therefore we detect the blended, broad \HI\ signal.

\section{The \HI\ properties of compact and extended sources}\label{sec:CompExt}

In \mbox{Paper \small{I}} we found that on average compact sources have higher $\tau$, FWHM, and column density than extended sources. Here, using the BF parameters we expand on these results by examining in more detail the difference in the \HI\ properties of the two types of radio sources.

In Fig. \ref{CompExt_BlueRedshift} and Fig. \ref{CompExt} we show the shift and asymmetry of the lines vs. their width, for compact and extended sources.
As also expected from the stacking results, compact sources tend to have broader, more asymmetric and more commonly blueshifted/redshifted profiles than extended sources. 
In \mbox{Paper \small{I}} we suggested that the larger width in compact sources is due to the presence of unsettled gas. In Sec. \ref{Discussion} we show that unsettled gas is typically traced by asymmetric lines; furthermore redshifted/blueshifted lines can also indicate non-rotational gas motions.

In Fig. \ref{CompExt}, among broad lines with FW20 $>$ 300 \kms, almost all sources are classified as compact (8 out of 9), while only one source is extended. The latter extended source is 3C 305, which object appears resolved in our classification scheme on the scales of the FIRST survey. However, it is interesting to mention that 3C 305 is a small, $\sim$ 3 kpc radio source \citep{Morganti2005}. Therefore, in fact, all broad \HI\ lines detected in our sample reside in rather compact radio sources.
The largest asymmetry of $|\Delta v_{\rm{CP}}|$ $\sim$ 250 \kms\ is measured in the compact source 4C +52.37, one of our \HI\ outflow candidates. 
Furthermore, except for one case (source no. 3), all blueshifted detections with $v_{\rm{HI \ Peak}}$ $<$ -100 \kms\ are compact sources in Fig. \ref{CompExt_BlueRedshift}.

\begin{figure}
\begin{center}
\includegraphics[trim = 0 10 10 0,clip,width=.45\textwidth]{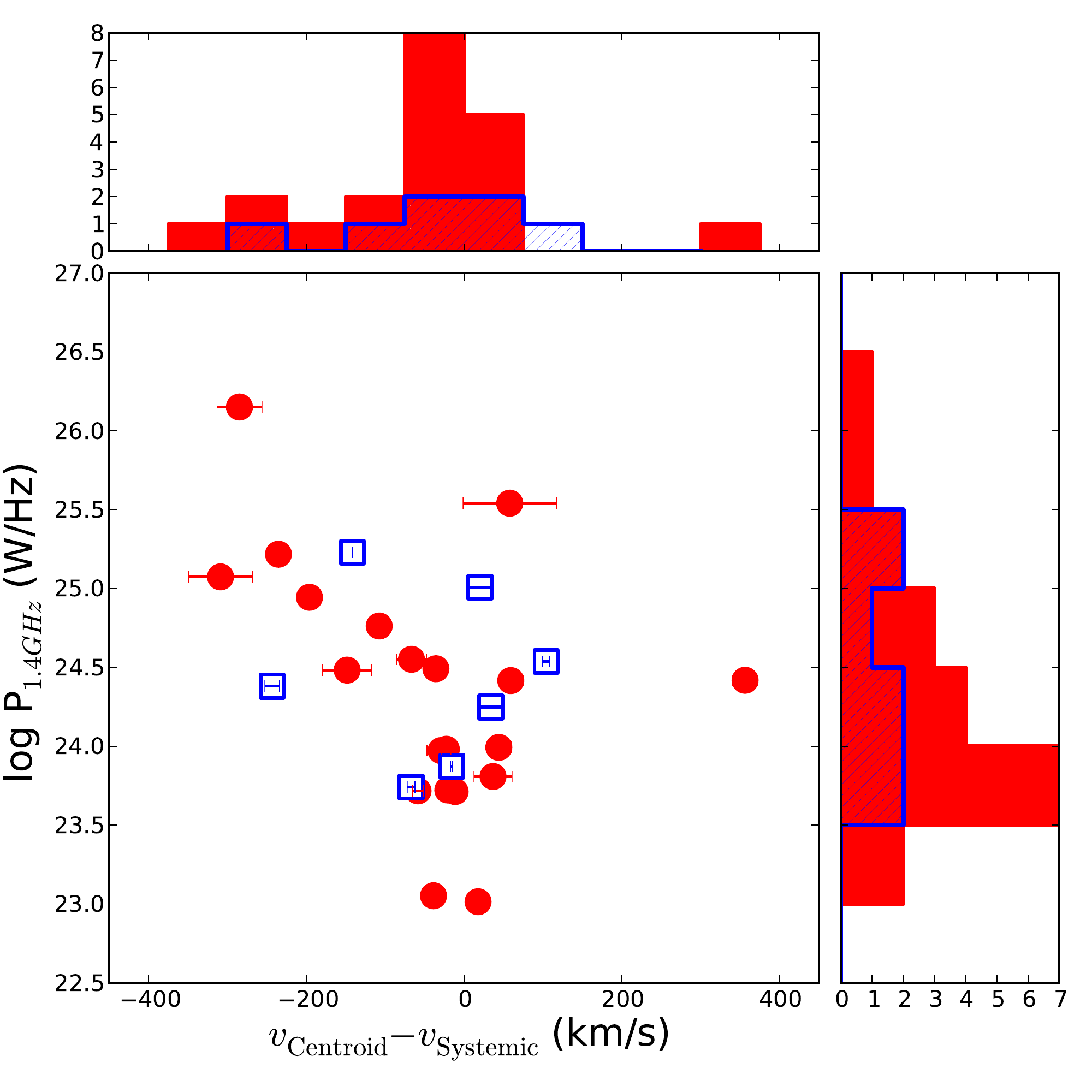}
\caption{Blueshift/redshift distribution of the \HI\ line centroid vs. the radio power of compact (red circles) and extended sources (blue empty squares).}\label{CompExt_RadoPow}
\end{center}
\end{figure}

Fig. \ref{CompExt_BlueRedshift} and Fig. \ref{CompExt} show that the traces of unsettled gas, e.g., blueshifted and broad/asymmetric lines, are found more often among compact sources. This suggests a link between the morphology of the radio source and the kinematics of the surrounding gas. In fact, all three \HI\ outflow candidates (no. 15, no. 17, and no. 29) from Sec. \ref{Outflows} are classified as compact, see Table \ref{table:detections}. Fig. \ref{CompExt_RadoPow} shows the radio power of the sources against the shift of the line, for compact and extended groups. Considering only the compact ones, we measure $\tau=0.57$ and a $p=0.0004$ for the Kendall-tau test between the two variables. Hence, we measure a correlation between the $|v_{\rm{Centroid}} - v_{\rm{Systemic}}|$ and the radio power of the sources at the $>3-\sigma$ significance level. 

All these properties suggest that the traces of unsettled \HI\ are more often associated with compact sources, hinting  that the interactions between small radio sources and the rich ambient medium are likely to occur in the young, compact phase of AGN, providing favorable sites for powerful jet-cloud interactions.

As we mention in \mbox{Paper \small{I}}, nine of our AGN are part of the COmpact RAdio sources at Low redshift (CORALZ) sample \citep{Snellen2004, Vries2009}, a collection of young CSS and GPS sources. The \cite{Vries2009} observations provided high resolution MERLIN, EVN, and global VLBI observations of the CORALZ sample at frequencies between 1.4 - 5 GHz, along with radio morphological classification and source size measurements.

\begin{figure}
\begin{center}
\includegraphics[trim = 20 180 20 200, clip,width=.52\textwidth]{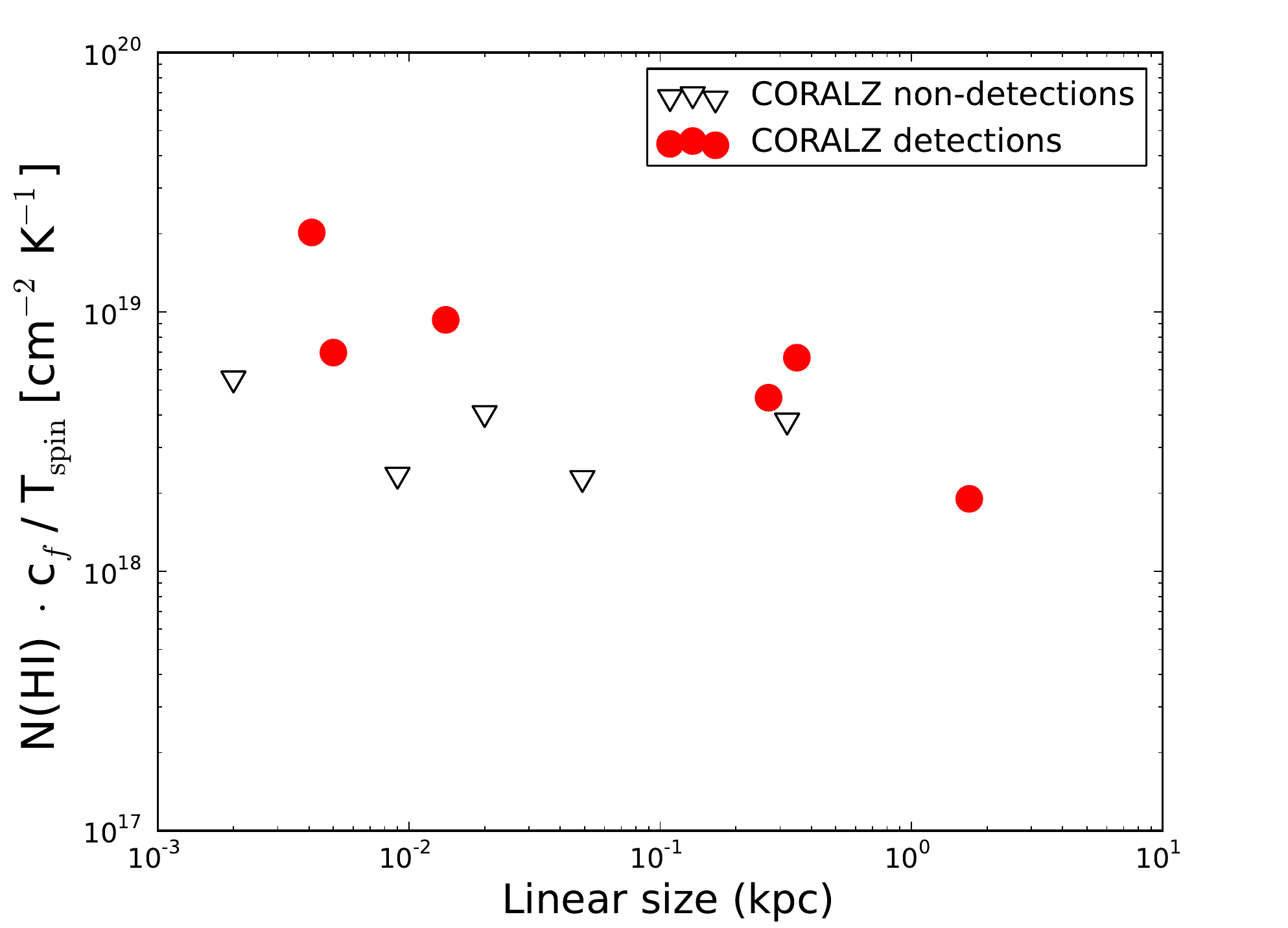}
\caption{Radio source size vs. column density in GPS and CSS sources from the CORALZ sample}\label{fig:size_NHI}
\end{center}
\end{figure}

Previously, \HI\ observations of 18 CORALZ sources were obtained by \cite{Chandola}, yielding a 40\% \HI\ detection rate. Our sample includes fewer, 11 objects, and we find a 55\% detection rate (six objects). We observed three CORALZ sources which were not studied by \cite{Chandola}. Among the three sources we have two new detections: no. 20, no. 21, and one non-detection: no. 52.

\cite{Pihlstrom,Gupta2006}, and \cite{Chandola} reported a column density vs. radio source size anti-correlation for CSS and GPS sources, accounting for the fact that at larger distances from the nucleus, lower opacity gas is probed in front of the continuum \citep{Fanti1995, Pihlstrom}. More recently, \cite{Curran2013} pointed out that the N(\HI)-radio-size inverse correlation is driven by the fact that the optical depth is anti-correlated with the linear extent of the radio source. In Fig. \ref{fig:size_NHI} we plot the largest projected linear size (LLS) of the sources, reported by \cite{Vries2009}, against the column densities measured from our \HI\ profiles. For the non-detections (\emph{white triangles}) we consider the upper limits measured from our observations. The column density of the CORALZ detections seems to decrease as a function of radio source size. A Kendall-tau test between the two variables measures $\tau=-0.73$ and $p=0.038$. The result, though, is affected by the small number of sources considered. Including also the N(\HI) upper limits for the non-detections, the trend becomes much less clear.
In fact, high frequency peakers (HFPs) were also found not to be following the inverse correlation \citep{Orienti2006}. HFP galaxies are thought to be recently triggered, 10$^{3}$-10$^{5}$ yr old, small radio sources of a few tens of pc. \cite{Orienti2006} measured low optical depth in these tiny sources, and they explain this by a combination of orientation effects and the small size of the sources. In this scenario, our line of sight intersects the inner region of the circumnuclear torus against the tiny radio source, therefore in absorption we can only detect this high $T_{\rm{spin}}$, (and therefore) low column density gas close to the nucleus. 

The above results suggest that both orientation effects and radio source size can affect the measured optical depths. The combination of these effects may be responsible for deviations from the N(\HI)-to-radio-size inverse correlation in Fig. \ref{fig:size_NHI}.

\section{Summary}

In this paper we presented the results of an \HI\ absorption study of a sample of 101 AGN. 
The relatively large sample of 32 detections has been parametrized using the BF \citep{Westmeier2014}.
The total sample was selected and used for stacking purposes in \mbox{Paper \small{I}}, and here we carry out a detailed analysis of the individual profiles.
Detections in our sample display a broad range of line shapes and kinematics. The BF is efficient in fitting almost all of the spectra, except for a few peculiar cases with multiple lines and \HI\ emission features.

In \mbox{Paper \small{I}} we find that \HI\ disks and unsettled gas structures are both present in our sample. Here we attempt to disentangle different \HI\ morphological structures using the BF parameters.
We find that the complexity of the lines increases with increasing profile width. Based on the line shapes we separate three groups of objects with different kinematical properties. The narrowest lines with FWHM $<$ 100 \kms\ are most likely produced by \HI\ rotating in large scale disks or \HI\ clouds. Relatively broad lines (100 \kms\ $<$ FWHM $<$ 200 \kms) may be produced by similar morphological structures with more complex kinematics. Broad lines with FWHM $>$ 200 \kms, however, may be produced not by simple rotation, but by the most unsettled gas structures, e.g., gas-rich mergers and outflows.

We detect three new profiles with broad, blueshifted \HI\ wings. Since their radio sources are powerful $\log(\rm{P}_{\rm 1.4 \ GHz})\gtrsim$ 25 W Hz$^{-1}$, these lines are the best candidates for tracing jet-driven \HI\ outflows. Considering certain and tentative cases, the detection rate of \HI\ outflows is 5\% in our total sample. 
This fraction represents a lower limit due to caveats concerning the spin temperature and covering factor of the outflowing gas, and it should be confirmed by more detailed \HI\ observations. However, if outflows are a characteristic phenomenon of all radio AGN, the relatively low detection rate would suggests that the depletion timescale of \HI\ outflows is short compared to the lifetime of radio AGN. This would be consistent with what is derived from observations of molecular outflows.

In \mbox{Paper \small{I}} we show that the stacked profile of compact sources is broader than the stacked width of extended sources. Here we confirm this result using the BF parameters of the individual detections. \HI\ in compact sources often shows the characteristics of unsettled gas, e.g., blueshifted lines and broad/asymmetric profiles. Such \HI\ line properties suggest that strong interactions between AGN and their rich circumnuclear medium are likely to occur in compact AGN, as young radio jets are clearing their way through the ambient medium in the early phases of the nuclear activity.

%We are currently increasing the observed sample to

\section{Acknowledgements}
We thank the referee for the useful and detailed comments that
helped us to improve the manuscript.
We thank Tobias Westmeier and Russell Jurek for the useful suggestions on the BF fitting module.
The WSRT is operated by the ASTRON (Netherlands Foundation for Research in Astronomy) with support from the Netherlands Foundation
for Scientific Research (NWO). This research makes use of the SDSS Archive, funding for the creation
and distribution of which was provided by the Alfred P. Sloan
Foundation, the Participating Institutions, the National Aeronautics
and Space Administration, the National Science Foundation, the
U.S. Department of Energy, the Japanese Monbukagakusho, and
the Max Planck Society.
RM and FMM gratefully acknowledge support from the European Research Council under the European Union's Seventh Framework Programme (FP/2007-2013) /ERC Advanced Grant RADIOLIFE-320745.

\newpage 

\appendix \label{App:AppendixA}

\section{Notes on the individual detections}\label{notes}
\noindent
{\bf 1: B3 0754+401}\\
This source is part of the low luminosity CSS sample presented by \cite{Kunert2010a}. The largest linear size of the source is 0.25 kpc, and the morphology remains unresolved in the multi-element radio linked interferometer network (MERLIN) observations. \cite{Kunert2010a} classified this object as a High Excitation Galaxy (HEG). We detect \HI\ in this source at the systemic velocity. The width and asymmetry parameters of the \HI\ suggest that the gas in this galaxy is not entirely settled.\\
\noindent
{\bf 2: 2MASX J08060148+1906142} \\
This source has never been studied individually before. The relatively large width (FW20 = 266\kms) and double-peaked nature of the \HI\ profile suggest that unsettled gas is present in this galaxy.  \\
\noindent
{\bf 3: B2 0806+35}\\
This source shows radio emission both on the kpc and pc scale. It has been observed with VLBA at $5$ GHz as part of a polarization survey of BL-Lac objects \citep{Bondi2004} as a possible BL-Lac candidate. On the parsec scale (beam size of $3.2\times1.7$ mas) the source reveals a radio core with an extended jet towards the south. The jet extends for about $10$ milliarcseconds (mas) ($\sim15$ pc at $z=-0.0825$). Among the BL-Lac candidates, this source is the weakest object, and has the steepest radio spectrum. It is also the only object not showing polarized emission neither in the jet or the core.  In this galaxy, the narrow \HI\ line is detected at blueshifted velocities with respect to the systemic. \\
\noindent
{\bf 4: B3 0833+442}\\
In the CORALZ sample this source is classified as a CSO. However, the 1.6 GHz VLBI image \citep{Vries2009} shows a C-shaped radio structure. The LLS of the source is 1.7 kpc. \cite{Chandola} did not find \HI\ in this source. In our observations, we detect a narrow \HI\ profile, slightly redshifted from the systemic velocity. The data cube also reveals (faint) \HI\ emission.\\
\noindent
{\bf 5: B3 0839+458}\\
This source has been observed as part of the  Combined Radio All-Sky Targeted Eight GHz Survey (CRATES) sample \citep{Healey2007}. It has been classified as a point source with flat spectrum (spectral index $\alpha=-0.396$). The VLBA observations at $5$ GHz, as part of the VLBA Imaging and Polarimetry survey (VISP, \cite{Helmboldt2007}), classify it as a core-jet source. The lobes have sizes of approximately $3.5$ mas and are separated by $6$ mas ($=0.2$ kpc at $z=0.1919$). The radio power of the source is $P_{1.4\rm{GHz}}\sim3\cdot10^{25}$ W Hz$^{-1}$.  In our work, we detect the \HI\ at the systemic velocity of the galaxy ($\Delta v=57$~\kms). The line is narrow and deep, suggesting that the \HI\ is settled in a rotating disk around the host galaxy. The line has a slight asymmetry along the blueshifted edge not fitted by the BF. \\
\noindent
{\bf 6: Mrk 1226}\\
This source has been observed as part of the CRATES sample \citep{Healey2007}. It has been classified as a point source with flat spectrum (spectral index $\alpha=0.284$). The object has also been observed as part of the VISP survey at $5$ GHz \citep{Helmboldt2007}. The approximate size of the radio source is $15$ mas ($\sim8$ pc at $z=0.0279$).  We detect \HI\ absorption close to the systemic velocity of the host galaxy. This, along with the symmetry of the line, suggests that we are tracing neutral hydrogen rotating in a disk. The host galaxy may have experienced a gas rich merger which has formed the \HI\ disk.\\
\noindent
{\bf 7: UGC 05101 - IRAS F09320+6134}\\
This radio source is hosted by an ultra-luminous far-infrared (ULIRG) Galaxy. The galaxy is undergoing a merger event, as also suggested by its optical morphology. The source has also been observed with a $\sim11.6\times9.9$ mas resolution using VLBI ~\citep{Lonsdale2003}. These observations show three compact ($\lesssim3-4$ pc) cores connected by a fainter component. The size of the overall structure is $48\times24$ pc. These VLBI observations also show that the radio continuum is dominated by the AGN and not by the starburst activity. The radio power of the source is $P_{1.4\rm{GHz}}\sim3\cdot10^{23}$ W Hz$^{-1}$. At the resolution of our observations, we are not able to disentangle the absorption seen against the different components: we detect a broad, blended line. The profile is also multi-peaked, reflecting the unsettled state of the neutral hydrogen disk and of the overall host galaxy. The \HI\ has been detected in emission via Effelsberg Telescope observations, through the study of Polar-Ring Galaxy candidates \citep{PolarRing}. The emission line is broad and asymmetric, with a peak flux of $+2.2$ \mJybeam. The low sensitivity of the spectrum does not allow us to set further constraints. The detection in emission has been confirmed by observations with the Nan\c{c}ay decimetric radio telescope, with higher sensitivity \citep{vanDriel2000}.\footnote{The galaxy belongs to the HYPERLEDA catalogue \cite{Paturel2003}}\\
\noindent
{\bf 8: 4C +48.29}\\
This extended AGN is an X-shaped radio source \citep{Jaegers1987, Mezcua2011, Landt2010}. We detect a double-peaked \HI\ profile close to the systemic velocity. Before Hanning smoothing, the two peaks are more separated, suggesting the presence of two \HI\ components (one at the systemic and one blueshifted).\\
\noindent
{\bf 9: J105327+205835}\\
In the literature there are no records of individual observations of this radio source. The NVSS and FIRST images suggest that it is a compact source. In our observations, we detect a broad profile peaked at the systemic velocity and slightly asymmetric towards blueshifted velocities. The SDSS image shows that the host galaxy of this source is very close to a companion. Past interaction with a companion could explain the presence of the \HI\ in the system.\\
\noindent
{\bf 10: 2MASX  J112030+273610}\\
There are no individual radio observation of this source reported in the literature. According to our classification, it is a compact source. The detected \HI\ profile is narrow and blueshifted with respect to the optical velocity. This may indicate that we are tracing neutral hydrogen which is not settled in a rotating disk.\\
\noindent
{\bf 11: 2MASX J12023112+1637414} \\
This source has never been studied individually before. According to our classification it is a compact source, showing a shallow, blueshifted profile, indicative of outflowing gas. \\
%Araya, E. D.; R... VLBA Observations of H I in the Archetype Compact Symmetric Object B2352+495 \\
\noindent
{\bf 12: NGC 4093 - MCG +04-29-02}\\
VLA observations reveal compact radio morphology in this source \citep{Burns1987, Castillo1988}. We detect a regular \HI\ component at the systemic velocity, likely tracing the kinematics of a rotating disk.\\
\noindent 
{\bf 13: B3 1206+469}\\
This radio source has been selected as part of the Cosmic Lens All Sky Survey (CLASS) as a possible BL-Lac object and then classified as a lobe dominated steep spectrum source. This radio source is extended with a central core and two symmetric lobes oriented in the north-south direction. The distance between the lobes is $\sim4$ arcminutes ($\sim550$ kpc at $z=0.100$). The spectral index has been measured in the wavelength intervals $1.4-4.8$ GHz and $1.4-8$ GHz: $\alpha^{4.8}_{1.4}=0.04$, $\alpha^{8}_{1.4}=0.39$;~\citep{Marcha2001}. Being extended, the steepness of the spectrum can be explained by the fact that  some of the flux is missed by the VLA observations at $8.4$ GHz. In our observations, we detect a narrow and shallow absorption line close to the systemic velocity. \\
\noindent
{\bf 14: B2 1229+33}\\
This extended source was classified as an FR II by \cite{Cohen2004}. Based on the SDSS optical spectrum and image, it appears to be a High Excitation Radio Galaxy (HERG). Optically, the galaxy appears to be blue ($g - r  = $ 0.6), hence it has been excluded from the analysis of the overall sample presented in Sec.\ref{Discussion} and  Sec.\ref{sec:CompExt}. The \HI\ profile shows a narrow detection at the systemic velocity and, a second, redshifted component is also seen. These \HI\ properties suggest the presence of a disk and infalling gas in this object. \\
\noindent
{\bf 15: 4C +49.25} \\
The size of this CSS source is 6 kpc \citep{Saikia2003, Fanti2000}. The 5 GHz VLA map reveals a core and two jets on the opposite sides \citep{Saikia2003}. It was suggested by \cite{Saikia2003} that the higher degree of polarization asymmetry in CSS objects, including 4C +49.25, could be the result of interactions with clouds of gas which possibly fuel the radio source. Indeed, we find a blueshifted, shallow \HI\ component in this source. This could be the result of outflowing gas, induced by jet-ISM interactions. At the systemic velocity, however, we do not detect \HI. \\
\noindent
{\bf 16: 2MASX  J125433+185602}\\
The source belongs to the CRATES sample \citep{Healey2007}. It has been classified as a point source with flat spectrum (spectral index $\alpha=0.282$). Observations at  5 GHz, as part of the VISP survey (~\citealt{Helmboldt2007}), identify this source as a CSO. Its lobes are separated by $7.3$ mas ($\sim15$ pc at $z\sim0.0115$). We detect \HI\ at redshifted velocities compared to the systemic. The line is narrow and asymmetric, with a broader wing towards lower velocities. The redshift of the line, along with the compactness of the source, suggests that the neutral hydrogen may have motions different from simple rotation in a disk.\\
\noindent
{\bf 17: 2MASX J13013264+4634032} \\
According to \cite{Augusto2006}, this radio source is a point source. We detect a faint, blueshifted \HI\ profile, which can indicate interactions between the AGN and the surrounding gaseous medium. 
The radio source is a Blazar candidate in the CLASS and CRATES surveys \citep{CaccianigaOpt, Healey2007}. However, it remains classified as an AGN by \cite{CaccianigaOpt}.\\
\noindent
{\bf 18: B3 1315+415}\\
VLBI observations of the CORALZ sample~\citep{Vries2009} reveal complex radio morphology in this object. The source has a small size of LLS = 5 pc. From the lobe expansion speed analysis~\citep{Vries2010}, a dynamical age of $130$ yr is estimated in this source.~\cite{Chandola} detected \HI\ absorption redshifted by $+77$~\kms\ relative to the systemic velocity, indicating in-falling gas towards the nuclear region. Our observations confirm the \HI\ detection.\\
\noindent
{\bf 19: IC 883 - ARP 193 - UGC 8387}\\
The host galaxy of this radio source is undergoing a major merger. The galaxy is a luminous infrared galaxy (LIRG) where $L_{\rm{IR}}=4.7\cdot10^{11}L_\odot$ at $z=0.0233$ \citep{Sanders2003}. The radio source has been observed with e-Merlin (beam size $=165.23\times88.35$ mas) and VLBI e-EVN (beam size $=9.20\times6.36$ mas)~\citep{Romero2012}. The radio source consists of $4$ knots and extends for about $\sim750$ pc. The innermost 100 pc of the galaxy show both nuclear activity and star formation. The nuclear activity originates in the central core, while the radio emission from the other knots is attributed to transient sources.
This galaxy has already been observed in \HI\ in the study of Polar Ring galaxy candidates~\citep{PolarRing}. Two complementary observations have been performed using the Green Bank Telescope and the Effelsberg Telescope. Because of the different sensitivity of the instruments, the \HI\ has been detected in emission only in the Green Bank observations \citep{Richter1994}, with a peak flux $=2.4$ mJy. In IC 883, \couno~and \cotre~are detected by \cite{Yao2003} in the same range of velocities as the \HI\ emission. The resolution of our observations does not allow the disentanglement of different absorption components. Hence, the \HI\ line in our observations is blended, spanning the same velocity range of the \HI\ seen in emission, and of the molecular gas. The morphology of the absorption line, along with the overall properties of the cold gas detected in emission, suggests that in this galaxy the cold gas is rotating in a disk, which is unsettled because of the ongoing merger event.\footnote{The galaxy belongs to the HYPERLEDA catalogue \cite{Paturel2003}}\\
\noindent
{\bf 20: SDSS J132513.37+395553.2} \\
In the CORALZ sample this source is classified as a compact symmetric object (CSO), with largest (projected) linear size (LLS) of 14 pc \citep{Vries2009}. Our observations show two \HI\ components, one blueshifted, and the other redshifted relative to the systemic velocity. The newly detected \HI\ profiles suggest that unsettled gas structures are present in this galaxy, e.g., infalling clouds, outflowing gas.  \\ 
\noindent
{\b 21: IRAS F13384+4503} \\
This galaxy is optically blue ($g - r  = $ 0.6), and the SDSS image revels a Seyfert galaxy with late-type morphology. This object is not included in the analysis presented in Sec.\ref{Discussion} and  Sec.\ref{sec:CompExt}. In the CORALZ sample, the radio source is classified as a compact core-jet (CJ) source with two components which are significantly different in flux density and/or spectral index \citep{Vries2009}. The largest linear size of the source is 4.1 pc. Against the small continuum source, a very narrow \HI\ absorption profile is detected at the systemic velocity, indicative of a gas disk. \\
\noindent
{\bf 22: Mrk 273} \\
This object is the host of an ongoing merger. The optical morphology shows a long tidal tail extending 40 kpc to the south (Iwasawa et al. 2011, and references therein). Low-resolution 8.44 GHz radio maps by \cite{Condon1991} show three radio components, a northern (N), south-western (SW), and a south-eastern (SE) region. The origin of the SE and SW component is unclear \citep{Knapen1997, Carilli2000}. The N radio component is slightly resolved in the observations of \cite{Knapen1997, Carilli2000, Bondi2005}, showing two peaks embedded in extended radio emission. It is thought that the northern component is hosting a weak AGN, however it is also the site of very active star-formation. Using Very Long Baseline Array (VLBA) observations, \cite{Carilli2000} detected \HI\ absorption against the N component supposedly coming from a disk (showing velocity gradient along the major axis), and estimated an \HI\ gas mass of 2 $\times$ 10$^{9}$ M$_{\odot}$. Molecular CO gas of similar amount (10$^{9}$ M$_{\odot}$) was also detected by \cite{Downes1998}. \cite{Carilli2000} also detect extended gas and an infalling gas cloud towards the SE component, suggesting that the SE component is indeed an AGN. Our low-resolution observations cannot distinguish between the different absorbing regions, we detect the blended signal, coming from all the \HI\ absorbing regions. The broad \HI\ absorption was also detected with the single dish Green Bank Telescope (GBT) by \cite{Teng}. \\
\noindent
{\bf 23: 3C 293}\\
This object is a compact steep spectrum (CSS) radio source, whose emission is divided in multiple knots \citep{Beswick2004}. It is a restarted radio source, possibly activated by a recent merger event ~\citep{Heckman1986}. \cite{Massaro2010} classify the radio source as FRI. A rotating \HI\ disk has been detected in absorption by~\cite{Baan1981}. WSRT observations~\citep{Morganti2005} show an extremely broad absorption component at blueshifted velocities ($FWZI=1400$\kms). VLA-A array observations, with $1.2\times1.3$ arcsec of spatial resolution, identify this feature as a fast \HI\ outflow pushed by the western radio jet, located at $500$ pc from the core~\citep{Mahony2013}. The radio jet is thought to inject energy into the ISM, driving the outflow of \HI\ at a rate of $8-50$~\msunyr. The broad shallow outflowing component is also detected. The fit of the spectrum with the BF identifies the rotating component, while it fails in fitting the shallow wings, highlighting the different nature of these clouds.\\
\noindent
{\bf 24: 2MASX  J142210+210554}\\
There are no individual observations of this source available in the literature. In our classification, the radio source is compact. The SDSS observations show that it is hosted by an early-type galaxy. We detect an absorption line, blueshifted with respect to the systemic velocity. The line is broad and asymmetric with a smoother blueshifted edge.\\
\noindent
{\bf 25: 2MASX J14352162+5051233 }\\ 
This is an unresolved CORALZ AGN, the size of the radio source is estimated to be 270 pc. This galaxy has been observed in \HI\ by \cite{Chandola}, however no components were detected. Our observations show a shallow, broad, blueshifted \HI\ profile without deep/narrow component at the systemic velocity. Likely we are seeing gas interacting with the radio source. \\
\noindent
{\bf 26: 3C 305	- IC 1065}\\ 
The \HI\ profile of this source shows a deep, narrow component, which could be associated with rotating gas. Furthermore, \cite{Morganti2005} reported the presence of a jet-driven \HI\ outflow in this galaxy. The outflow is also detected in our observations, and it is successfully fitted by the BF. The column density of the outflow is N(\HI) = 2 $\times 10^{21}$, assuming $T_{\rm{spin}}$ = 1000 K, and the corresponding \HI\ mass was estimated to be M(\HI) = $1.3 \times 10^{7}$ M$_{\odot}$ \citep{Morganti2005}. X-ray observations suggest that the power supplied by the radio jet to the \HI\ outflow is $\sim 10^{43}$ erg s$^{-1}$ \citep{Hardcastle2012}. Molecular \HII\ gas was also detected in this source by \cite{Guillard2012}. However the molecular phase of the gas is inefficiently coupled to the AGN jet-driven outflow.\cite{Massaro2010} classified this source as a high excitation galaxy (HEG).
\\
\noindent
{\bf 27: 2MASX  J150034+364845}\\
In the literature, there is no record of targeted observations of this radio source. According to our classification it is a compact source. We detect a deep absorption line. The line lies at the systemic velocity of the host galaxy and traces a regularly rotating \HI\ disk. The line is slightly asymmetric in the blueshifted range of velocities. This asymmetry is not recovered by the BF fit, suggesting that non-circular motions characterize the neutral hydrogen.\\
\noindent
{\bf 28: 2MASX J15292250+3621423}\\
We find no individual observations of this source in the literature. In our sample it is classified as a compact source. We detect \HI\ in this object close to the systemic velocity. However, similarly to the case of source $\#9$ the profile is not entirely smooth.   \\ 
\noindent
{\bf 29: 4C +52.37} \\
This source is classified as a compact symmetric object (CSO) in the CORALZ sample.
High-resolution observations reveal a core, and jet-like emission on the opposite sides \citep{Vries2009}. The main \HI\ absorption component in 4C +52.37 was detected by \cite{Chandola}, using the Giant Metrewave Radio Telescope (GMRT). Besides the main \HI\ line, we detect a broad, shallow profile of blueshifted \HI\ absorption. The broad component was not detected by \cite{Chandola}, most likely because of the higher noise of the GMRT spectra. The kinematical properties of the newly detected blueshifted wing are indicative of a jet-driven \HI\ outflow in this compact radio source.\\
\noindent
{\bf 30: NGC 6034}\\
This radio source is hosted by a S0 optical galaxy, which belongs to cluster A2151 of the Hercules Supercluster. The radio source is extended, with two jets emerging toward the north and the south \citep{Mack1993}. The spectrum is flat with no variation of the spectral index ($\alpha=-0.65$). 
The line is very narrow and it is centred at the systemic velocity. This suggests that the \HI\ may form a rotating disk in the host galaxy. 
The neutral hydrogen in NGC 6034 has been first detected in absorption by VLA observations \citep{Dickey1997}.\footnote{the galaxy belongs to the HYPERLEDA catalogue \cite{Paturel2003}}\\
\noindent
{\bf 31: Abell 2147} \\
Based on the SDSS optical images, the host galaxy of this source is an early-type galaxy with a very red bulge. \cite{Taylor2007} classified this object as a flat-spectrum radio quasar. The size of the radio source is about 10 mas ($\sim$20 pc at z = 0.1), and the morphology remains unresolved in the 5 GHz VLBA images. Therefore, it is intriguing that we find a broad \HI\ detection against this very compact radio source. It is likely that along the line of sight the \HI\ has non-circular motions. \\
\noindent
{\bf 32: 2MASX  J161217+282546}\\
The radio source is hosted by an S0 galaxy and has been observed  with the VLA-A configuration by \cite{Feretti1994}. At the resolution of the VLA-A observations ($1.4\times1.1$ arcseconds, $\sim0.7$ kpc at z=0.0320), the radio source is unresolved. We detect an absorption line at the systemic velocity of the host galaxy, indicative of neutral hydrogen rotating in a disk.\\

\section{Summary table of non-detections}\label{Table:NonDet}

\begin{table*}
   \begin{center}
 \scalebox{0.7}{  
     \begin{tabular}{l c c c c c c c c c c c c}
                       
          \hline
                                                      
no. & RA,  Dec      &  z                                &  Other name  & S$_{\rm{1.4 \ GHz}} $ & P$_{\rm{1.4 \ GHz}}$  &  Radio Morphology & $\tau_{peak}$ & N(\HI)  \\        
         &                    &                                 &                    & mJy                             & W Hz$^{-1}$              &               &   &$ 10^{18}$ ($T_{\rm{spin}}/c_{\rm{f}}$) cm$^{-2}$ \\ \\
                           
\hline	
33&07h56m07.1s +38d34m01s&0.215605&B3 0752+387&70&24.98&C&< 	0.041&< 	7.4	\\
34&07h58m28.1s +37d47m12s&0.040825&NGC 2484&243&23.98&E&< 	0.009&< 	1.7	\\
35&07h58m47.0s +27d05m16s&0.098745&&69&24.23&C&< 	0.056&< 	10.2	\\
36&08h00m42.0s +32d17m28s&0.187239&B2 0757+32&104&25.01&E&< 	0.024&< 	4.4	\\
37&08h18m27.3s +28d14m03s&0.225235&&47&24.84&C&< 	0.047&< 	8.6	\\
38&08h18m54.1s +22d47m45s&0.095831&&194&24.65&E&< 	0.011&< 	2.1	\\
39&08h20m28.1s +48d53m47s&0.132447&&124&24.76&E&< 	0.026&< 	4.8	\\
40&08h29m04.8s +17d54m16s&0.089467&&190&24.58&E&< 	0.012&< 	2.1	\\
41&08h31m38.8s +22d34m23s&0.086882&&93&24.24&E&< 	0.022&< 	4.0	\\
42&08h31m39.8s +46d08m01s&0.131065&&123&24.75&C&< 	0.022&< 	4.1	\\
43&08h34m11.1s +58d03m21s&0.093357&&46&24.01&C&< 	0.057&< 	10.4	\\
44&08h39m15.8s +28d50m47s&0.078961&B2 0836+29&126&24.29&E&< 	0.023&< 	4.2	\\
45&08h43m59.1s +51d05m25s&0.126344&&79&24.52&E&< 	0.041&< 	7.5	\\
46&09h01m05.2s +29d01m47s&0.194045&3C 213.1&1670&26.25&E&< 	0.001&< 	0.3	\\
47&09h03m42.7s +26d50m20s&0.084306&&81&24.16&C&< 	0.025&< 	4.5	\\
48&09h06m15.5s +46d36m19s&0.084697&B3 0902+468&273&24.69&C&< 	0.013&< 	2.3	\\
49&09h06m52.8s +41d24m30s&0.027358&&56&22.99&C&< 	0.044&< 	8.0	\\
50&09h12m18.3s +48d30m45s&0.117168&&143&24.71&E&< 	0.028&< 	5.1	\\
51&09h16m51.9s +52d38m28s&0.190385&&63&24.81&E&< 	0.054&< 	9.8	\\
52&09h36m09.4s +33d13m08s&0.076152&&61&23.94&C&< 	0.030&< 	5.5	\\
53&09h43m19.1s +36d14m52s&0.022336&NGC 2965&104&23.07&C&< 	0.018&< 	3.2	\\
54&09h45m42.2s +57d57m48s&0.22893&&89&25.14&E&< 	0.030&< 	5.5	\\
55&10h09m35.7s +18d26m02s&0.116479&&44&24.19&E&< 	0.088&< 	16.0	\\
56&10h37m19.3s +43d35m15s&0.024679&UGC 05771&142&23.3&C&< 	0.013&< 	2.4	\\
57&10h40m29.9s +29d57m58s&0.090938&4C +30.19&407&24.93&C&< 	0.007&< 	1.4	\\
58&10h46m09.6s +16d55m11s&0.206868&&63&24.89&C&< 	0.039&< 	7.0	\\
59&11h03m05.8s +19d17m02s&0.214267&&99&25.12&E&< 	0.024&< 	4.4	\\
60&11h16m22.7s +29d15m08s&0.045282&&73&23.55&C&< 	0.025&< 	4.6	\\
61&11h23m49.9s +20d16m55s&0.130406&&102&24.66&E&< 	0.035&< 	6.3	\\
62&11h31m42.3s +47d00m09s&0.125721&B3 1128+472&100&24.62&E&< 	0.027&< 	4.9	\\
63&11h33m59.2s +49d03m43s&0.031625&IC 0708&168&23.59&E&< 	0.011&< 	2.1	\\
64&11h45m05.0s +19d36m23s&0.021596&NGC 3862&777&23.92&E&< 	0.003&< 	0.6	\\
65&11h47m22.1s +35d01m08s&0.062894&CGCG 186-048&276&24.42&E&< 	0.006&< 	1.2	\\
66&12h03m03.5s +60d31m19s&0.065297&SBS 1200+608&153&24.2&C&< 	0.012&< 	2.1	\\
67&12h13m29.3s +50d44m29&0.030757&NGC 4187&110&23.38&C&< 	0.019&< 	3.5	\\
68&12h21m21.9s +30d10m37s&0.183599&FBQS J1221+3010&60&24.76&C&< 	0.044&< 	8.0	\\
69&12h28m23.1s +16d26m13s&0.229959&&117&25.26&E&< 	0.023&< 	4.3	\\
70&12h30m11.8s +47d00m23s&0.039099&CGCG 244-025&94&23.52&C&< 	0.019&< 	3.5	\\
71&12h33m49.3s +50d26m23s&0.206843&&135&25.22&E&< 	0.015&< 	2.7	\\
72&12h52m36.9s +28d51m51s&0.195079&B2 1250+29&430&25.67&E&< 	0.005&< 	0.8	\\
73&13h03m46.6s +19d16m18s&0.06351&IC 4130&70&23.83&E&< 	0.038&< 	6.9	\\
74&13h06m21.7s +43d47m51s&0.202562&B3 1304+440&137&25.21&E&< 	0.021&< 	3.8	\\
75&13h08m37.9s +43d44m15s&0.035812&NGC 5003&61&23.26&C&< 	0.031&< 	5.6	\\
76&13h14m24.6s +62d19m46s&0.130805&&72&24.51&E&< 	0.042&< 	7.6	\\
77&14h00m26.4s +17d51m33s&0.050559&CGCG 103-041&88&23.73&C&< 	0.022&< 	4.0	\\
78&14h00m51.6s +52d16m07s&0.117887&&176&24.8&C&< 	0.021&< 	3.8	\\
79&14h08m10.4s +52d40m48s&0.082875&&176&24.48&E&< 	0.018&< 	3.4	\\
80&14h09m35.5s +57d58m41s&0.179874&&114&25.02&C&< 	0.035&< 	6.5	\\
81&14h11m49.4s +52d49m00s&0.076489&&393&24.75&E&< 	0.005&< 	1.0	\\
82&14h47m12.7s +40d47m45s&0.195146&B3 1445+410&348&25.58&E&< 	0.006&< 	1.1	\\
83&15h04m57.2s +26d00m54s&0.053984&VV 204b&190&24.12&E&< 	0.012&< 	2.2	\\
84&15h09m50.5s +15d57m26s&0.187373&&424&25.62&C&< 	0.005&< 	0.9	\\
85&15h16m41.6s +29d18m10s&0.129882&&72&24.51&E&< 	0.059&< 	10.7	\\
86&15h21m16.5s +15d12m10s&0.214777&&405&25.74&C&< 	0.007&< 	1.3	\\
87&15h23m49.3s +32d13m50s&0.10999&&182&24.75&C&< 	0.016&< 	3.0	\\
88&15h25m00.8s +33d24m00s&0.081559&&59&23.99&E&< 	0.032&< 	5.9	\\
89&15h32m02.2s +30d16m29s&0.065335&&33&23.53&C&< 	0.064&< 	11.7	\\
90&15h39m01.6s +35d30m46s&0.077837&&91&24.13&C&< 	0.019&< 	3.5	\\
91&15h49m12.3s +30d47m16s&0.111582&4C +30.29&914&25.47&E&< 	0.004&< 	0.8	\\
92&15h53m43.6s +23d48m25s&0.117606&4C +23.42&168&24.78&E&< 	0.034&< 	6.2	\\
93&15h56m03.9s +24d26m53s&0.042538&&92&23.59&E&< 	0.021&< 	3.9	\\
94&15h56m11.6s +28d11m33s&0.207927&&84&25.02&E&< 	0.024&< 	4.4	\\
95&15h59m54.0s +44d42m32s&0.04173&B3 1558+448&58&23.37&C&< 	0.036&< 	6.6	\\
96&16h04m26.5s +17d44m31s&0.040894&NGC 6040B&73&23.46&C&< 	0.040&< 	7.3	\\
97&16h06m16.0s +18d15m00s&0.0369&NGC 6061&225&23.85&E&< 	0.010&< 	1.8	\\
98&16h08m21.1s +28d28m43s&0.050182&IC 4590&113&23.83&E&< 	0.016&< 	3.0	\\
99&16h14m19.6s +50d27m56s&0.060258&&81&23.85&E&< 	0.025&< 	4.5	\\
100&16h15m41.2s +47d11m12s&0.198625&B3 1614+473&135&25.18&E&< 	0.023&< 	4.2	\\
101&16h24m24.5s +48d31m42s&0.057104&&67&23.72&C&< 	0.030&< 	5.5	\\
\end{tabular}}
    \caption{\HI\ non-detections. The N(\HI) upper limit is calculated from the 3-$\sigma$ rms, assuming a velocity width of 100 \kms.}\label{table:non-detections}
\end{center}
\end{table*}

\end{document}